\documentclass[sigconf, nonacm]{acmart}
\usepackage{graphicx}
\usepackage[ruled, vlined, linesnumbered]{algorithm2e}
\usepackage{balance}
\usepackage{caption}
\usepackage{subcaption}
\usepackage{multirow}	
\usepackage{enumerate}
\usepackage{bbm}
\usepackage{url}
\usepackage{xcolor}
\usepackage{xspace}
\usepackage{afterpage}

\newcommand{\nil}{\textit{NIL}}

\newcommand{\eg}{\emph{e.g.},\xspace}
\newcommand{\ie}{\emph{i.e.},\xspace}

\newcommand\figref[1]{Fig.~\ref{#1}}

\newcommand\tabref[1]{Table~\ref{#1}}

\newcommand\secref[1]{Section.~\ref{#1}}
\newcommand\equref[1]{Eq.~(\ref{#1})}

\newcommand\algoref[1]{Algo.~\ref{#1}}

\newcommand\theref[1]{Theorem~\ref{#1}}

\theoremstyle{definition}
\newtheorem{theorem}{Theorem}

\newtheorem{definition}{Definition}

\newcommand{\fakeparagraph}[1]{\vspace{1mm}\noindent\textbf{#1.}}

\ifodd 1

\else

\fi

\begin{document}
\title{Querying Shortest Path on Large Time-Dependent Road Networks with Shortcuts}


\author{Zengyang Gong \quad
        Yuxiang Zeng \quad 
        Lei Chen \quad}
\affiliation{
 \institution{The Hong Kong University of Science and Technology, Hong Kong SAR, China}
 \institution{zgongae@connect.ust.hk, \{yzengal, leichen\}@cse.ust.hk}
}

\begin{abstract}
Querying the shortest path between two vertexes is a fundamental operation in a variety of applications, which has been extensively studied over static road networks. However, in reality, the travel costs of road segments evolve over time, and hence the road network can be modeled as a time-dependent graph. In this paper, we study the shortest path query over large-scale time-dependent road networks. Existing work focuses on a hierarchical partition structure, which makes the index construction and travel cost query inefficient. To improve the efficiency of such queries, we propose a novel index by decomposing a road network into a tree structure and selecting a set of shortcuts on the tree to speed up the query processing. Specifically, we first formally define a shortcut selection problem over the tree decomposition of the time-dependent road network. This problem, which is proven to be NP-hard, aims to select and build the most effective shortcut set. We first devise a dynamic programming method with exact results to solve the selection problem. To obtain the optimal shortcut set quickly, we design an approximation algorithm that guarantees a 0.5-approximation ratio. Based on the novel tree structure, we devise a shortcut-based algorithm to answer the shortest path query over time-dependent road networks. Finally, we conduct extensive performance studies using large-scale real-world road networks. The results demonstrate that our method can achieve better efficiency and scalability than the state-of-the-art method.
\end{abstract}

\maketitle

\section{Introduction}	\label{sec:introduction}
Querying the shortest distance between two locations over static road networks has been extensively studied over the past decades 
\cite{ICDE2012, SIGMOD2013, ICDE2020, ICDE2021, h2h, p2h, GTree, GstarTree},
due to its wide applications \cite{DBLP:journals/pvldb/TongZZCYX18,DBLP:journals/tods/TongZZCX22,DBLP:journals/pvldb/ZengTSC20,TongJoS17}. However, in real road networks, the travel cost of a road segment varies at different times of the day. For example, traffic flow patterns of roads near the central business district (CBD) of a city may change dramatically between day and night. Therefore, many applications place time-dependent travel cost functions on edges to represent travel cost, which leads to the study of shortest path queries over formalized time-dependent graphs  \cite{shortestquery}.

On the one hand, a straightforward approach is to use non-index methods. $\textit{Dijkstra-}$ or $\textit{A*-}$ based algorithms 
\cite{nonindex1, nonindex2, DBLP:conf/icde/KanoulasDXZ06, nonindex4}
visit the travel cost functions of edges during the traversal. Thus for each query, these methods need more than several seconds to get the final result. For example, the TWO-STEP-LTT algorithm \cite{nonindex2} takes about 10 seconds to answer the query over the road network with 16,326 vertices and
26,528 edges. The unsatisfied processing performance of non-index methods can hardly meet the real-time requirements for shortest path queries, especially when the time-dependent road network is large-scale and the travel pattern between queried vertexes changes rapidly.

On the other hand, index-based techniques are proposed to make up for the query deficiency. TD-G-tree \cite{shortestquery} is one of the state-of-the-art solutions that belong to index-based algorithm family and outperforms others significantly. Intuitively, TD-G-tree splits a time-dependent road network in a hierarchical way, each tree node corresponds to a partition and maintains a travel-time matrix, and the shortest travel cost functions between all pairs of borders of this partition are cached in the matrix. When a query is issued, TD-G-tree traverses the cached border information in an assembly-based way from the bottom to the top of the tree structure. Obviously, the performance of TD-G-tree heavily depends on 
the number of visited tree nodes and the number of borders associated with each node. However, the assembly-based methods could perform even worse than the basic Dijkstra algorithm if the queried vertexes are close to each other in the road network where are far away in the index \cite{GstarTree}. Besides that, TD-G-tree also suffers from data redundancy and expensive construction problems. Due to the hierarchical partitions structure, a tree node is clearly a sub-graph of its parent node and the optimal travel cost function of each partition's border pairs may not always be in the same partition. Considering road networks in the real world are typically more than millions of edges, in the meanwhile, the travel pattern changes frequently, and TD-G-tree could be inefficient under a large-scale time-dependent road network. For example, in our experiments, TD-G-tree takes about 4 hours to construct the index, and more than 0.5 second to answer the query over the road network of Florida with 1,070,376 vertices and 2,712,798 edges.

In this work, to address the weakness of the existing methods, we study applying tree decomposition based index to time-dependent road networks for answering shortest path queries efficiently. Tree decomposition based methods have achieved great success in querying the shortest path on static road networks \cite{h2h, p2h, lsd}, because it can decompose a road network into a tree-like structure, and at the same time, the treeheight and treewidth are both small even if the road network is very large.

However, extending the tree decomposition technique into the time-dependent road network is non-trivial. The biggest challenge is to select and build shortcuts between each tree node to several ancestors of it. Since storing shortcuts is quite costly, we need to choose shortcuts wisely to maximize the usage of limited in-memory space. In this paper, we first prove that our shortcut selection problem over tree decomposition is NP-hard. Then, we aim to select and build the most effective shortcuts to accelerate query processing over the tree decomposition. Finally, we design solutions with theoretical guarantees and also propose a novel shortest path query algorithm based on selected shortcuts. 

We summarize our contributions in this work as follows:
\begin{itemize}
  \item
  We apply the tree decomposition method to support the shortest path query over time-dependent road networks. We first decompose a time-dependent road network into a tree structure. Meanwhile, the travel cost information is preserved in this tree.
  \item
  We formulate the shortcut selection problem over the tree structure to accelerate the query processing. It chooses the most effective shortcuts under limited space cost. We also prove this problem is NP-hard.
  \item
  To solve the shortcut selection problem, we design a dynamic programming based selection algorithm, and propose an effective and efficient approximation algorithm with an approximation ratio of 0.5.
  \item
  Based on selected shortcuts, we design a novel algorithm to answer shortest path queries which is more efficient than the state-of-the-art algorithms. 
  \item
  Extensive experiments on real-world large-scale road networks demonstrate the efficiency of our solution outperforms the state-of-the-art algorithms by a large margin in terms of query responding.
\end{itemize}

The rest of this paper is organized as follows. We first present the problem statement in Section \ref{sec:problem}. Then, our algorithms are elaborated in Section \ref{sec:methodology}. We evaluate our algorithms in Section \ref{sec:experiment}. We review the related works in Section \ref{sec:related} and conclude in Section \ref{sec:conclusion}.


\section{Problem Statement}\label{sec:problem}

\begin{table}[t]
	\centering
	\caption{Summary of major notations.}
	\vspace{-2ex}
	\label{table:notations}
    \begin{small}
    \resizebox{0.45\textwidth}{!} {%
	\begin{tabular}{|c|c|}
		\hline
		Notation & Description \\
		\hline
		$G(V,E,W)$ & Time-dependent road network \\
		$w_{u,v}(t)$ & Weight function of edge $e_{u,v}$ \\
		$f_{s,d}(t)$ & Shortest Travel cost function from $s$ to $d$ \\
		$Compound()$ & Compound operator function \\
		\hline
		$T_G$ & Tree decomposition of $G$ and tree node \\
		$X(i)$ & Tree node of $T_G$ \\ 
		$w(T_G), h(T_G)$ & Treewidth and Treeheight of $T_G$ \\
		$s_{\langle i, j \rangle}(t)$ & Shortcut between $X(i)$ to $X(j)$ \\
		\hline
	\end{tabular}
	}
	\end{small}
  \vspace{-2ex}
\end{table}

\begin{figure}[t]
	\centering
    \begin{subfigure}[b]{0.36\textwidth}
		\includegraphics[width=\textwidth]{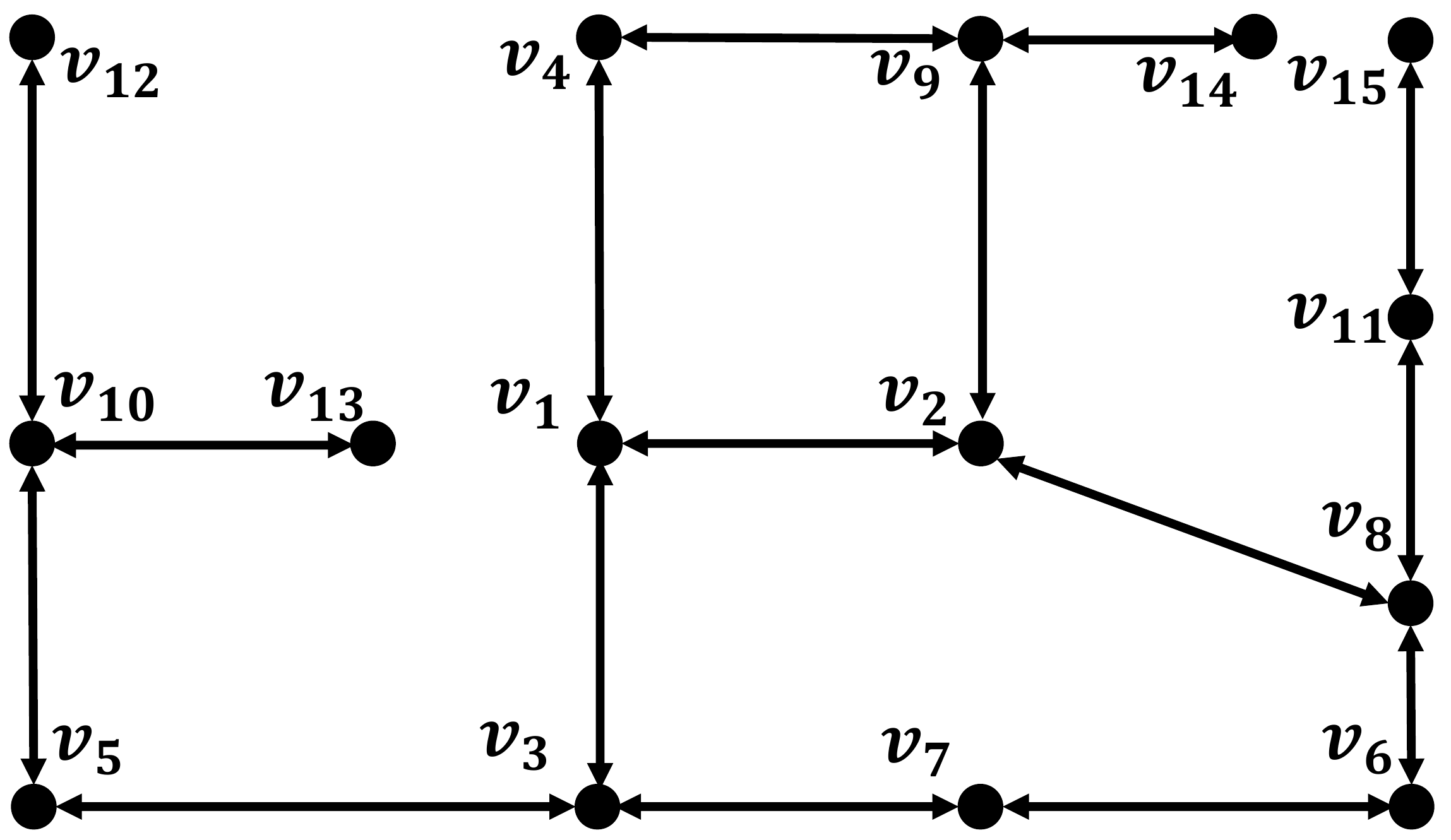}
		\vspace{-4ex}
 		\caption{\footnotesize{Road network}}
		\label{fig: example1a}
	\end{subfigure}
	
	\begin{subfigure}[b]{0.35\textwidth}
		\includegraphics[width=\textwidth]{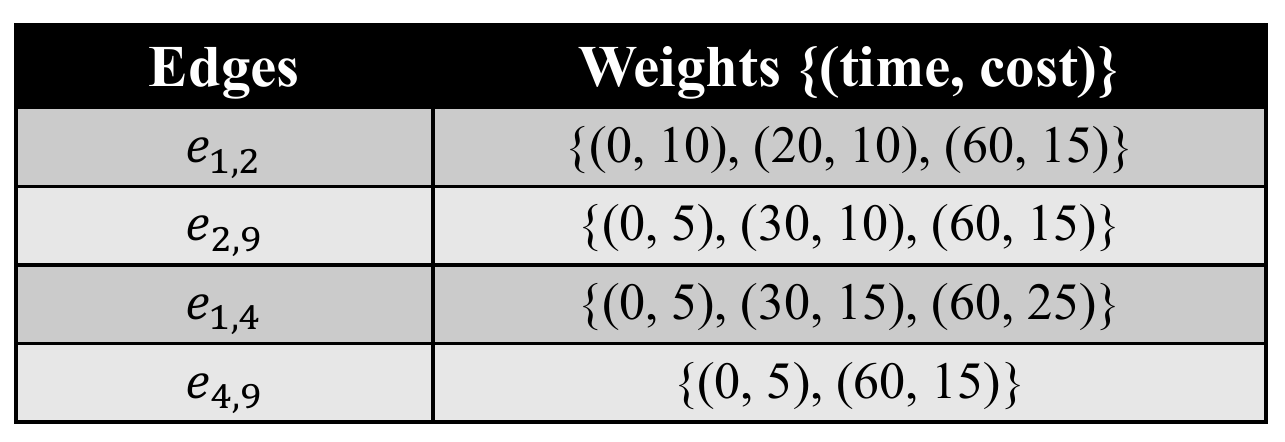}
		\vspace{-4ex}
		\caption{\footnotesize{Weights of $e_{1,2}, e_{2,9}, e_{1,4}$ and $e_{4,9}$}}
		\label{fig:example1b}
	\end{subfigure}
	\vspace{-2ex}
	\caption{Time-dependent road networks.($w_{u,v}(t) = w_{v,u}(t)$)}
	\label{fig:example1}
	\vspace{-3ex}
\end{figure}

In this section, we first give a definition of the time-dependent directed graph which is utilized to model the road network and then the travel cost function of the path in the graph is formulated. Finally, we give the problem definition. Major notations are summarised in \tabref{table:notations}.

\begin{definition}[Time-dependent Directed Graph]
Let a directed graph $\textit{G(V,E,W)}$ model a time-dependent road network where $V$ is the set of vertexes and $E$ is the set of edges. We use $n = |V|$ and $m = |E|$ to denote the number of vertexes and edges respectively. Specifically, a weight function $w_{u,v}(t) \in \textit{W}$ is associated with the edge $e_{u,v} \in \textit{E}$, the value of this function denotes the travel cost from $u$ to $v$ when starting at time $t$, which is always non-negative. Following the previous work \cite{shortestquery, yuanyeicde2019, yuanyeicde2021},
each function $w_{u,v}(t)$ is represented by a set of interpolation points $I_{\langle u,v \rangle} = \{(t_1, c_1), (t_2, c_2), \dots, (t_k, c_k)\}$.
\end{definition}
Then the weight function associated with $e_{u,v}$ can be formalized as    
\begin{equation}
\label{equ:weightFunction}
    w_{u,v}(t) = \begin{cases}
                    c_1, & t = t_1 \\
                    c_1 + (t - t_1) \frac{c_2 - c_1}{t_2 - t_1}, & t_1 < t \leq t_2\\
                    \cdots\\
                    c_{k-1} + (t - t_{k-1}) \frac{c_k - c_{k-1}}{t_{k} - t_{k-1}}, & t_{k-1} < t \leq t_k\\
                    c_k, &   t = t_k
                    \end{cases} 
\end{equation}
In \equref{equ:weightFunction}, $t_1$ and $t_k$ are the earliest departure time from the beginning of the edge and the last available start time, respectively.

\fakeparagraph{Example 2.1} 
\textit{ \figref{fig: example1a} shows a time-dependent road network with 15 vertexes and 17 edges. \figref{fig:example1b} shows the weight function of four edges on the path from $v_1$ to $v_9$. For edge $e_{1,2}$, the weight function $w_{1,2}(t)$ is fit by 3 (time, cost) pairs, $I_{\langle 1, 2 \rangle} = \{(0,10), (20,10), (60,15)\}$. Consider pair $(0, 10)$, it indicates that, at time 0, it takes 10 minutes to travel from $v_{1}$ to $v_2$.} 

\begin{definition}[Shortest Travel Cost Function]
Consider a path $(e_{1,2}, e_{2,3}, \dots$ $e_{k-1,k})$ travel from $v_1$ to $v_k$ with the minimal cost, it is a sequence of edges $e_{i-1,i} \in E$ for all $0 \leq i \leq k$. The shortest travel cost function for the path, starting at time $t$, denoted by $f_{1,k}(t)$, is recursively defined as $f_{1,k}(t) = Compound(f_{1,k-1}(t), f_{k-1,k}(t))$, where $f_{1,k-1}(t)$ is the shortest travel cost function of sub-path from $v_1$ to $v_{k-1}$, $f_{k-1,k}(t)$ is the weight function of edge $e_{k-1,k}$, \eg{$f_{v_{k-1},v_k}(t) = w_{v_{k-1},v_k}(t)$}. $Compound()$ is defined as the function that calculates the travel cost function of the connected path and the intermediate vertex is also recorded in the function, \eg{  $Compound(f_{1,k-1}(t),$ $f_{k-1,k}(t)) = f_{k-1,k}(t+f_{1,k-1}(t))$, and intermediate vertex $k-1$ is recorded}.

\end{definition}

\begin{figure}[t]
	\centering
	\begin{subfigure}[b]{0.35\textwidth}
		\includegraphics[width=\textwidth]{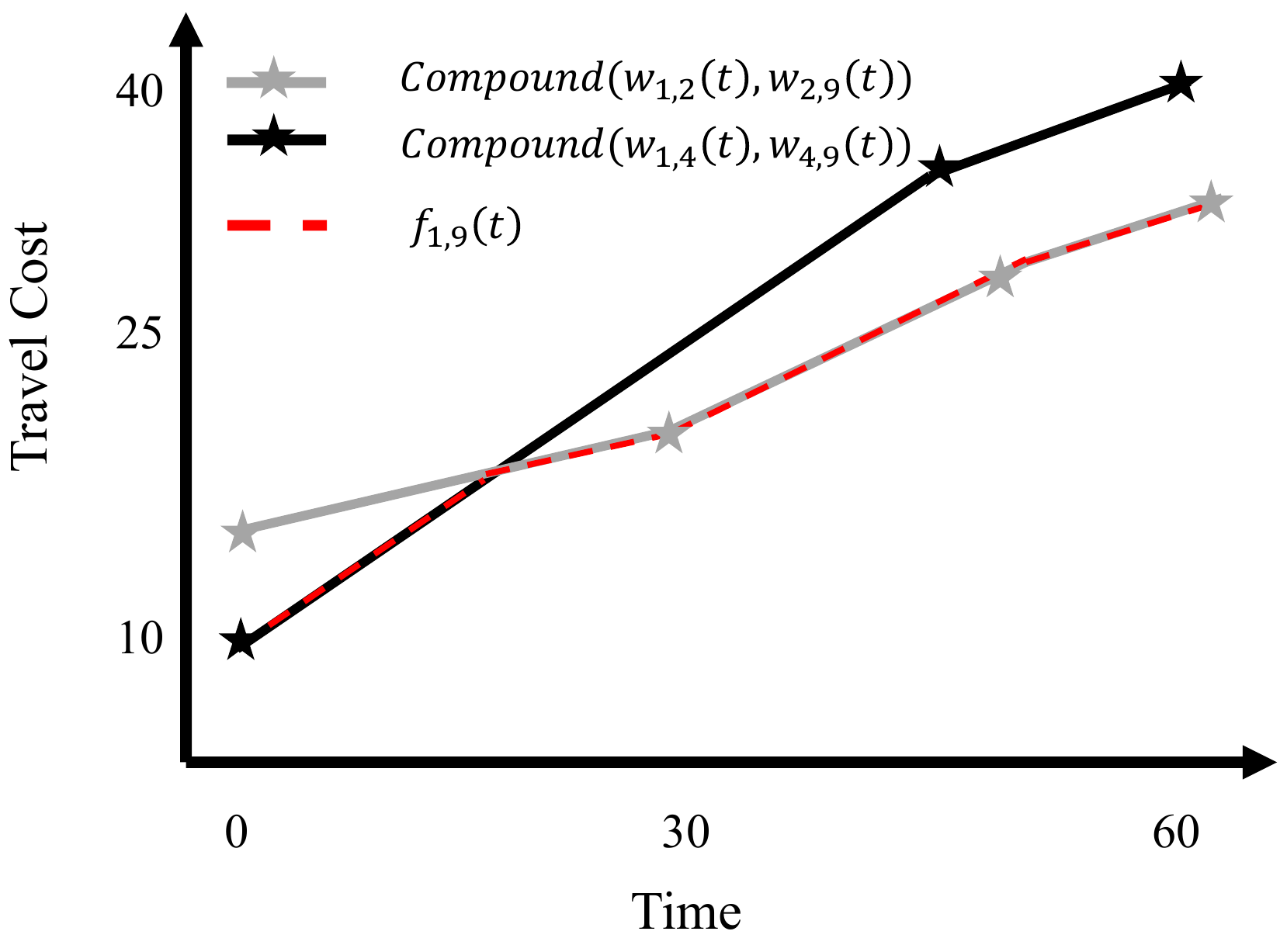}
		\vspace{-4ex}
	\end{subfigure}
	\vspace{-2ex}
	\caption{Shortest Travel Cost Function from $v_1$ to $v_9$}
	\label{fig:example2}
	\vspace{-3ex}
\end{figure}

\fakeparagraph{Example 2.2} 
\textit{ From $v_1$ to $v_9$, there exists two paths in \figref{fig: example1a}, ($(e_{1,4},e_{4,9})$ and $((e_{1,2},e_{2,9})$). \figref{fig:example2} shows the compounded travel cost function between these two vertices. For $(e_{1,4},e_{4,9})$, the travel cost along this path is calculated by compounding weight functions of these two edges, 
$Compound(w_{1,4}(t),w_{4,9}(t)))$. Similarly, the travel cost for path $(e_{1,2},e_{2,9})$ is $Compound(w_{1,2}(t),w_{2,9}(t)))$. Therefore, from $v_1$ to $v_9$, the shortest travel cost function $f_{1,9}(t) = min \{ Compound(w_{1,4}(t),$ $ w_{4,9}(t)))$ $, Compound(w_{2,9}(t),w_{1,2}(t))) \}.$}

\fakeparagraph{Problem Definition \textbf{(Shortest Path Queries over the Time-dependent Road Network)}}  Given a time-dependent road network $G$, two query vertices and a specific start time $Q(s,d,t)$, the shortest path query problem aims to find the minimum cost from vertex $s$ to vertex $d$ when starting at time $t$ and the associated shortest travel cost function $f_{s,d}(t)$.

\fakeparagraph{Example 2.3} 
\textit{ Considering the path from $v_1$ to $v_9$ in \figref{fig: example1a}, as the shortest travel cost function $f_{1,9}(t)$ shown in \figref{fig:example2}, we can get that, at the beginning the shortest path is $(e_{1,4},e_{4,9})$ and the intermediate vertex is recorded as $v_4$, and as time goes the travel cost of path $(e_{1,2},e_{2,9})$ is much lower than the previous one. Thus the shortest path from $v_1$ to $v_9$ will change to $(e_{1,2},e_{2,9})$, and the intermediate vertex is updated to $v_2$ in $f_{1,9}(t)$.}

\section{Our Tree Decomposition based shortest path query algorithm}\label{sec:tree}
In this section, we first introduce the tree decomposition approach over time-dependent road networks, which organizes the vertices in a tree structure efficiently. Then, based on the properties of tree decomposition, we present a basic query algorithm based on tree decomposition for the shortest path queries over time-dependent road networks.

\subsection{Tree Decomposition} \label{subsec:tree decomposition}
Tree decomposition \cite{treedecomposition} is an efficient approach to solving static graph related problems \cite{tlindex, h2h, p2h, lsd}, since it can decompose a large-scale graph into a tree-like structure. The formal definition is given in the following. 

\begin{definition}[Tree Decomposition]
A tree decomposition of the time-dependent graph $\textit{G (V,E,W)}$ is denoted as $T_G$, it is a rooted tree with a tree node set $\{X(v_1), X(v_2) \dots X(v_n)\}$. For each vertex $v$ of the graph, there is a tree node denoted as $X(v)$, and each tree node is a subset of $V$ (\ie{ $X(v) \subseteq V$}). Thus $T_G$ satisfies the following properties:
\begin{enumerate}
        \item $X(v_1) \cup X(v_2) \cup \dots X(v_n) = V$. 
        \item $\forall e_{u,v} \in E$, $\exists X(v_i)$ such that $u \in X(v_i)$ and $v \in X(v_i)$.
        \item $\forall v_i \in V$, $\{ X(v_i)|v \in X(v_i) \}$ is a connected subtree of $T_G$.
\end{enumerate}
\end{definition}

For a given graph, building a tree decomposition with the minimized treewidth is an NP-Complete problem \cite{tlindex}. Therefore, a sub-optimal tree decomposition \cite{treedecomposition} method is adopted in this paper. The time complexity of this method is $O(n \cdot (w^2 + logn))$, this method has been used in several research works on static road network \cite{h2h, lsd, tlindex}. In each iteration, the algorithm greedily picks the vertex with the smallest degree and creates a corresponding tree node. We can get the tree decomposition of the given graph after processing all vertices, therefore, there is a one-to-one correspondence from graph vertices to tree nodes. Given a vertex $v$, $X(v)$ denotes the corresponding tree node in $T_G$. For each tree node $X(v)$, $height(X(v))$ and $Anc(X(v))$ denote the height and the set of ancestors of $X(v)$ in the tree decomposition respectively. 

\begin{figure}[t]
	\begin{subfigure}[b]{0.4\textwidth}
		\includegraphics[width=\textwidth]{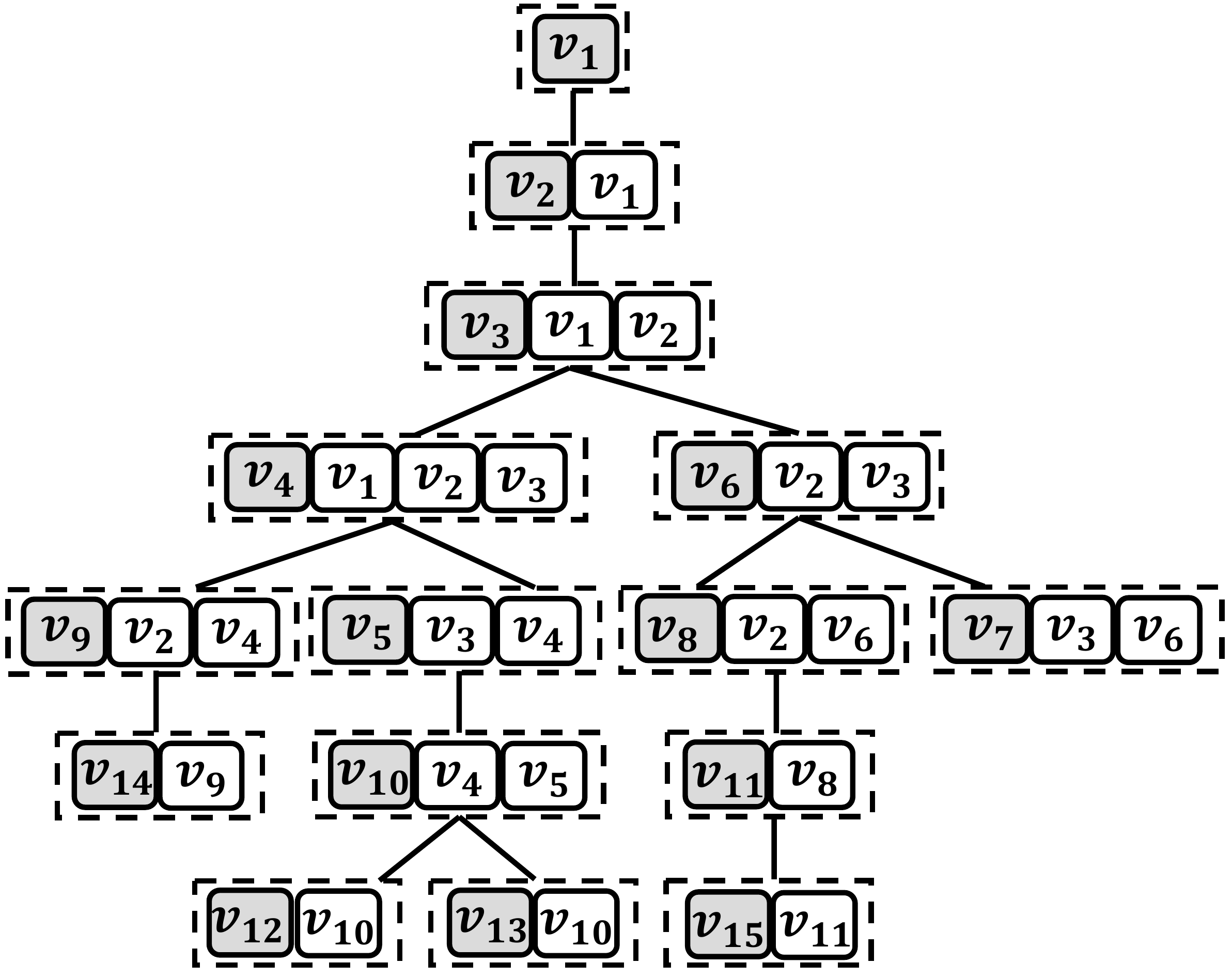}
		\vspace{-4ex}
	\end{subfigure}
	\vspace{1ex}
	\caption{Tree Decomposition of $G$}
	\label{fig:TD}
	\vspace{-3ex}
\end{figure}

\fakeparagraph{Example 3.1} 
\textit{ \figref{fig:TD} shows the tree decomposition $T_G$ for the road network in \figref{fig: example1a}. $T_G$ has 15 nodes. In each elimination step, we pick the vertex with the lowest degree and then create the corresponding tree node containing its neighbours. The elimination order is $\{ v_{15}, v_{11}, v_{14}, v_{13}, v_{12},$ $v_{10}, v_{9}, v_{8}, v_{7}, v_{6}, v_{5}, v_{4}, v_{3}, v_{2}, v_{1} \}$. For example, consider the vertex $v_{10}$ the corresponding tree node of is $X(v_{10}) = \{ v_{10}, v_4, v_5 \}$. The set of ancestors of  $X(v_{10})$ is $Anc(X(v_{10}))$ = $(v_1, $ $ v_2, $ $ v_3, v_4, v_5)$, and the height of the tree node $height(X(v_{10})) = 6$.
}

\begin{definition}[Treewidth and Treeheight]
Given a generated tree $T_G$, the treewidth and tree height are denoted as $w(T_G)$ and $h(T_G)$ respectively. The treewidth $w(T_G)$ is one less than the maximum size of all tree nodes in $T_G$, $w(T_G) = \max_{X(v) \in T_G} |X(v)|-1 $. The treehight $h(T_G)$ is the maximum height of all tree nodes in $T_G$, $h(T_G) = \max_{X(v) \in T_G} height(X(v)) $, where the height $height(X(v))$ denotes the distance of $X(v)$ to the root node.

\end{definition}

\fakeparagraph{Example 3.2} 
\textit{ For the tree decomposition shown in \figref{fig:TD}, the treewidth $(w(T_G))$ of $T_G$ is 3 since tree node $X(v_4)$ contains at most 4 vertices, the treeheight $h(T_G)$ is 7 since the tree nodes $X(v_{12})$, $X(v_{13})$ and $X(v_{15})$ have the maximum height 7.}

\subsection{Tree Construction using TFP-Graph}
In order to design the query algorithm based on the tree structure, we first give the definition of the shortest travel cost function preserved graph in the following.

\begin{definition}[TFP-Graph]
\label{def:TFP-Graph}
Given a time-dependent graph $G(V,E,W)$, a graph $G'(V',E',W')$ is called a Travel cost Function Preserved Graph (TFP-Graph), if $V' \subseteq V$, and for any pair of vertices $u \in V'$ and $v \in V'$, when $e_{u,v} \in G'$ associated weight function $w'_{u,v}(t) = f_{u,v}(t)$, otherwise $f'_{u,v}(t)$ denotes the shortest travel cost function of the path from $u$ to $v$ in $G'$ and $f'_{u,v}(t) = f_{u,v}(t)$. Here, we use $G' \sqsubseteq G $ to denote that $G'$ is a TFP-Graph of original graph $G$.
\end{definition}

\begin{algorithm}
\caption{Reduction Operator $\ominus$}
\label{alg:reduction}
\LinesNumbered 
\KwIn{Time-dependent graph $G (V,E,W)$ and vertex $v$}
\KwOut{TFP-Graph $G'$}
$G' \gets G$ \;
\For{$v_i \in N(v)$}{
    \For{$v_j \in N(v)$}{
        \textbf{if} $i \neq j$ and $e_{i,j}, e_{j,i} \notin G$ \textbf{then} \
        insert both edge $e_{i,j}$, $e_{j,i}$ and associated weight functions $w'_{i,j}(t) \gets Compound(w_{i,v}(t),w_{v,j}(t))$, \
        $w'_{j,i}(t) \gets Compound(w_{j,v}(t),w_{v,i}(t))$
        into $G'$; \
        
        \textbf{else}\
        
        update $w_{i,j}(t)$ and $w_{j,i}(t)$ in $G'$ to \\ 
        $w'_{i,j}(t) \gets min \{Compound(w_{i,v}(t),w_{v,j}(t)), w_{i,j}(t)\}$ \\
        $w'_{j,i}(t) \gets min \{Compound(w_{j,v}(t),w_{v,i}(t)),w_{j,i}(t)\}$;

    }
}
\KwRet{$G'(V',E',W')$}\;
\end{algorithm}

\fakeparagraph{Time-dependent Graph Reduction}
Let the operation $G \ominus v$ denote the graph reduction operator for vertex $v$ in a time-dependent graph $G$. \algoref{alg:reduction} shows the algorithm for the details of the $\ominus$ operator. This operation removes $v$ from $G$ and processes the edges associated with this vertex, and thus it transforms $G$ into another TFP-Graph $G'$. The detailed procedures are introduced as follows. Given the graph $G$ and vertex $v$, let $N(v)$ denote the neighbour vertices of $v$. For every pair of neighbours $v_i, v_j \in N(v)$, if edges $e_{i,j}$ and $e_{j,i}$ do not exist in $G$, new edges $e_{i,j}$ and $e_{j,i}$ which take $v$ as the bridge vertex are built and inserted into $G'$. Thus, the weight function of $e_{i,j}$ is calculated by $e_{i,v}$ and $e_{v,j}$, $w'_{i,j}(t) = Compound(w_{i,v}(t),w_{v,j}(t))$. Otherwise, the weight of edge $e_{i,j}$ is updated as $w'_{i,j}(t) = min \{Compound(w_{i,v}(t),w_{v,j}(t)), w_{i,j}(t)\}$. For $e_{j,i}$, which has a different direction with $e_{i,j}$, $w'_{j,i}(t) = Compound$ $(w_{j,v}(t),w_{v,i}(t))$ if $e_{j,i}$ is not an edge of $G'$. Otherwise $w'_{j,i}(t) = min \{Compound$ $(w_{j,v}(t),w_{v,i}(t)), w_{j,i}(t)\}$.

\begin{algorithm}
\caption{TFP Tree decomposition}
\label{alg:tree decomposition}
\LinesNumbered 
\KwIn{A time-dependent graph $G(V,E,W)$}
\KwOut{TFP tree decomposition $T_G$}
$G' \gets G; T_G \gets \emptyset $ \;
\For{$i \gets $ 1 to $n$}{
    $v \gets $ pick the vertex with the smallest degree in $G'$ \;
    $X(v) \gets$ add $v$ \;
    \For{$u \in N(v)$}{
        $X(v) \gets$ add $u$ \;
        $X(v).W^s_{u} \gets w'_{v,u}(t)$, $X(v).W^d_{u} \gets w'_{u,v}(t)$  \;
    }
    $G' \gets G' \ominus v$ \;
    $\pi(v) \gets i$ \;
}
\For{ $v \in V$}{
    \If{$|X(v)| > 1$}{
    $u \gets$ the vertex in $X(v)\backslash \{v\}$ with smallest order value given in $\pi(u)$ \;
    Node $X(u)$ is set as the parent node of $X(v)$ in $T_G$ \;
    }
}
\KwRet{$T_G$}\;
\end{algorithm}

\begin{figure}[t]
	\centering
	\begin{subfigure}[b]{0.33\textwidth}
		\includegraphics[width=\textwidth]{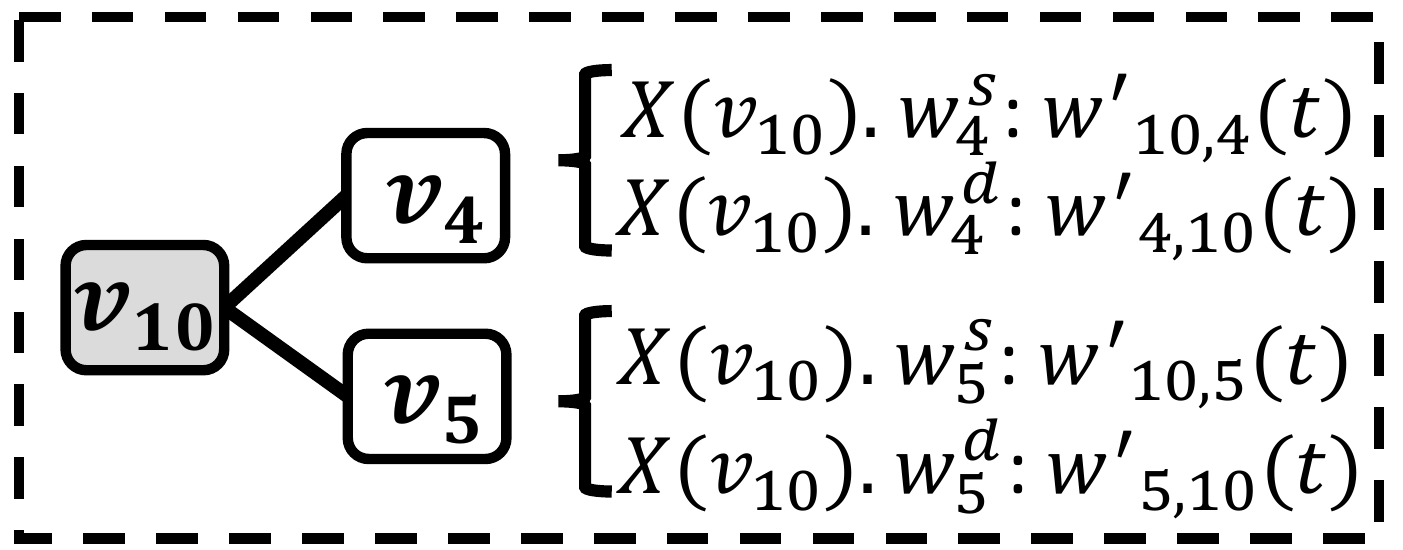}
		\vspace{-4ex}
	\end{subfigure}
	\vspace{-1ex}
	\caption{Travel functions preserved in $X(v_{10})$}
	\label{fig:TFPExmaple}
	\vspace{-3ex}
\end{figure}

\fakeparagraph{Travel Function Preserved Tree Decomposition} For a time-dependent graph, we want to organize vertices in the tree structure introduced in \secref{subsec:tree decomposition}, at the meantime, the weight functions between $v$ to each other vertex in $X(v)$ are preserved. For example, \figref{fig:TFPExmaple} shows weight functions preserved in tree node $X(v_{10})$. There are two function lists $X(v_{10}).W^s$ and $X(v_{10}).W^d$. The list $X(v_{10}).W^s$ preserves two weight functions start from $v_{10}$ to $v_4$ and $v_5$ respectively. The weight functions from $v_4$ and $v_5$ to $v_{10}$ are preserved in the list $X(v_{10}).W^d$. We can apply a graph reduction operator over each vertex until all vertices are removed to achieve this aim. We call the procedure Travel Function Preserved (TFP) Tree Decomposition. 

\algoref{alg:tree decomposition} presents the details of this procedure. In line 1, \algoref{alg:tree decomposition} picks vertex $v$ with the smallest degree and builds the associated tree node $X(v)$ in each iteration. $X(v)$ contains not only vertex $v$, but also all neighbour vertices in lines 4-6. Besides that, each tree node $X(v)$ also stores information related to travel cost in line 7, a list $X(v).W^s$ is utilized to preserve the weight functions from $v$ to other vertices in $X(v)$. Another list $X(v).W^d$ preserves the weight functions from vertices in $X(v)$ to $v$. Then, in line 8, we apply the time-dependent graph reduction operator on $v$ to eliminate this vertex and update the edges associated with it, and in line 9 the process order of $v$ is recorded in $\pi(v)$. Based on the elimination order of vertices, the tree structure is built in lines 10-13. For each tree node $X(v)$, $X(v)$ is not the root node if $|X(v)|$ is larger than 1. When $X(v)$ is a non-root node, we assign $X(u)$ as the parent node of $X(v)$, if $u$ in $X(v)\backslash\{v\}$ and has the smallest $\pi$ value.

\fakeparagraph{Complexity Analysis} For time complexity, given any picked vertex $v$ in line 8 of \algoref{alg:tree decomposition}, one reduction operator is invoked in \algoref{alg:reduction}. For each reduction operator, lines 2-3 of \algoref{alg:reduction} take $O(w(T_G)^2)$ time to process the neighbour vertices, then $Compound()$ function in line 4 and line 6 takes $O(c)$ time to compute the new weight function, where $c$ is a constant parameter to indicates the average number of $(time, cost)$ pairs of edges. Therefore, the time cost of one reduction operator in line 8 of \algoref{alg:tree decomposition} is $O(w(T_G)^2 \cdot c)$. Lines 2-4 of \algoref{alg:tree decomposition} take $O(m+ n\cdot log(n))$ time to pick the vertex with the smallest degree, thus the overall time complexity of \algoref{alg:tree decomposition} is $O( n \cdot (w(T_G)^2 \cdot c + log(n)))$. For space complexity, each weight function takes $O(c)$ space, and each tree node maintains $O(w(T_G))$ weight functions, therefore the space complexity of \algoref{alg:tree decomposition} is $O(n \cdot w(T_G) \cdot c)$.

\subsection{Basic Query Algorithm}
In this section, we introduce our basic query algorithm based on the Travel Function Preserved Tree Decomposition of a time-dependent graph. Before presenting the details, we first introduce several properties that the tree structure holds.

\textbf{\textsc{Property 1}} \textit{If $X$ is the lowest common ancestor (LCA) of $X(s)$ and $X(d)$ in $T_G$, then $X$ is a vertex cut of $s$ and $d$ in $G$.}

Refer to property 1 and definition 4.7 in \cite{h2h}. Given a tree decomposition $T_G$, if there are two query vertices $s$ and $d$ in $G$, suppose the associated tree node $X(s)$ is not an ancestor or descent node of $X(d)$ in $T_G$, let $VC$ denote the lowest common ancestor (LCA) of $X(s)$ and $X(d)$ in $T_G$, then $VC$ is a vertex cut of $s$ and $d$ in $G$. The vertex cut $VC$ means that, for any path from $s$ to $d$, it should contain at least one vertex of $VC$. Therefore we can calculate the shortest travel cost function from $s$ to $d$ as: $f_{s,d}(t) = min_{v \in VC} \{ Compound(f_{s,v}(t), f_{v,d}(t)) \}$.

\textbf{\textsc{Property 2}} \textit{For any tree node $X(v) \in T_G$, $X(v)\backslash \{v\} \subseteq Anc(X(v))$}.

Refer to property 2 in \cite{h2h}. Given a tree decomposition $T_G$ and each tree node $X(v) \in T_G$, for any $u \in X(v)\backslash \{v\}$, $X(u)$ is an ancestor of $X(v)$.

\textbf{\textsc{Property 3}} \textit{For any vertex $v \in V$, let $G(v)$ denotes the graph which contains the vertices from $X(v)$ to the root of $T_G$, and $G(v) \sqsubseteq G $}.

Refer to definition 6.6 and lemma 6.8  in \cite{h2h}. Given a vertex $v \in V$, $G(v)$ is a sub-graph of $G$ and the shortest travel cost is preserved at the same time. Thus, if $\forall s, d \in G(v)$, $f'_{s,d}(t)$ is the shortest travel cost function between $s$ and $d$ in $G(v)$, we can get $f'_{s,d}(t) = f_{s,d}(t)$. 

\fakeparagraph{Example 3.3} 
\textit{Let us back to the previous example \figref{fig:TD}, for the query processing $Q(v_{12}, v_{15}, t)$, we first locate the associated tree nodes $X(v_{12})$ and $X(v_{15})$ in $T_G$. Consider $X(v_{12})$, it maintains the weight function to $X(v_{12})\backslash \{ v_{12} \} = v_{10}$, and $v_{10}$ is one of ancestors in $Anc(X(v_{12})) = (v_1, v_2, v_3, v_4, v_5, v_{10})$. For $X(v_{12})$ and $X(v_{15})$, the LCA is $X(v_3)$ in $T_G$, thus $X(v_3) = \{ v_3, v_1, v_2\}$ is a vertex cut between $v_{12}$ and $v_{15}$, $f_{12,15}(t) = min_{u \in \{3, 1, 2\}} Compound(f_{12,u}(t)$ $,f_{u,15}(t))$. To compute the shortest travel cost functions $f_{12,u}(t)$ where $u \in \{3, 1, 2\}$, the traditional Dijkstra based algorithm can be directly applied over $G(12)$. The reason is that $G(12)$ is a TFP-graph which preserves the shortest travel cost from $X(v_{12})$ to all its ancestors. Similarly, we can compute the shortest travel cost functions from $X(v_{15})$ to all vertices in the vertex cut.}

\begin{algorithm}
\caption{Basic Query Algorithm}
\label{alg:basicquery}
\LinesNumbered 
\KwIn{Tree decomposition $T_G$ and query $Q(s,d,t)$}
\KwOut{Travel cost function $f_{s,d}(t)$}

\textbf{For} $ w \in G(s) $ \textbf{then} $cost_s[w] \gets +\infty$ \;

\textbf{For} $ u \in X(s) \backslash \{ s \}$, $cost_s[u] \gets X(s).W^s_u$ \;

$v \gets$ parent node of $X(s)$ \;

\While{$v \neq T_G.root$}{
\For{$u \in X(v) \backslash \{ v \}$}{

    \textbf{if} $cost_s[u] \neq +\infty$ \textbf{then} \\
    $cost_s[u] = min\{ cost_s[u], Compound(cost_s[v], X(v).W^s_u) \}$ \;
    \textbf{else}\\
    $cost_s[u] = Compound(cost_s[v], X(v).W^s_u)$ \;
}
$v \gets$ parent node of $X(v)$ \;
}
Repeat lines 1-8 for calculating $cost_d[\cdot]$ over $G(d)$ \;
$VC \gets LCA(X(s), X(d))$ \;
\textbf{For} $ w \in VC$, $f_{s,w}(t) = cost_s[w], f_{w,d}(t) = cost_d[w]$ \;
$f_{s,d}(t) = min_{w \in VC} \{Compound(f_{s,w}(t), f_{w,d}(t)) \}$ \;

\KwRet{$f_{s,d}(t)$}\;
\end{algorithm}

With properties 1-3, we are ready to design our query algorithm over $T_G$. The details of the query algorithm are shown in \algoref{alg:basicquery}. Given the query $Q(s,d,t)$, the algorithm returns the shortest travel cost functions from source $s$ to destination $d$ with the time parameter $t$. Line 1 initialize the list $cost_s[]$, and this list maintains the shortest travel cost functions from source $s$ to vertices in $G(s)$. We apply the modified Dijkstra based method over $G(s)$. It traverses $T_G$ from $X(s)$ to the root of the tree in lines 4-8. Specifically, for a node $X(v)$, we check the weight functions it maintains to $X(v)\backslash \{v\}$ in line 5. If the vertex $u$ has not been visited before processing $X(v)$ in line 8, the shortest travel cost function from $s$ to $u$ can be calculated considering $v$ as a bridging vertex. Otherwise, we check whether we can get a shorter travel cost through $v$ in line 7. Then we repeat the same procedure over $G(d)$ to calculate the list $cost_d[]$, which caches shortest travel cost functions from all vertices in $G(d)$ to $d$. Finally, based on the vertex cut property, we get the $LCA$ of $X(s)$ and $X(d)$ as $VC$ in line 11, we can calculate the result of the shortest travel cost function in lines 12-13 by checking the stored shortest travel cost functions to vertices in the set $VC$.

\fakeparagraph{Time Complexity} The query algorithm traverses the whole tree from bottom to up in lines 3-9, it takes $O(h(T_G))$ time. For each tree node $X(v)$, it contains at most $2 \cdot w(T_G)$ weight functions, thus each tree node takes  $O(w(T_G))$ time to calculate the shortest travel cost functions to vertices in $X(v)$. Therefore, the total time complexity of the query algorithm is $O(h(T_G) \cdot w(T_G) \cdot c)$, where the time complexity of $Compound()$ function is $O(c)$, and $c$ is a constant factor.

\begin{figure}[t]
	\centering
	\begin{subfigure}[b]{0.34\textwidth}
        \includegraphics[width=\textwidth]{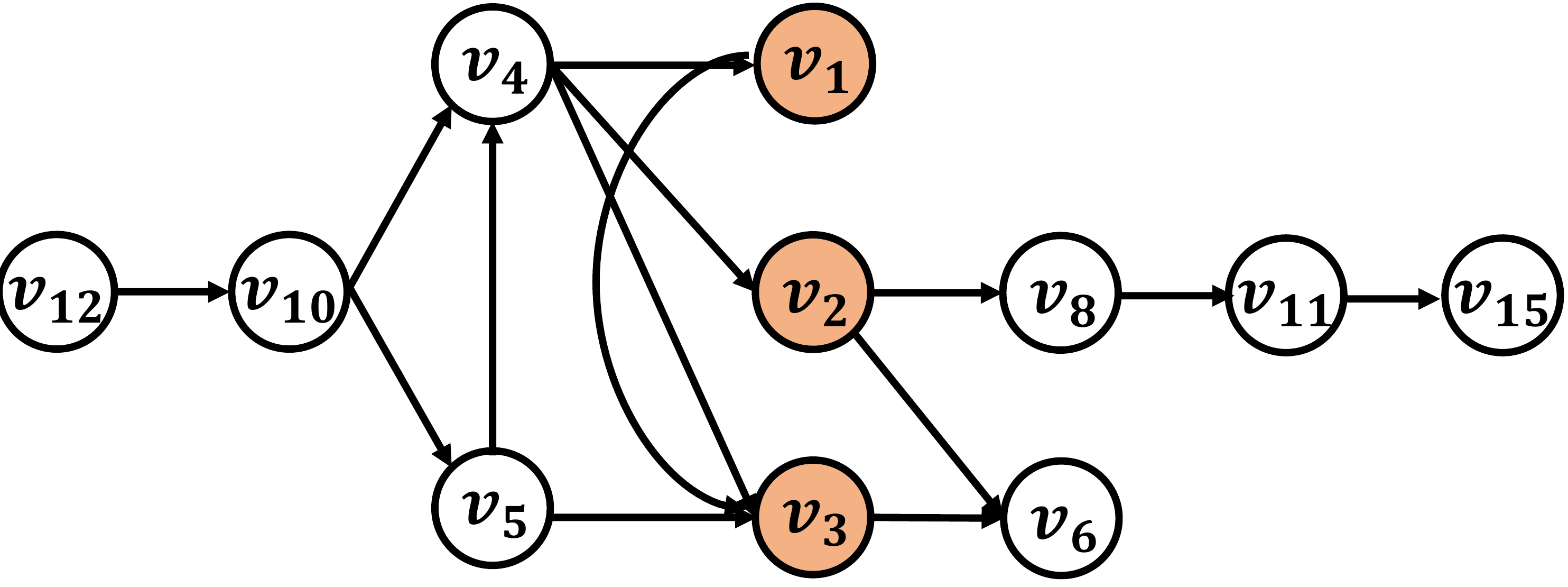}
		\vspace{-4ex}
	\end{subfigure}
	\vspace{-1ex}
	\caption{Query Processing between $v_{12}$ and $v_{15}$}
	\label{fig:basicquery}
	\vspace{-3ex}
\end{figure}

\fakeparagraph{Example 3.4} 
\textit{ Consider the query $Q(v_{12},v_{15},t)$, \figref{fig:basicquery} shows the processing procedure between $v_{12}$ and $v_{15}$ based on the $T_G$. The vertices marked in yellow are the vertex cut between query vertices. From left side shows the modified Dijkstra based algorithm processing over graph $G(12)$, and from the right side, it is the processing over graph $G(15)$. Consider the left side, at the beginning, the shortest travel cost function from source $v_{12}$ to vertex $v_{10}$ is stored in $cost_{12}[v_{10}] = X(v_{12}).W^s_{10} $. Next, we can get the parent node of $X(v_{12})$ is $X(v_{10})$, because $X(v_{10})$ contains two vertices which have not been visited, we can get $cost_{12}[v_{4}] = Compound(cost_{12}[v_{10}], X(v_{10}).W^s_{4})$ and $cost_{12}[v_{5}]$ $=  Compound(cost_{12}[v_{10}], X(v_{10}).W^s_{5})$. Furthermore, $X(v_5)$ is the parent node of $X(v_{10})$, and it contains one visited vertex $v_4$ and another vertex $v_3$. For $v_4$, we can update the shortest travel cost function as $cost_{12}[v_4] = min\{ cost_{12}[v_4], Compound(cost_{12}[v_5],$ $ X(v_5).W^s_4) \}$ which tries to find a shorter way through $v_5$. The algorithm stops until it reaches the root of $T_G$. We repeat the same procedure for $X(v_{15})$. Then we can get the shortest travel cost functions from the source and destination to the vertex cut respectively, and the final result is calculated. 
}

\section{Shortcut Selection}\label{sec:methodology}
In this section, we first define the shortcut over $T_G$, next we formulate the shortcut selection problem, and then we prove the NP-hardness of this problem. We first use the dynamic programming based method to solve the shortcut selection problem in \secref{subsection:dp}. Then, in \secref{subsection: appro} we propose an approximation algorithm which guarantees a $0.5$-approximation ratio. Finally, in \secref{subsection:query}, we propose a more efficient query algorithm based on the selected shortcuts.

\subsection{Preliminary}
In \algoref{alg:basicquery}, the algorithm traverses the tree structure from bottom to up frequently, which hinders the efficiency of queries dramatically. Intuitively, based on the properties 1-3 of $T_G$ introduced in the previous section, we can skip the traversing if a set of shortcuts  $\{s_{\langle i, j \rangle}(t) | v_j \in Anc(X(v_i)) \}$ is built for each  $X(v_i)$. 
The set of shortcuts is a set of shortest travel cost functions between $X(v_i)$ and all its ancestors in $T_G$. However, since the space of the main memory is limited, it is prohibitive to build and maintain all shortcuts between each tree node to all ancestors of it in $T_G$. We will show the results in the experiment section. Therefore, we need to select the shortcuts wisely to improve the query efficiency under the constrained memory space.

\begin{definition}[Shortcut]
Given a tree node $X(v_i) \in T_G$, $Anc(X(v_i))$ $= \{l_1, l_2 \dots l_m\}$ is an ancestor nodes list of $X(v_i)$ and is sorted in increasing order based on $height(X(v_j)), v_j \in \{l_1, \dots l_m\}$. Refer to Property 2, for any $v \in X(v_i) \backslash \{ v_i\}$, $v \in Anc(X(v_i))$. For ease of presentation, given a tree node $X(v_i)$ and one ancestor node $X(v_j)$ of it, one pair $\langle i,j \rangle$ indicates that there are shortcuts between these two tree nodes. For one pair instance $\langle i,j \rangle$, there are two shortcuts $s_{\langle i,j \rangle}(t)$ and $s_{\langle j,i \rangle}(t)$ selected and built in $T_G$. $s_{\langle i,j \rangle}(t)$ is the shortest travel cost function from $X(v_i)$ to $X(v_j)$, and $s_{\langle i,j \rangle}(t)$ is the shortest travel cost function from $X(v_j)$ to $X(v_i)$. Each shortcut is modeled as a piecewise linear function.
\end{definition}

\begin{definition}[Utility and Weight]
The utility value of one shortcuts pair instance $\langle i,j \rangle$ is presented as $u_{\langle i,j \rangle} = (height(X(v_i)) - height(X(v_j)) \cdot w(T_G) \cdot p_{\langle i,j \rangle})$, it is utilized to measure the benefit for querying, if the shortcut $s_{\langle i,j \rangle}(t)$ is selected and built. Specifically, $p_{\langle i,j \rangle}$ is the probability that shortcuts pair instance $ \langle i,j \rangle$ improves querying efficiency, can be defined as $\frac{vertexes_{k,j}}{|V|}$, $\forall v_k \in vertexes_{k,j}, LCA(X(v_i),X(v_k)) = X(v_j)$. For one selected shortcuts pair instance $\langle i,j \rangle$, the shortcuts $s_{\langle i,j \rangle}(t)$ and $s_{\langle j,i \rangle}(t)$ are also modeled as a PLF function. As discussed before, two sets of interpolation points $I_{ \langle i,j \rangle} = \{ (t_1, w_1), (t_2, w_2) \cdots $ $(t_z, w_z)\}$ and $I_{ \langle j,i \rangle} = \{ (t'_1, w'_1), (t'_2, w'_2) \cdots $ $(t'_z, w'_z)\}$ are cached to represent two shortcuts respectively. The weight of this shortcuts pair instance is
$(|I_{\langle i,j \rangle}|+|I_{\langle j,i \rangle}|)$.
\end{definition}

\begin{figure}[t]
	\begin{subfigure}[b]{0.2\textwidth}
		\includegraphics[width=\textwidth]{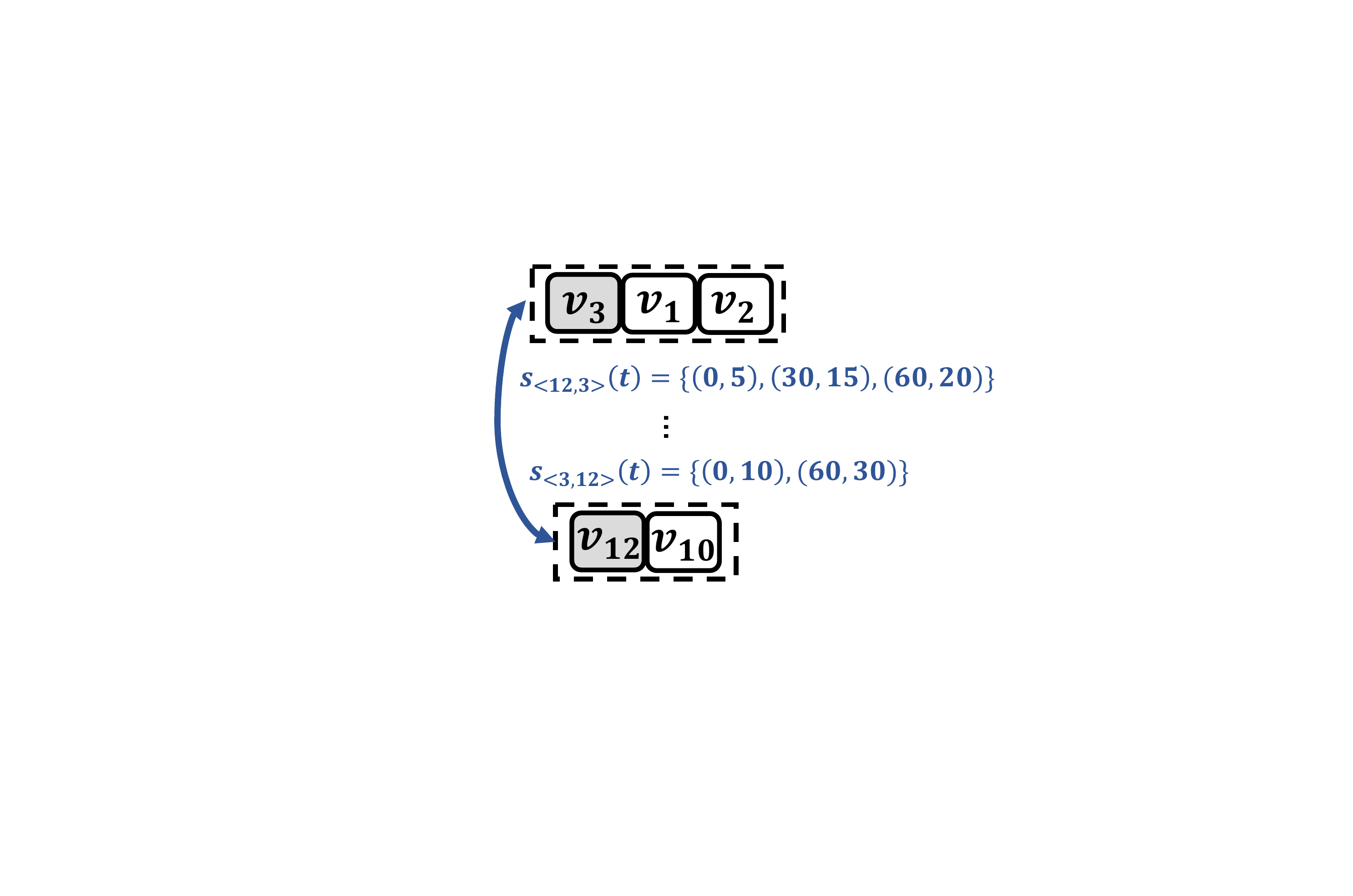}
		\vspace{-4ex}
	\end{subfigure}
	\vspace{-1ex}
	\caption{Shortcuts between $X(v_{12})$ and $X(v_3)$}
	\label{fig:example4}
	\vspace{-2ex}
\end{figure}

\fakeparagraph{Example 4.1} 
\textit{Take shortcuts pair instance $\langle 12,3 \rangle$ of $T_G$ in  \figref{fig:TD} as an example. In \figref{fig:example4}, it shows the shortcuts between $X(v_{12})$ and $X(v_3)$, we first assume that there are 3 (time, weight) pairs to represent the shortest travel cost function from $v_{12}$ to $v_3$ and 2 (time, weight) pairs to represent the shortest travel cost function from $v_{3}$ to $v_{12}$. Thus we can get the weight of shortcut $\langle 12, 3 \rangle$ is $5$. For the utility value, $\{ v_{15}, v_{11}, v_8, v_6, v_7 \}$ have the same LCA node $X(3)$ with $X(12)$, in other words shortcut $\langle 12,3 \rangle$ could help improve the efficiency of querying $v_{12}$ to these 5 vertexes. So we can get $p_{\langle 12,3 \rangle} = \frac{1}{3}$. Besides that, this shortcut could help avoid at most $(height(X(12)) - height(X(3))) \cdot w(T_G)$ times PLF compound function operation, thus the utility value $u_{\langle 12, 3 \rangle} = 3 \cdot 4 \cdot \frac{1}{3}$.} 

For each tree node $X(v_i)$, based on Def. \ref{def:TFP-Graph}, $G(v_i)$ contains all ancestors of $X(v_i)$ in $T_G$. To collect all shortcuts in $T_G$, for every tree node $X(v_i)$, we need to construct the graph $G(v_i)$ first, then a $Dijkstra$-based method can be implemented to calculate the shortest travel cost functions between $v_i$ to other vertices in $G(v_i)$. However, computing the shortcuts for every tree node independently is computation-consuming. Based on property 2, $\forall v \in X(v_i)\backslash \{v_i\}$, $X(v)$ is an ancestor node of $X(v_i)$, and the shortest travel cost functions between $v$ and $v_i$ is preserved in $X(v_i)$. Besides that, refer to Lemma 6.11 in \cite{h2h}, if $v_j \in Anc(X(v_i))$, then $G(v_j)$ is a supergraph of $G(v_i)$. Therefore, a top-down manner can be utilized to calculate the shortcut set of $T_G$. Specifically, when computing shortcuts for $X(v_i)$, the shortcuts for nodes in $X(v_i)\backslash \{v_i\}$ can be reused. Given a tree node $G(v_i)$, the shortcuts building lemma is introduced in the following:

\textbf{Fact 1 (Lemma 6.11 [24]).}
For any $v_j \in Anc(X(v_i))$, we calculate the shortcuts for $X(v_i)$ as:

$s_{\langle i,j \rangle}(t) = min_{v \in X(v_i)\backslash \{v_i\}} \{Compound(X(v_i).W^s_v, s_{\langle v,j \rangle}(t))) \}$

$s_{\langle j,i \rangle}(t) = min_{v \in X(v_i)\backslash \{v_i\}} \{Compound(s_{\langle j,v \rangle}(t), X(v_i).W^d_v))$

Given a tree decomposition $T_G$, the shortcut selection problem is to select the ``best'' set of pairs of tree nodes and their ancestor nodes, under the limited memory space. As discussed in the previous section, the ``best'' set of shortcuts should improve the querying efficiency most. In other words, we want to select a set of shortcuts with the highest utility value under a specific weight constraint. 

\begin{definition}[Shortcut Selection]
Given the set of all possible shortcuts within $T_G$,  $\mathbb{S} = \{\langle i,j \rangle | X(v_i) \in T_G, v_j \in Anc(X(v_i)) \}$. Let $x_{i,j} \in \{0,1\}$ indicate a shortcuts pair instance $\langle i,j \rangle$ is selected or not, if $x_{i,j} = 1$, it means that shortcuts between $X(v_i)$ and $X(v_j)$ are selected, vice versa. Thus, the shortcut selection problem can be formulated as: 

\begin{align*}
    \max \sum_{ \langle i,j \rangle \in \mathbb{S}} u_{\langle i,j \rangle} \cdot x_{i,j}
\end{align*}
\begin{align*}
    \text{subject to} . \sum_{\langle i, j \rangle \in \mathbb{S}} (|I_{\langle i, j \rangle}| + |I_{\langle j, i \rangle}|) \cdot x_{i,j} \leq N
\end{align*}
\end{definition}
\subsection{NP-Hardness of Shortcut Selection}
In this subsection we prove the NP-hardness of the shortcut selection Problem.

\begin{theorem}
\label{the:nphardTheorem}
    The shortcut selection problem is NP-hard
\end{theorem}

\vspace{-3ex}
\begin{proof}
We prove the theorem by reducing the 0-1 knapsack problem to the shortcut selection problem. A 0-1 knapsack problem can be formalized as follows. Given a set $C = {1,2,\cdots n}$, there are $n$ items numbered from 1 up to $n$, each item $i$ associated with a value $y_i$ and a weight $w_i$. Under the maximum weight capacity $W$, the 0-1 knapsack problem is to find a subset $C'$ of $C$ that maximizes the total values: $\max \sum_{i \in C'} y_i$, subjected to $\sum_{i \in C'} w_i \leq W$ 

For a given 0-1 knapsack problem instance, an instance of shortcut selection problem can be constructed as follows. Given a time-dependent road network $G=(V, E, F)$, we generate a spanning tree $T$ over $G$ in polynomial time, furthermore, we renumber the vertexes. We map $v_i$ to the $i$-th level of the tree, thus $v_0$ is the leaf node and $v_{|V|}$ is the root node (\eg{ For graph in \figref{fig: example1a}, we can easily generate a tree which the height is 15 due to it has 15 vertices, and the leaf node is $v_0$ and the root node is $v_{15}$ }); For $v_i$, $v_{i+1}$ is the parent node, and the potential shortcuts can be built to its ancestors $\{ v_{i+2}, \cdots v_{|V|} \}$. Then we can get $\sum_{0 \leq i \leq (|v|-1)} i$ possible shortcuts pair instances, such that for each shortcuts instance $ \langle i,j \rangle$, the utility value $u_{\langle i,j \rangle} = y_i$, the shortcuts instance weight $(|I_{\langle i,j \rangle}|+|I_{\langle j,i \rangle}|) = w_i$. Also, the memory cost constrain $N = W$. Thus, for this shortcut selection instance, we want to achieve an optimal selection set $\mathbb{I}$ that maximizes the overall utility $\max \sum_{\langle i,j \rangle \in \mathbb{I}} u_{\langle i,j \rangle}$ subjected to $\sum_{\langle i,j \rangle \in \mathbb{I}} (|I_{\langle i,j \rangle}|+|I_{\langle j,i \rangle}|) \leq N$. 

Given this mapping, we derive that the 0-1 knapsack problem instance can be solved, if and only if the transformed shortcut selection problem can be solved.

Thus, the shortcut selection problem can be reduced from 0-1 knapsack problem, and it is an NP-hard problem.
\end{proof}

\subsection{Dynamic Programming Algorithm}
\label{subsection:dp}
To search all possible shortcuts, in this subsection, we first illustrate the details of our dynamic programming based selection algorithm and analyze its time complexity.

We first introduce some useful notations before describing the algorithm. We use $U(\langle i, l_k \rangle,w)$ to represent the maximum utility value of selected shortcuts when take the instance $\langle i,l_k \rangle$ into consideration and weight constraint is $w$. Instance $\langle i,l_k  \rangle$ denotes the shortcuts between $X(v_i)$ and its $k$-th ancestor in $Anc(X(v_i))$. Let $\textit{root}$ denote the root of the tree decomposition $T_G$, the algorithm iteratively appends each tree nodes and ancestor nodes associated to it into the shortcut selection set until the $root$. Thus we define the state transition rules as : 

\begin{small}
\begin{equation}
\begin{split}
\label{equ:dp}
U(\langle i, l_k \rangle, w) =
        & \begin{cases}
        u_{\langle i,l_k \rangle}+U(\langle i,l_{k-1} \rangle, w - (|I_{\langle i,l_{k-1} \rangle}|+|I_{\langle l_{k-1},i \rangle}|), \textit{case 1} \\
        U(\langle i, l_{k-1} \rangle, w), \textit{case 2} \\
        \end{cases}
\end{split}
\end{equation}
\end{small}

In \equref{equ:dp}, if $U(\langle i, l_k \rangle, w)$ is updated to $u_{\langle i,l_k \rangle}+U(\langle i,l_{k-1} \rangle, w - (|I_{\langle i,l_{k-1} \rangle}|+I_{\langle l_{k-1},i \rangle}|))$ in case 1, it indicates that the maximum utility value can be updated to a higher value if the shortcuts pair instance $\langle i,l_{k} \rangle$ is selected. Otherwise, in case 2, the maximum utility value is the same as the value in the previous iteration, when checking the shortcuts between $v_i$ and its $(k-1)$-th ancestor. Therefore, under the weight constraint $w$ in case 2, the instance $\langle i,l_{k} \rangle$ is not selected.

\algoref{alg:dp} shows the detailed algorithm of dynamic programming based shortcut selection algorithm. In lines 3-4, we enumerate each tree node and each ancestor of it to update the maximum utility value. We update the constraint value $w$ from 1 to $N$, the enumeration stops when we finish updating value $U(\langle i, l_k \rangle, N)$. In each iteration in lines 6-8, if the value is updated based on case 1 in \equref{equ:dp}, we select the shortcuts pair instance $\langle i, l_k \rangle$ and put it into the set $\mathbb{S}$. Finally, we return the set $\mathbb{S}$ as the selected shortcut set.

\begin{algorithm}
\caption{DP selection}
\label{alg:dp}
\LinesNumbered 
\KwIn{Tree decomposition $T_G$ and weight constraint $N$}
\KwOut{Selected shortcut set $\mathbb{S^*}$}
$\mathbb{S} \gets \{\langle i,j \rangle | X(v_i) \in T_G, v_j \in Anc(X(v_i)) \}$ \;
$\mathbb{S^*} \gets \emptyset$ \;
\For{$X(v_i) \in T_G$}{
    \For{$l_k \in Anc(X(v_i))$}{
        \For{$w \gets $ 1 to $N$}{
        \If{case 1 in \equref{equ:dp}}{
            Shortcuts pair instance $\langle i,l_k \rangle$ is selected \;
            $\mathbb{S} \gets \mathbb{S} + \langle i,l_k \rangle$ \;
        }

        }

    }
}
\KwRet{$\mathbb{S^*}$}\;
\end{algorithm}

\fakeparagraph{Complexity Analysis} The number of tree nodes is $O(n)$ in line 3. For each tree node, we check all ancestors of it in line 4, the number of ancestors is bounded by the treeheight $O(h(T_G))$. Therefore, the overall time complexity of \algoref{alg:dp} is $O(n \cdot h(T_G) \cdot N)$.

\fakeparagraph{Example 4.2} Let us back to the tree $T_G$ in Example 3.1. We first make the assumption that $N = 100$, and we start from $X(v_{15})$. \algoref{alg:dp} checks the shortcut instances from $\langle v_{15},v_1 \rangle$, $\langle v_{15},v_2 \rangle \dots$ until $\langle v_{15}, v_{11} \rangle$ and calculate the associated values $U(\langle v_{15}, v_1 \rangle, 1)$, $U(\langle v_{15},v_2 \rangle,1)\dots$ ,$U(\langle v_{15}, v_{11} \rangle, 1)$. Then we update the value $w$ from 1 to the constraint $100$. We check other tree nodes in a similar way. Finally, we calculate the value $U(\langle v_{2}, v_{1} \rangle, 100)$, and get the final selected shortcut set $\mathbb{S}$.

\subsection{Approximation Algorithm}
\label{subsection: appro}

\fakeparagraph{Basic Idea} To improve the efficiency of index construction, we design an approximation algorithm that simultaneously considers two different greedy strategies. For the first greedy strategy, we consider the shortcuts instances in order of their utility values and add the instance into the selection set until the total weights exceed the constraint. For the second strategy, we consider the shortcuts instances in order of the ``density value'' ($\frac{u_{\langle i,j \rangle}}{|I_{\langle i, j \rangle}|}$) in the same way. We may get bad performance (\eg{ arbitrarily bad approximation guarantees}) if we implement these two greedy strategies independently. On one hand, one shortcut from the leaf node to the root may take huge space but with less utility with several shortcuts from the leaf node to some ancestors near to it. On the other hand, one shortcut from the leaf node to the root may benefit many queries from other vertices which have the same LCA with the leaf node. Therefore, we jointly consider both two greedy strategies. We first implement the greedy strategies respectively and then pick the selected set with the highest utility value.   

\algoref{alg:appro} shows the detailed algorithm of approximation selection algorithm. Two priority queues $Q_1$ and $Q_2$ are initialized in line 2, $Q_1$ is prioritised by the utility value and $Q_2$ is prioritised by the ``density value''. In line 3, we define $\mathbb{S^*}_1$ as the selected shortcut set from queue $Q_1$, and the sum of the weights is $W_1$. Correspondingly, $\mathbb{S^*}_2$ is the selected set from $Q_2$, and $W_2$ is the total weights of shortcuts in $\mathbb{S^*}_2$. We first collect all shortcuts pair instances into two queues in line 4. Then we implement two greedy selection strategies in lines 5-8 and lines 9-12, respectively. Finally we compare the total utility values of two selected set, and we return the set with the bigger value as the final selected set.  

\vspace{-2.8ex}
\begin{algorithm}
\caption{Approximation selection}
\label{alg:appro}
\LinesNumbered 
\KwIn{Tree decomposition $T_G$ and weight constrain $N$}
\KwOut{Selected shortcut set $\mathbb{S^*}$}
$\mathbb{S} \gets \{\langle i,j \rangle | X(v_i) \in T_G, v_j \in Anc(X(v_i)) \}$ \;
$Q_1,Q_2 \gets$ empty queues prioritized by  $u_{\langle i,j \rangle}$ and $\frac{u_{\langle i,j \rangle}}{|I_{\langle i, j \rangle}|}$ \;
$\mathbb{S^*}_1,  \mathbb{S^*}_2 \gets \emptyset; W_1, W_2 \gets 0$ \;

\textbf{For} $\langle i,j \rangle \in \mathbb{S}$, $Q_1.push(\langle i,j \rangle), Q_2.push(\langle i,j \rangle)$ \;

\While{$Q_1$ is not empty and $W_1 < N$}{

    $\langle i,j \rangle \gets Q_1.pop()$ \;
    \textbf{if} $W_1+(|I_{\langle i, j \rangle}| + |I_{\langle j, i \rangle}|)|>N$ \textbf{break$;$} \\
    $\mathbb{S^*}_1 \gets \langle i,j \rangle, W_1 += (|I_{\langle i, j \rangle}| + |I_{\langle j, i \rangle}|)$ \;
}

\While{$Q_2$ is not empty and $W_2 < N$}{

    $\langle i',j' \rangle \gets Q_2.pop()$ \;
    \textbf{if} $W_2+(|I_{\langle i', j' \rangle}| + |I_{\langle j', i' \rangle}|)>N$ \textbf{break$;$} \\
    $\mathbb{S^*}_2 \gets \langle i',j' \rangle, W_2 += (|I_{\langle i', j' \rangle}| + |I_{\langle j', i' \rangle}|)$ \;
}
\If{$\sum_{\langle i,j \rangle \in \mathbb{S^*}_1}u_{\langle i,j \rangle} > \sum_{\langle i',j' \rangle \in \mathbb{S^*}_2}u_{\langle i',j' \rangle}$}{
    \KwRet{$\mathbb{S^*}_1$;}
}
\Else{
    \KwRet{$\mathbb{S^*}_2$;}
}
\end{algorithm}

\fakeparagraph{Performance Analysis} In \theref{the:bound}, we analyze the approximation ratio of \algoref{alg:appro}. This algorithm should be effective for large-scale tree decomposition based on the theoretical results.

\begin{theorem}
\label{the:bound}
    The approximation ratio of \algoref{alg:appro} is 0.5.
\end{theorem}

\vspace{-3ex}
\begin{proof}
We consider two greedy strategies in the algorithm, and let $U_1$ and $U_2$ the total utility value achieved by the selected set $\mathbb{S^*}_1$ and $\mathbb{S^*}_2$ respectively, and let $OPT$ denotes the maximum utility value of the selected shortcut set. We define two functions $utility_1(k)$ and $utility_2(k)$, these two functions return the utility value of $k$-th shortcuts pair instance of two different priority queues $Q_1$ and $Q_2$ respectively. For the ``density'' value based greedy based strategy, $\mathbb{S^*}_2$ is the set of selected shortcuts, and let the $y$-th shortcut in $Q_2$ be the first instance that did not fit into $\mathbb{S^*}_2$ due to the weight constrain $N$. Because for set $\mathbb{S^*}_1$, it takes the first greedy strategy, that start selecting the single shortcuts pair instance with the maximum utility value into set $\mathbb{S^*}_1$. Therefore, $\mathbb{S^*}_1$ contains the shortcuts pair instance with the maximum utility value definitely, the total utility value must be larger than the $y$-th instance in $Q_2$. We can have 
\begin{align*}
    U_1 \ge utility_2(k)
\end{align*}

For set $\mathbb{S^*}_2$, we choose the ``density'' value based strategy to select shortcut in $Q_2$. Let $d_y$ represent the ``density'' value of the $y$-th instance in $Q_2$. Because $y$-th instance is the first shortcut that didn't fit into $\mathbb{S^*}_2$, so for other shortcuts pair instances in $\mathbb{S^*}_2$, their ``density'' values are all larger than $d_y$.  

The main observation is that if we can cut a part of the $y$-th instance in $Q_2$ so as to exactly fill the weight constrain, that would clearly be the optimum solution if selecting a partial instance is allowed: it uses all shortcuts pair instances of ``density'' values are larger than $d_y$ and fills the remaining weight with ``density'' value is equal to $d_y$, and all other instances not selected have ``density'' value less than $d_y$. This shows that the optimum value is for the case when we take a fraction of the $y$-th instance in $Q_2$. Therefore, the true optimum can only be smaller, if we take the whole $y$-th instance in $Q_2$:
\begin{align*}
    \sum_{k=1}^{|\mathbb{S^*}_2|}utility_2(k) + utility_2(y) \ge OPT
\end{align*}

Based on the definition of $U_2$, it is the total utility value in $\mathbb{S^*}_2$, $U_2 = \sum_{k=1}^{|\mathbb{S^*}_2|}utility_2(k)$, therefore, we can get:
\begin{align*}
    U_1 + U_2 \ge \sum_{k=1}^{|\mathbb{S^*}_2|}utility_2(k) + utility_2(y) \ge OPT
\end{align*}

This implies that $U_1 + U_2 \ge OPT$, so the larger of $U_1$ and $U_2$ must be at least $\frac{1}{2}$ of $OPT$, the selected set $\mathbb{S^*}_1$ and $\mathbb{S^*}_2$ which has the larger total utility value must be at least $\frac{1}{2}$ of $OPT$.

Finally, we derive that the approximation ratio is 0.5.

\end{proof}

\fakeparagraph{Complexity Analysis} In line 2 of \algoref{alg:appro}, two priority queues are created and maintained in the whole selection procedure, thus the time complexity of this algorithm is $O(n \cdot h(T_G) \cdot log(nh(T_G)))$.

\begin{figure*}[t]
	\centering
	\begin{subfigure}[b]{0.21\textwidth}
		\includegraphics[width=\textwidth]{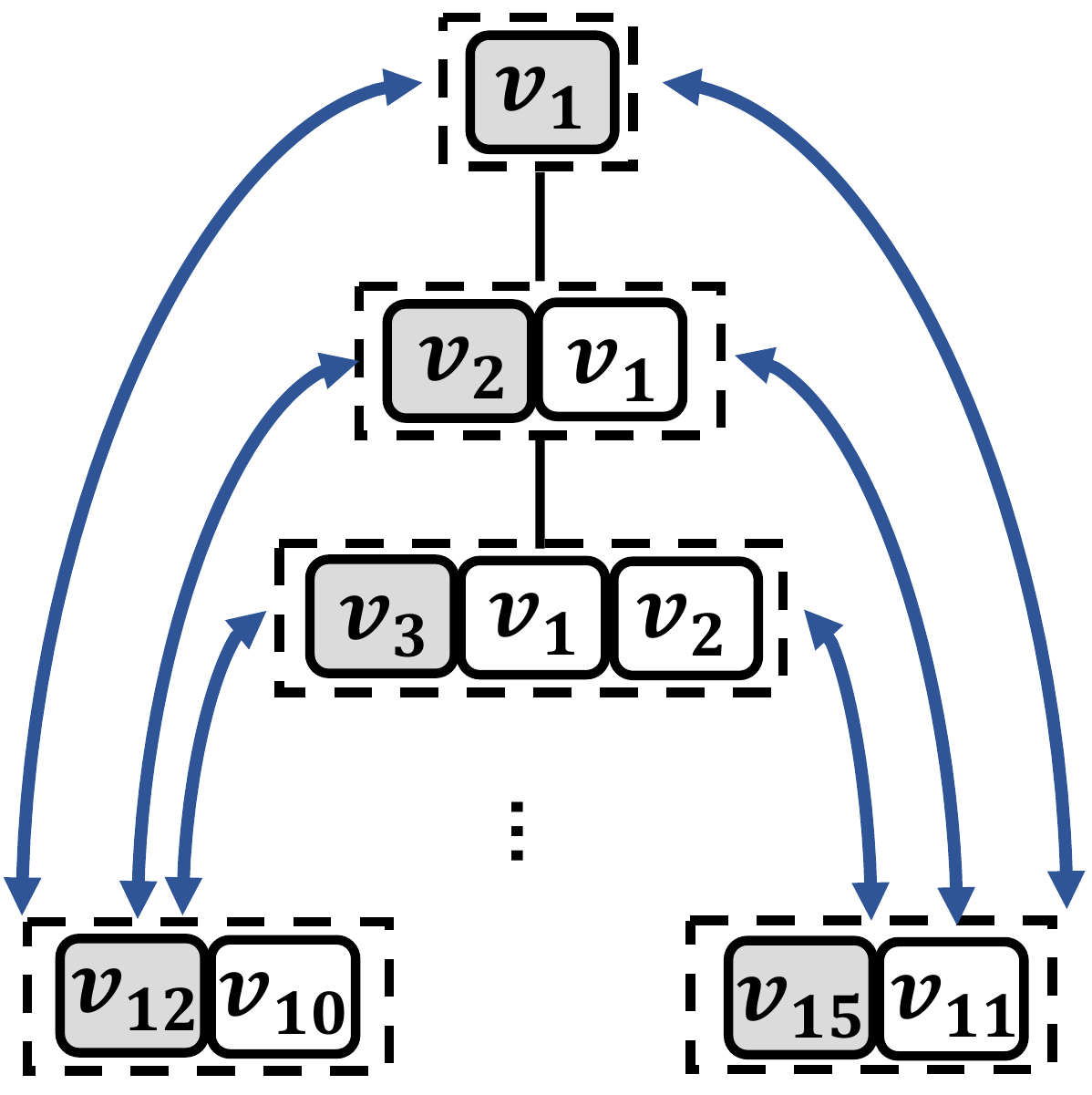}
		\vspace{-1ex}
		\caption{\footnotesize{Shortcuts to $LCA$}}
		\label{fig:query1a}
	\end{subfigure}
	~~
	\begin{subfigure}[b]{0.19\textwidth}
		\includegraphics[width=\textwidth]{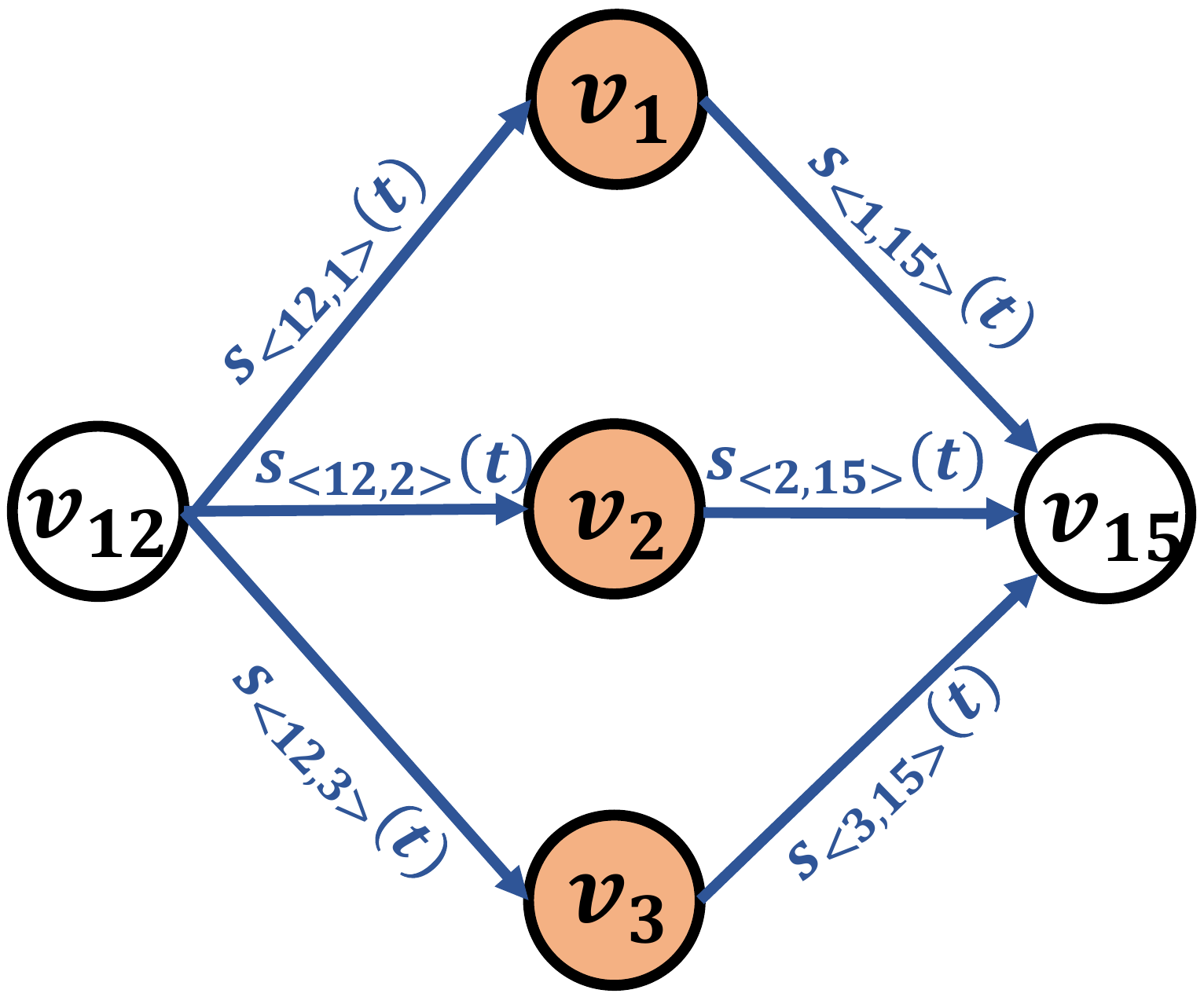}
		\vspace{2ex}
		\caption{\footnotesize{Query processing in $(a)$}}
		\label{fig:query1b}
	\end{subfigure} \hspace{10mm}
	\begin{subfigure}[b]{0.21\textwidth}
		\includegraphics[width=\textwidth]{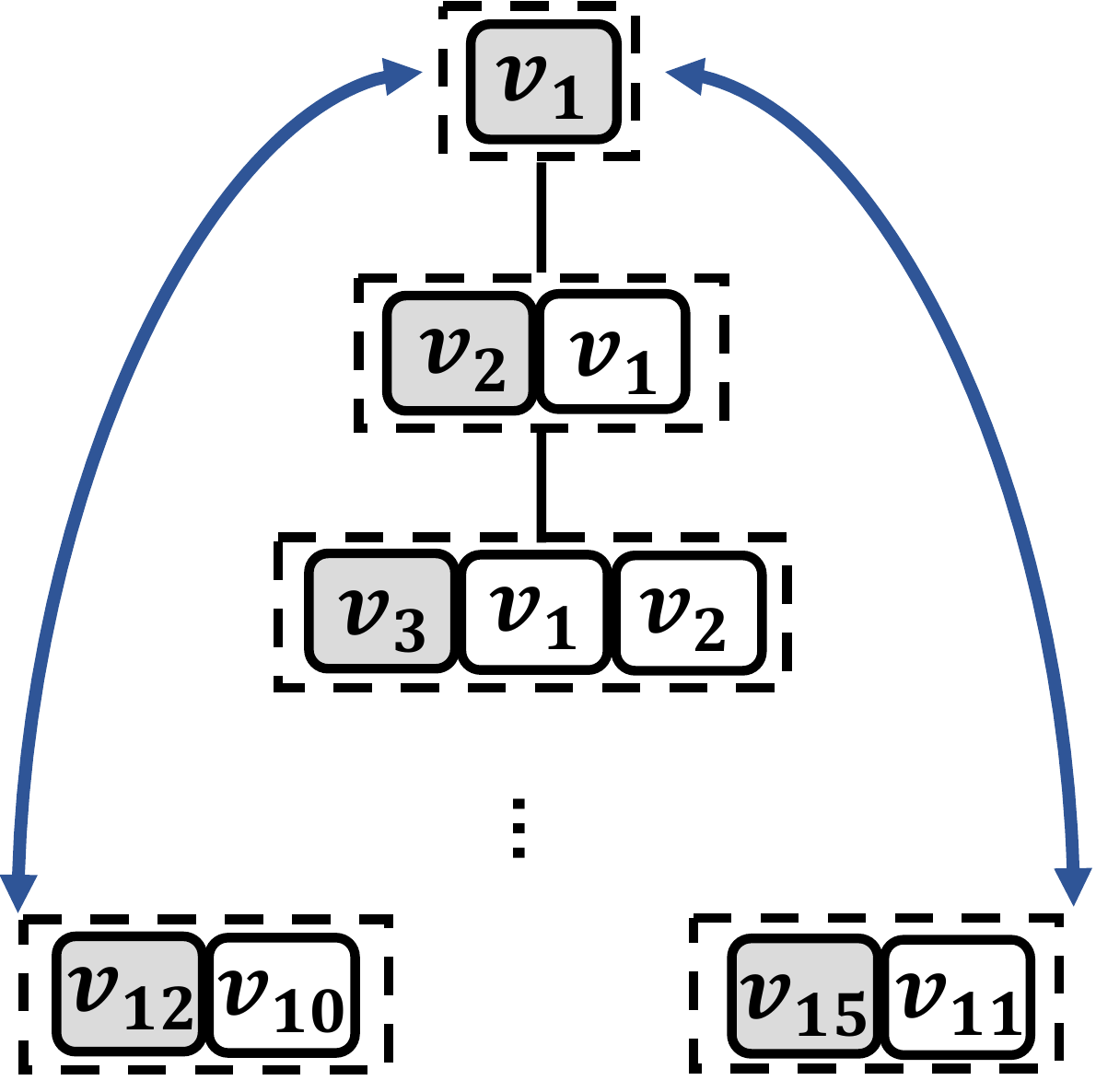}
		\vspace{-1ex}
		\caption{\footnotesize{Shortcuts to subset of $LCA$}}
		\label{fig:query2a}
	\end{subfigure}
	~~	
	\begin{subfigure}[b]{0.31\textwidth}
		\includegraphics[width=\textwidth]{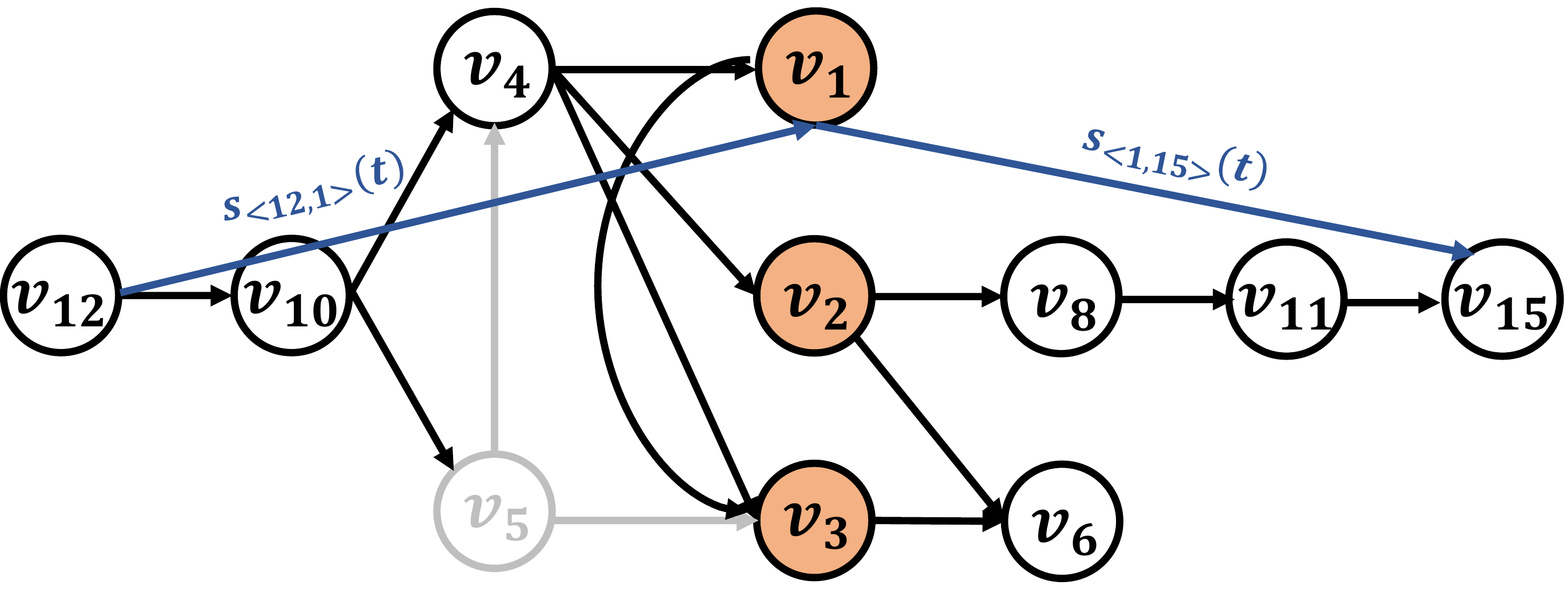}
  		\vspace{4ex}
		\caption{\footnotesize{Query processing in $(c)$}}
		\label{fig:query2b}
	\end{subfigure}
    \vspace{-1ex}
    \caption{Query processing from $v_{12}$ to $v_{15}$ with shortcuts.}
	\label{fig:QuerywithShortcuts}
	\vspace{-2ex}
\end{figure*}

\begin{algorithm}
\caption{Query with shortcuts}
\label{alg:query}
\LinesNumbered 
\KwIn{Query $Q(s,d,t)$, $T_G$ and selected shortcuts $\mathbb{S^*}$}
\KwOut{Travel cost function $f_{s,d}(t)$}
\If{$\forall w \in LCA(X(s),X(d)),$ $\langle s,w \rangle$ and $\langle w,d \rangle \in \mathbb{S^*}$}{
$f_{s,d}(t) = min_{w \in LCA(X(s),X(d))} \{ Compound(s_{\langle s,w \rangle}(t), s_{\langle w,d \rangle}(t)) \};$ \textbf{Return} $f_{s,d}(t)$ \;
}

$W_s \gets \{w|\forall w, \langle s,w \rangle \in \mathbb{S^*} \}$, $W_s \subseteq LCA$ \;
\For{$v \in G(s)$}{
    \textbf{if} $ v \in W_s $ \textbf{then} $cost_s[v] \gets s_{\langle s,v \rangle}(t)$ \;
    \textbf{else} $cost_s[v] \gets \infty$ \;
}

$W_d \gets \{w|\forall w, \langle w,d \rangle \in \mathbb{S^*} \}$, $W_d \subseteq LCA$ \;
\For{$v \in G(d)$}{
    \textbf{if} $ v \in W_d $ \textbf{then} $cost_d[v] \gets s_{\langle v,d \rangle}(t)$ \;
    \textbf{else} $cost_d[v] \gets \infty$ \;
}

$f_{s,d}^+(t) = min_{u \in W_s \cap W_d} \{ Compound(cost_s[u], cost_d[u]) \}$ \;
$ X(v) \gets$ parent node of $X(s)$, $cost_s[v] = X(s).W_v^s$ \;
\While{$v \neq T_G.root$}{
\For{$u \in X(v) \backslash \{ v \}$}{
    \textbf{if} $u \in W_s$ or $cost_s[u] = \nil$ \textbf{continue}; \\
    
    \textbf{if} $cost_s[u] \neq \infty$ \textbf{then} \\
    $cost_s[u] = min\{ cost_s[u], Compound(cost_s[v], X(v).W_u^s) \}$ \;
    \textbf{else}\\
    $cost_s[u] = Compound(cost_s[v], X(v).W_u^s)$ \;
    
    \textbf{if} $cost_s[u] > f_{s,d}^+(t)$ \textbf{then} $cost_s[u] \gets \nil$; \\
}
$v \gets$ parent node of $X(v)$ \;
}
Repeat lines 12-20 for calculating $cost_d[\cdot]$ over $G(d)$ \; 
$f_{s,d}(t) = min_{w \in LCA(X(s),X(d))} \{ Compound(cost_s[w], cost_d[w] \};$ \textbf{Return} $f_{s,d}(t)$ \;


\end{algorithm}

\subsection{Query Processing with Shortcuts}
\label{subsection:query}
\fakeparagraph{Basic Idea} Given a query $Q(s,d,t)$, based on the properties of $T_G$, the important procedure is to find the shortest travel cost from $s$ and $d$ to the vertices in $LCA(s,d)$ respectively. Based on the selected shortcuts pair instances set, there are 3 situations to speed up the query processing with the shortcuts: (1) When the shortcuts from $s$ and $d$ to the vertices in $LCA(s,d)$ are all selected, the query time can be bounded by $O(w(T_G))$; (2) When the subset of shortcuts from $s$ and $d$ to the vertices in $LCA(s,d)$ are selected, the selected shortcuts can be utilized to calculate an upper bound shortest travel cost from $s$ to $d$, then the upper bound value can guide the searching over $T_G$; (3) When non of the shortcuts from $s$ and $d$ to the vertices in $LCA(s,d)$ are selected, we need to invoke the basic query algorithm to answer the query.

\algoref{alg:query} shows the detailed query algorithm with the selected set of shortcuts. In lines 1-2, we check if all shortcuts from $s$ and $d$ to vertices in their $LCA$ are selected in $\mathbb{S^*}$, then we can derive the shortest travel cost function directly:  $f_{s,d}(t) = min_{w \in LCA(X(s),X(d))}$ $\{ Compound(s_{\langle s,w \rangle}(t),s_{\langle w,d \rangle}(t)) \}$. This procedure takes $O(w(T_G))$ time.    
Then we get $W_s$, a set of shortcuts from $s$ to a subset of vertices in $LCA$. Then we get the similar set $W_d$, from query vertex $d$. If the intersection of $W_s$ and $W_d$ is not empty, it means that there are shortcuts from $s$ and $d$ that contain at least one common vertex in $LCA$. Therefore, we can calculate an upper bound travel cost from $s$ to $d$ as $f_{s,d}^+(t)$ in line 11. Then this upper bound can guide our searching in $T_G$. In lines 12-20, the traversing procedure is similar as the basic query. However, when we check one vertex with the travel cost is larger than $f_{s,d}^+(t)$, we set the travel cost to it is $\nil$ in line 19. It means there is no need to check this vertex as a internal vertex in the final result, in line 15, if the searching procedure reaches vertex with the travel cost $\nil$, we can skip this vertex directly. Similar to the basic query algorithm, after calculating the shortest travel cost functions from $s$ and $d$ to every vertex in $LCA$, we can get the final query result in line 22.

\fakeparagraph{Complexity Analysis} Based on the selected shortcuts, the query algorithm can be bounded by $O(w(T_G))$.

\fakeparagraph{Example 4.3} \figref{fig:QuerywithShortcuts} shows an example of query with the selected shortcuts. Given a query $Q(v_{12}, v_{15},t)$, \figref{fig:query1a} and \figref{fig:query1b} show the query processing with shortcuts from $v_{12}$ and $v_{15}$ to all vertices in their $LCA$ node $X(v_3)$. Then \figref{fig:query2a} and \figref{fig:query2b} show the query processing with one shortcut from $v_{12}$ to $v_1$ and another shortcut from $v_{1}$ to $v_{15}$ respectively. In \figref{fig:query1b}, it is the first situation of the algorithm, we can get the final result based on the shortcuts directly $f_{12,15}(t) = min_{w \in \{ 1, 2, 3 \}} \{ Compound(s_{\langle 12,w \rangle}(t),$ $  s_{\langle w, 15 \rangle}(t)) \}$. In \figref{fig:query2a}, not all shortcuts to vertices in $X(v_3)$ are selected and built. However, we can easily get a travel cost upper bound  $f_{12,15}^+(t) =  Compound$ $(s_{\langle 12,1 \rangle}(t)), s_{\langle 1, 15 \rangle}(t)))$. In \figref{fig:query2b}, it shows the searching procedure under the guide of $f_{12,15}^+(t)$. From $X(v_{12})$, after we calculating the travel cost to $v_4$ and $v_5$ in tree node $X(v_{10})$, if we have $cost_{12}[5] >  f_{12,15}^+(t)$, we will set $cost_{12}[5]$ as $\nil$. There is no need to check vertex $v_5$ as an internal vertex, and there is also no need to traverse the tree node $X(v_5)$ to calculate the travel cost to $v_3$ and $v_4$ through $v_5$.

\section{Experiment Study}\label{sec:experiment}

\textbf{Algorithms.} We compare our proposed algorithms with the state-of-the-art algorithms for query processing in time-dependent road
networks. We implement and compare 5 algorithms:
\begin{itemize}
  \item TD-G-tree: The state-of-the-art algorithm for querying the shortest path in time-dependent graphs \cite{shortestquery};
  \item TD-H2H: Time-Dependent H2H index, which extends the H2H index to time-dependent scenario \cite{TD-KNN};
  \item TD-basic: Basic query algorithm over travel function preserved tree decomposition (\algoref{alg:basicquery});
  \item TD-dp: Query algorithm with shortcuts selected by dynamic programming strategy (\algoref{alg:dp} + \algoref{alg:query});
  \item TD-appro: Query algorithm with shortcuts selected by approximation strategy (\algoref{alg:appro} + \algoref{alg:query}).
\end{itemize}

\begin{table}[t]
\vspace{-1ex}
	\centering
	\caption{Statistics of datasets.}
    \vspace{-2ex}
	\label{table:dataset}
    \resizebox{0.45\textwidth}{!} {%
	\begin{tabular}{|c|c|c|c|c|c|}
		\hline
		Dataset & \#(Vertices) & \#(Edges) & \ $h(T_G)$ & $w(T_G)$ & $N$ \\
		\hline														
		\textit{CAL(California)} & $21,048$ & $43,386$ & $224$ & $18$ &10M \\
		\hline
		\textit{SF(San Francisco)} & $321,270$ & $800,172
$ & $529$ & $105$ & 20M \\
		\hline
		\textit{COL(Colorado)} & $435,666
$ & $1,057,066$ & $511$ & $122$ & 50M \\
		\hline
		\textit{FLA(Florida)} & $1,070,376$ & $2,712,798$ & $706$ & $89$ & 100M \\
		\hline
		\textit{W-USA(Western USA)} & $6,262,104
$ & $15,248,146
$ & $1041$ & $386$ & 200M \\
		\hline
	\end{tabular}
	}
\vspace{-2ex}
\end{table}

\begin{table}[t]
\vspace{-1ex}
	\centering
	\caption{Performance on $CAL$.}
    \vspace{-2ex}
	\label{table:exp1_cal}
    \resizebox{0.4\textwidth}{!} {%
	\begin{tabular}{|c|c|c|c|}
		\hline
		 & \ Query cost & \ Construction & \ Memory \\
		\hline														
		TD-G-tree & $0.16 ms$ & $0.006 h$ & $0.169G$ \\
		\hline
		TD-H2H & $0.0001 ms$ & $0.12h$ & $3.7G$ \\
		\hline
		TD-basic & $4.4 ms$ & $0.0002h$ & $0.089G$ \\
		\hline
	\end{tabular}
	}
\vspace{-2ex}
\end{table}

\begin{table}[t]
\vspace{-1ex}
	\centering
	\caption{Performance on $W-USA$.}
    \vspace{-2ex}
	\label{table:exp1_wusa}
    \resizebox{0.4\textwidth}{!} {%
	\begin{tabular}{|c|c|c|c|}
		\hline
		 & \ Query cost & \ Construction & \ Memory \\
		\hline														
		TD-G-tree & $30ms$ & $15h$ & $102G$ \\
		\hline
		TD-H2H & $N/A$ & $N/A$ & $N/A$ \\
		\hline
		TD-basic & $9,118ms$ & $1.18h$ & $66G$ \\
		\hline
	\end{tabular}
	}
\vspace{-2ex}
\end{table}

The experiments are conducted on a server with 40 Intel(R)
Xeon(R) E5 2.30GHz processors with 512GB memory. These five query algorithms are implemented in GNU C++. Following the setup in \cite{DBLP:conf/kdd/DuTZTZ18,DBLP:conf/waim/GaoTSSCX16,DBLP:conf/sigmod/SheT0S17}, each experiment is repeated 10 times and the average results are reported.

\textbf{Datasets.} 
We use 5 publicly available real road networks to conduct our experiments. These road networks are directed graphs and have been widely used in shortest path querying related works recently \cite{h2h} \cite{shortestquery} \cite{p2h}. We set the time domain as one day, \ie{  86400 seconds} which is the same as the setting in \cite{shortestquery}. We adopt the same strategy in \cite{TD-KNN} to build the piecewise linear function to model the weight of each edge in the graph. Furthermore, to evaluate the scalability of our algorithms, we vary the number of interpolation points of each edge from 2 to 6 (\eg{ the parameter $c \in \{2,3,4,5,6\}$}) which follows the parameter setting in \cite{TD-KNN}. We set the default value of $c$ as 3 (\eg{ the travel cost of one road segment could be 3 different values one day}). The detail of these datasets were shown in \tabref{table:dataset}.

For each dataset, we follow the same setting in \cite{shortestquery} to generate the queries. Specifically, we first randomly choose 1,000 pairs of vertices and uniformly generate the query time in 10 different time intervals, thus we have 10,000 queries for each dataset. The average query processing time for the 10,000 queries in the corresponding query set is recorded.

\begin{figure}[t]
	\centering
	\begin{subfigure}[b]{0.22\textwidth}
		\includegraphics[width=\textwidth]{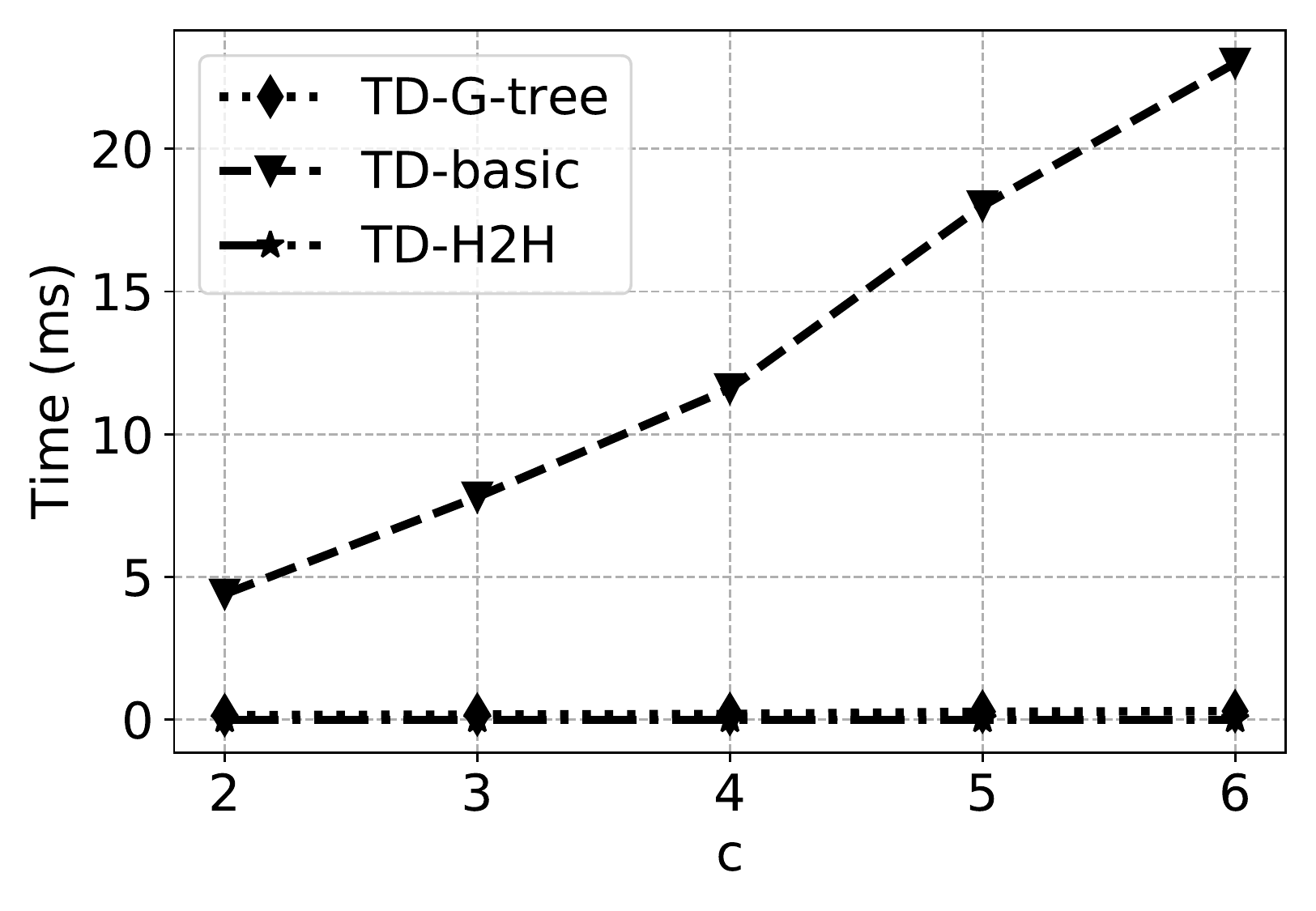}
		\vspace{-4ex}
		\caption{\footnotesize{Cost query on $CAL$}}
	\end{subfigure}
	~~
    \begin{subfigure}[b]{0.22\textwidth}
		\includegraphics[width=\textwidth]{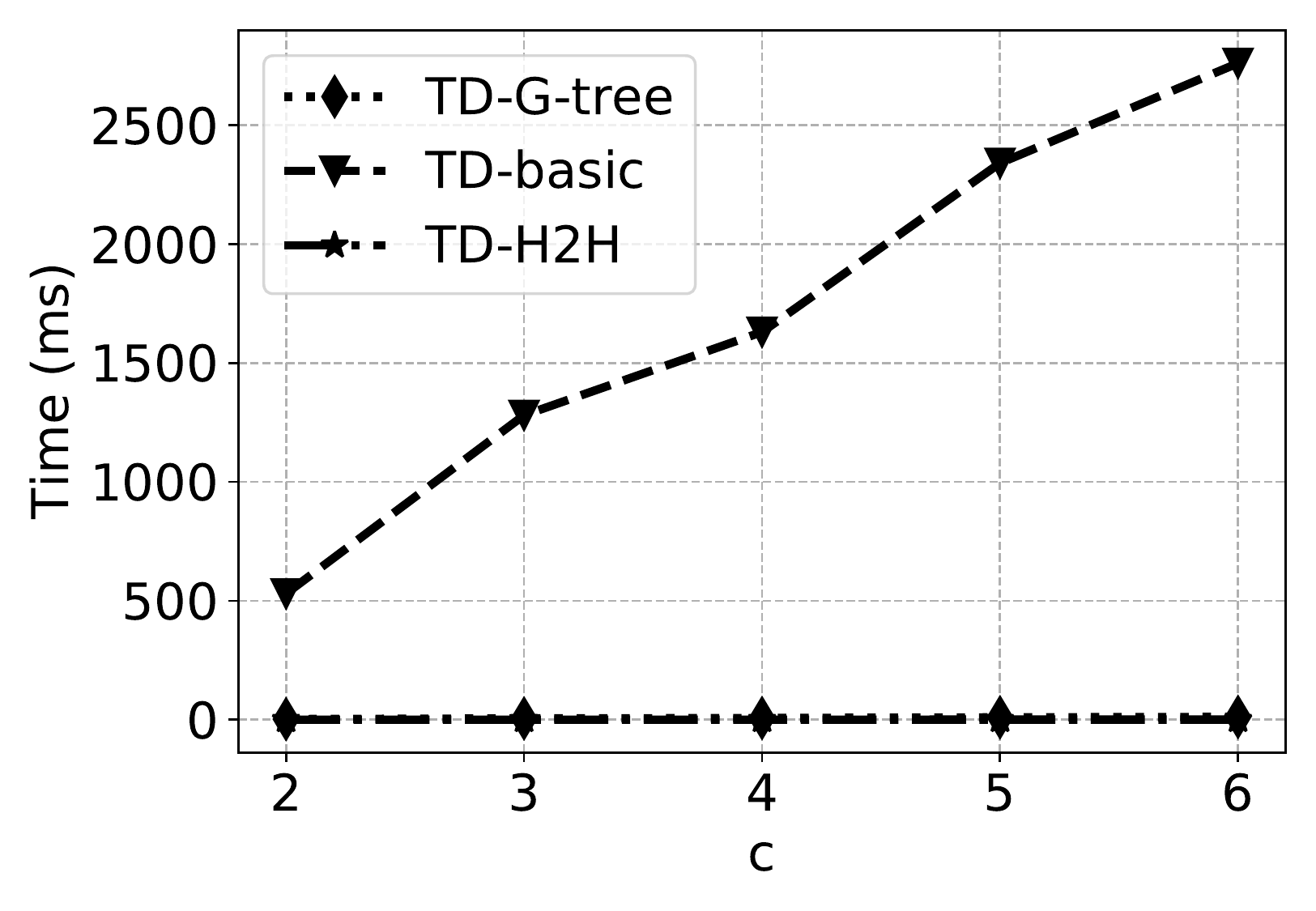}
		\vspace{-4ex}
		\caption{\footnotesize{Cost function query on $CAL$}}
	\end{subfigure}
	
	\begin{subfigure}[b]{0.22\textwidth}
		\includegraphics[width=\textwidth]{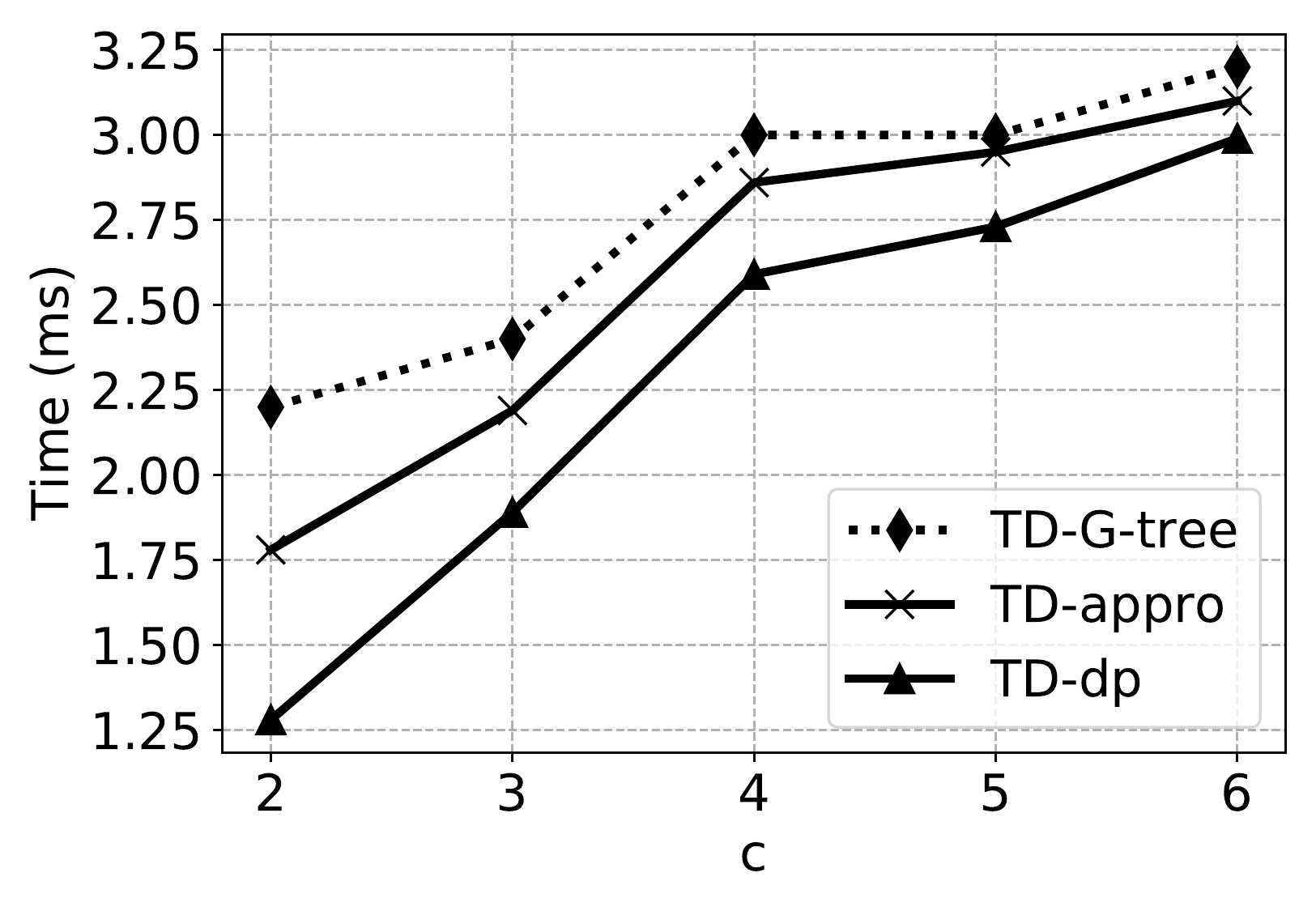}
		\vspace{-4ex}
		\caption{\footnotesize{Cost query on $SF$}}
	\end{subfigure}
	~~
    \begin{subfigure}[b]{0.22\textwidth}
		\includegraphics[width=\textwidth]{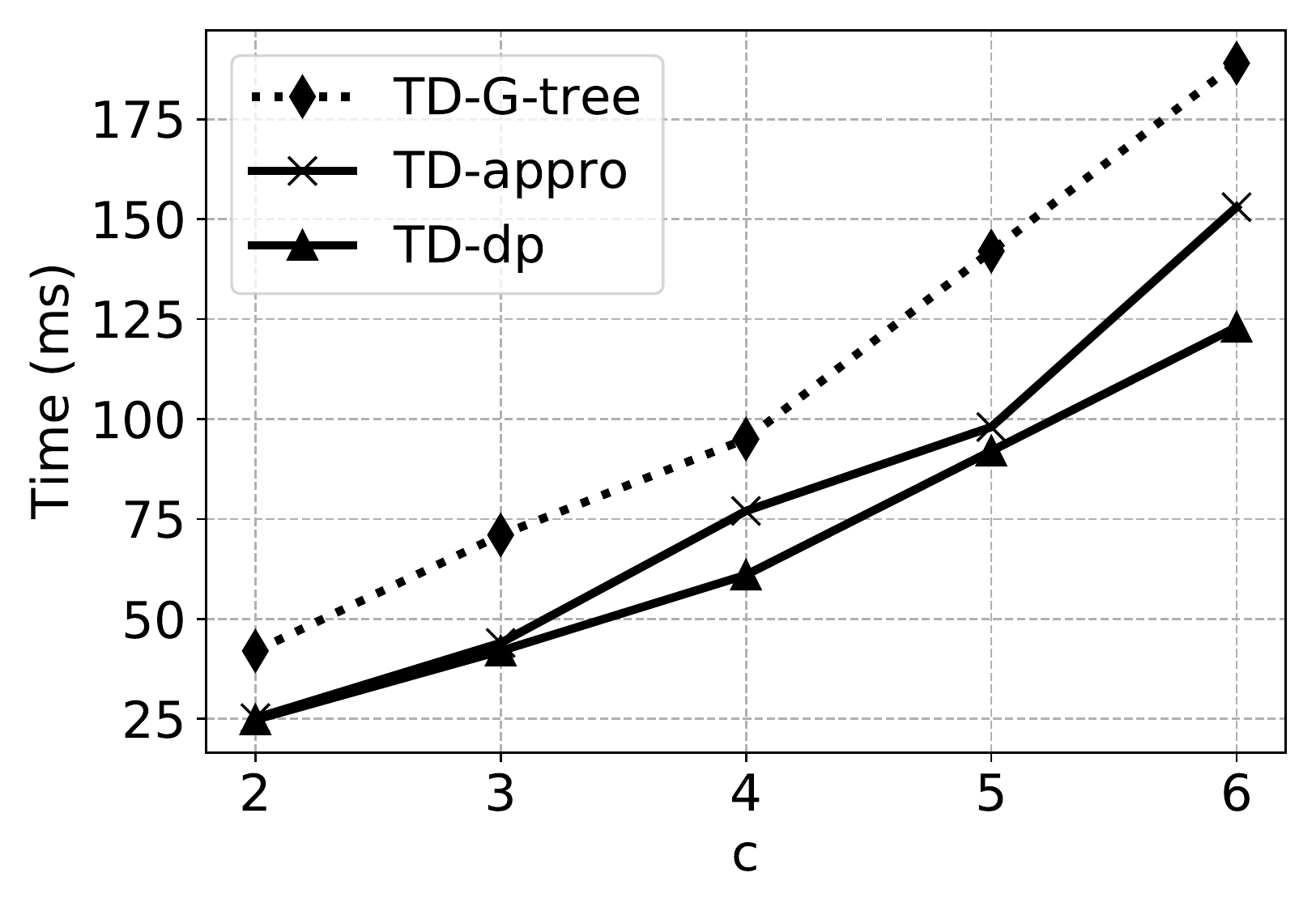}
		\vspace{-4ex}
		\caption{\footnotesize{Cost function query on $SF$}}
	\end{subfigure}
	
	\begin{subfigure}[b]{0.22\textwidth}
		\includegraphics[width=\textwidth]{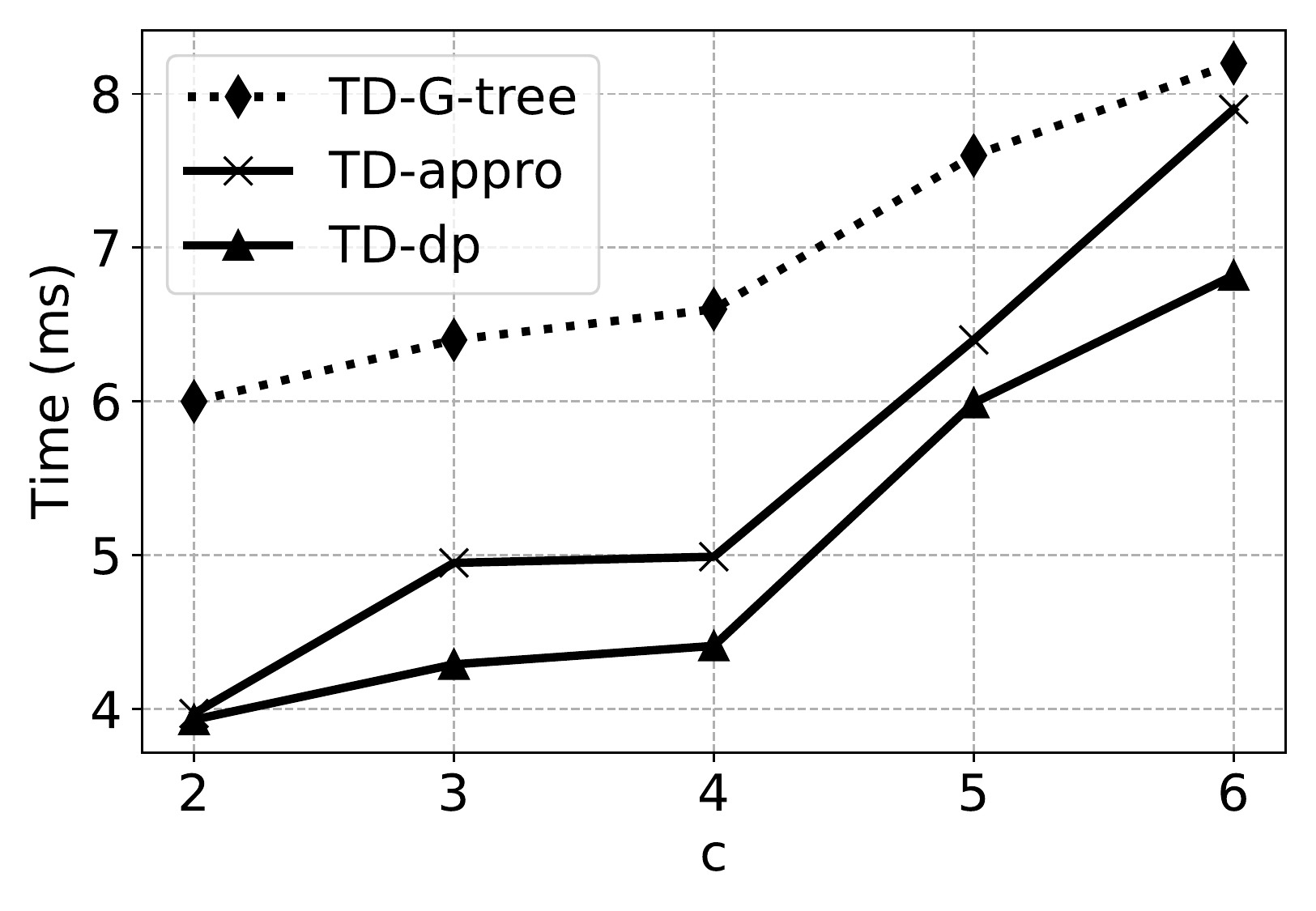}
		\vspace{-4ex}
		\caption{\footnotesize{Cost query on $COL$}}
	\end{subfigure}
	~~
    \begin{subfigure}[b]{0.22\textwidth}
		\includegraphics[width=\textwidth]{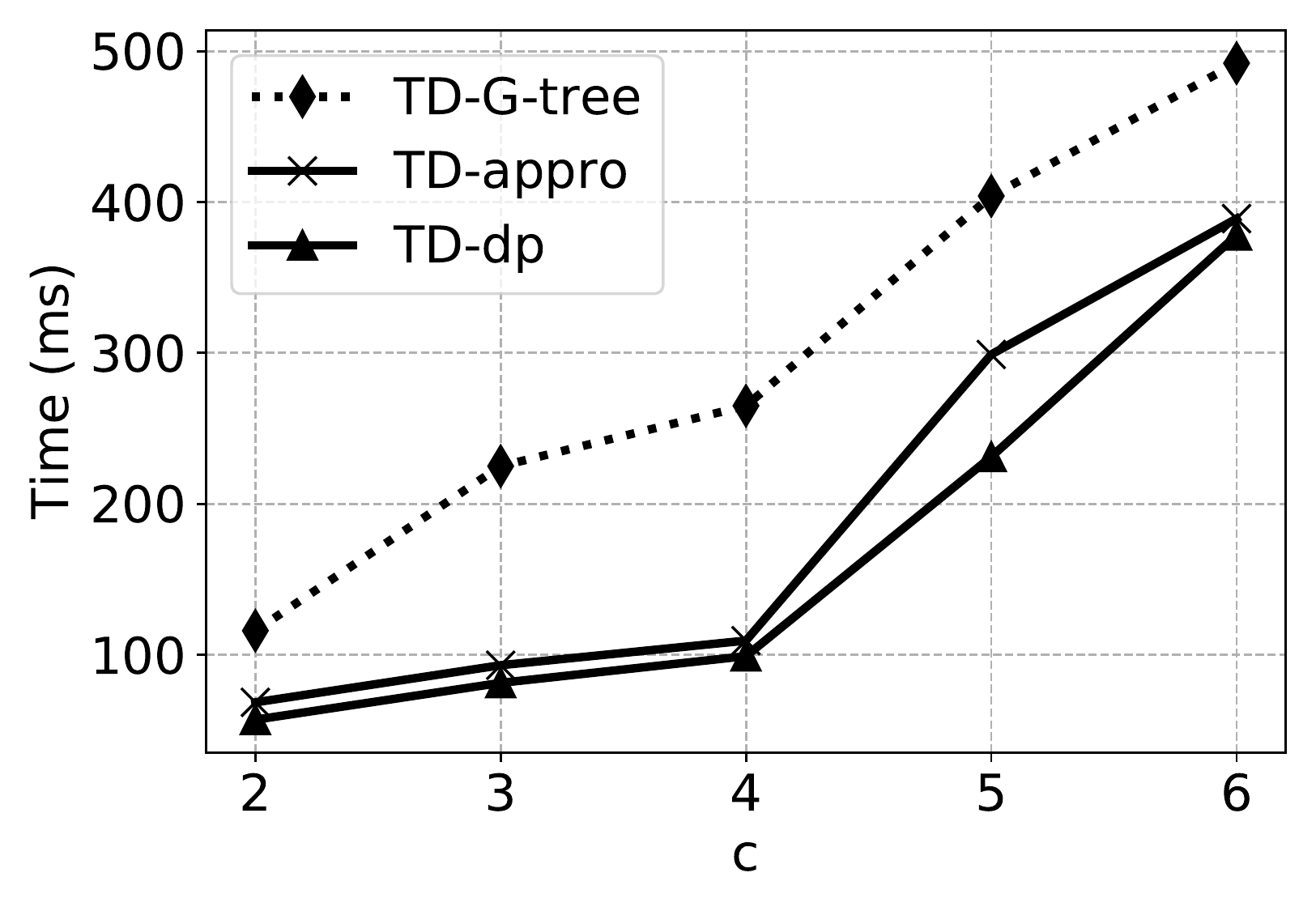}
		\vspace{-4ex}
		\label{exp1:13}
		\caption{\footnotesize{Cost function query on $COL$}}
	\end{subfigure}	
	
	\begin{subfigure}[b]{0.22\textwidth}
		\includegraphics[width=\textwidth]{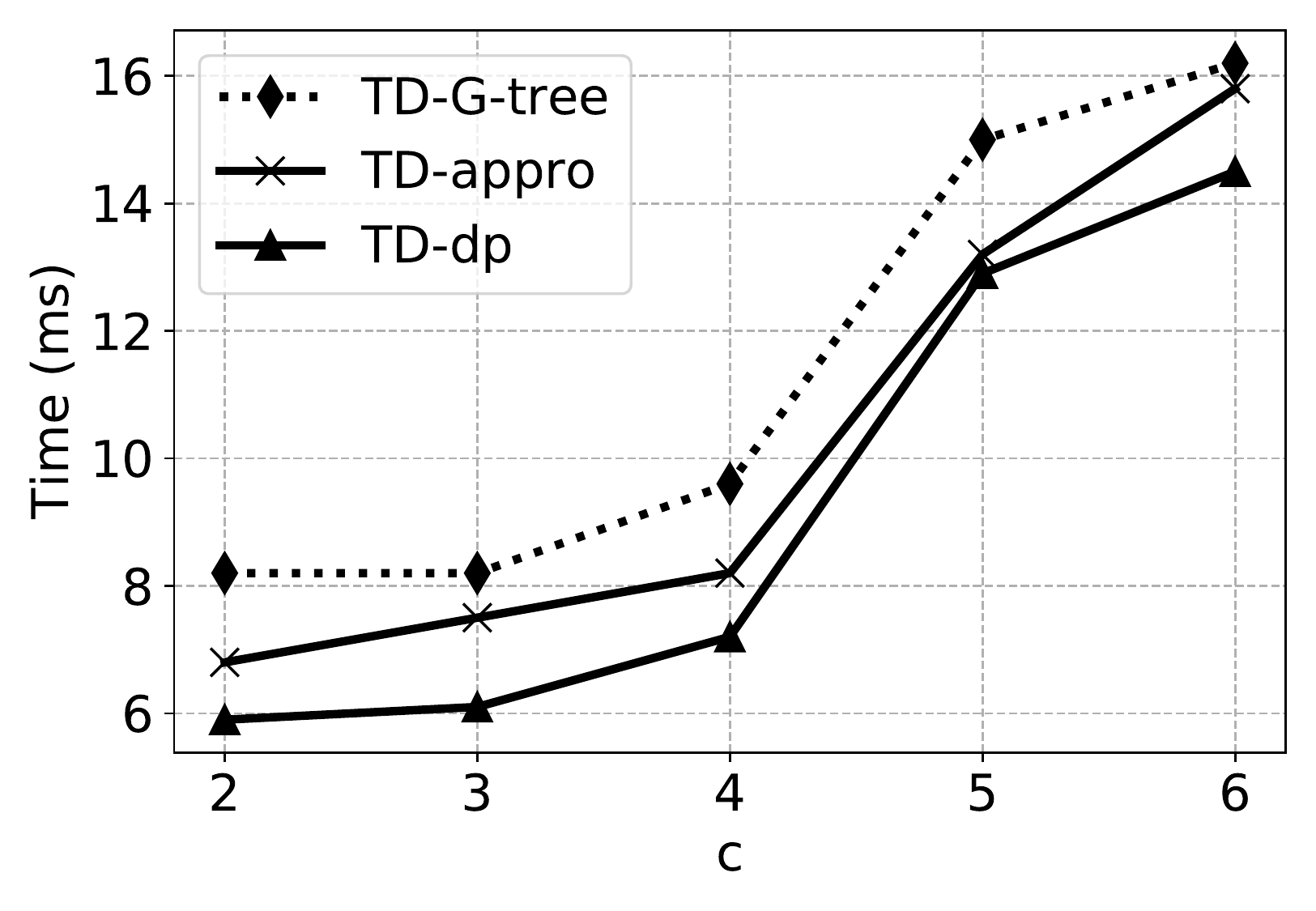}
		\vspace{-4ex}
		\caption{\footnotesize{Cost query on $FLA$}}
	\end{subfigure}
	~~
    \begin{subfigure}[b]{0.22\textwidth}
		\includegraphics[width=\textwidth]{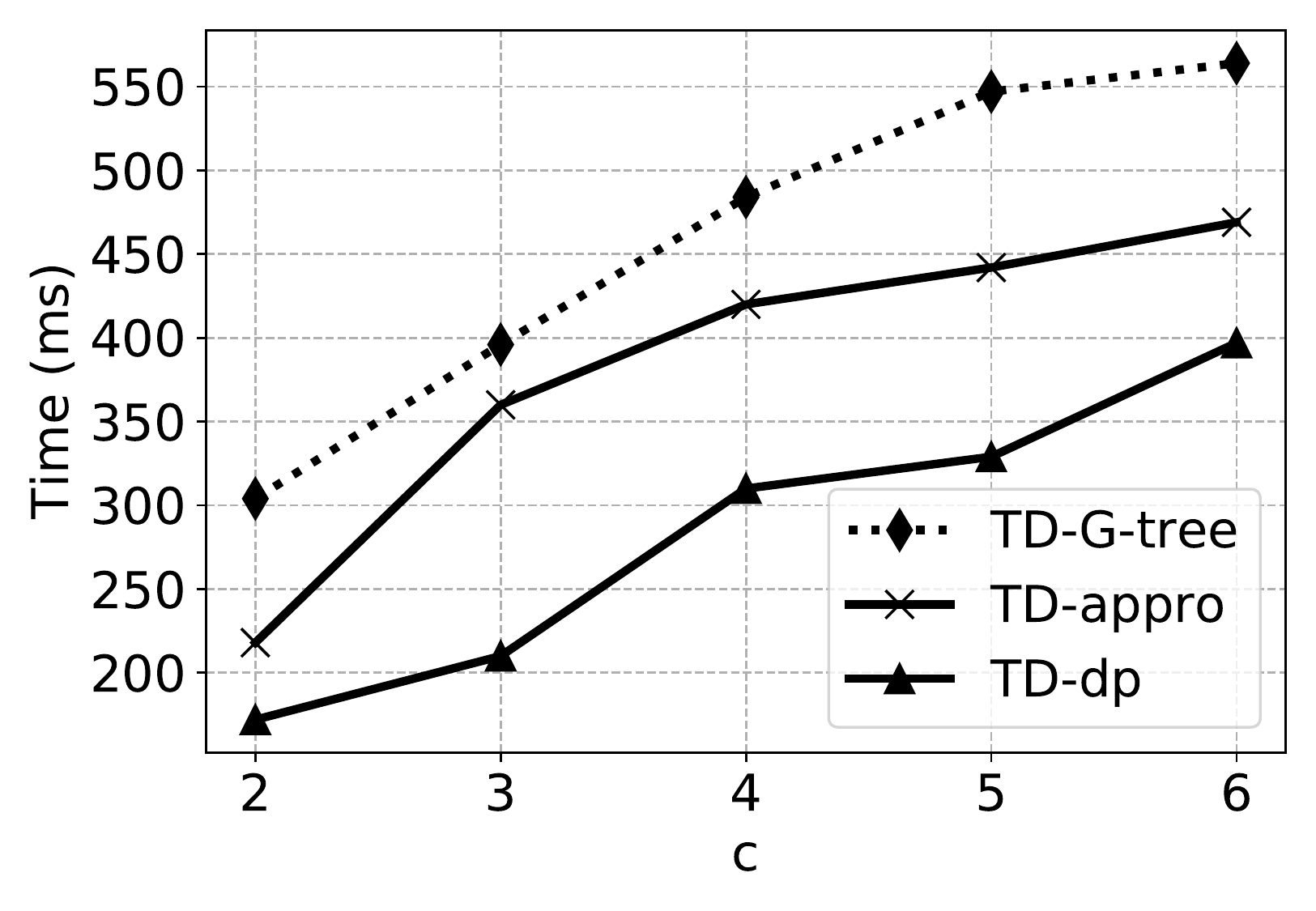}
		\vspace{-4ex}
		\caption{\footnotesize{Cost function query on $FLA$}}
	\end{subfigure}	
	
	\vspace{-2ex}
	\caption{Results of querying efficiency}
	\label{fig:exp1}
	\vspace{-3ex}
\end{figure}

\subsection{Evaluation on Query Efficiency}

\fakeparagraph{Evaluation on travel cost query} We evaluate the travel cost query efficiency and compare our proposed algorithms with TD-G-tree and TD-H2H. The results are shown in \figref{fig:exp1}.

We could see that the tree decomposition based methods outperform TD-G-tree by several orders of magnitude, besides that it is not impractical to expand TD-G-tree to the large-scale road network in the real world. For dataset $W-USA$, under the settings $c \ge 3$ (\eg{ the travel cost of each road changes more than twice in one day}), it takes more than 24 hours to build the index, so we do not report the results for these settings. This is because the hierarchical partition-building scheme suffers from data redundancy problems. When $c = 2$, the results for $W-USA$ are shown in \tabref{table:exp1_wusa}.
Although the TD-H2H index performs the best in the $CAL$ dataset in \tabref{table:exp1_cal}, this index cannot be extended to other road networks which have more than 100k vertices because of the huge memory cost (we will talk about this in the next subsection). Our methods TD-dp and TD-appro achieve taking nearly the same processing time as TD-H2H in the $CAL$ dataset, and they also perform well in large-scale road networks varied from 1,00,000 to more than 5,000,000 vertices. The travel cost time for both TD-dp and TD-appro grows slowly as the parameter $c$ become larger (\eg{ the travel cost changes frequently}), for all datasets. Comparing with the index TD-basic, although it is efficient to build and occupies the least memory, it performs worst in the query time. It is unacceptable to take more than 10 seconds to get the query result, this result also shows the necessity of building shortcuts over the index TD-basic.

\fakeparagraph{Evaluation on shortest travel cost function $f_{s,d}(t)$ query} We test the performance of the shortest travel cost function query on the generated query set of each dataset. \figref{fig:exp1} shows the result. 

The shortest travel cost function query takes much longer time compared with the travel cost query, this is because the shortest travel cost function query is required to calculate the $Compound()$ operator functions while the algorithm traverses the tree index. Moreover, with the increase of the parameter $c$ (the travel time of the road segment changes more times in one day), the query time of TD-G-tree increases significantly in each dataset. The results also show that the basic tree decomposition technique can not be implemented into the time-dependent road network directly, it takes more than 10 seconds to get the shortest travel cost function between two vertices. With the selected and build shortcuts over the basic tree decomposition, our algorithms TD-dp and TD-appro can perform much better. We also find that compared with  TD-G-tree, TD-dp and TD-appro reduce the query time by an average of 109 ms and 77 ms respectively in all datasets ($c = 2$).

\begin{figure}[t]
	\centering
	\begin{subfigure}[b]{0.22\textwidth}
		\includegraphics[width=\textwidth]{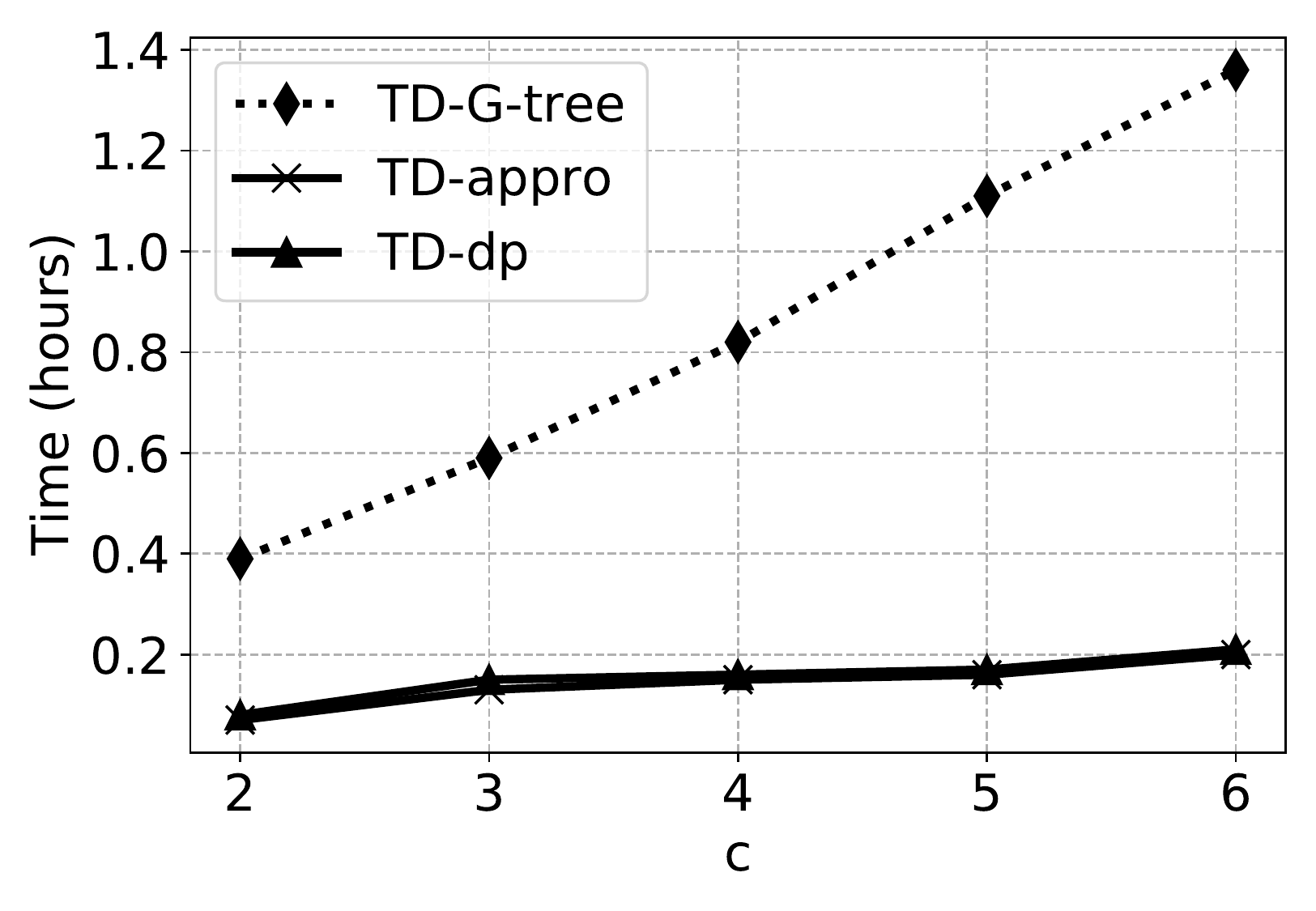}
		\vspace{-4ex}
		\caption{\footnotesize{Construction time on $SF$}}
	\end{subfigure}
	~~
    \begin{subfigure}[b]{0.22\textwidth}
		\includegraphics[width=\textwidth]{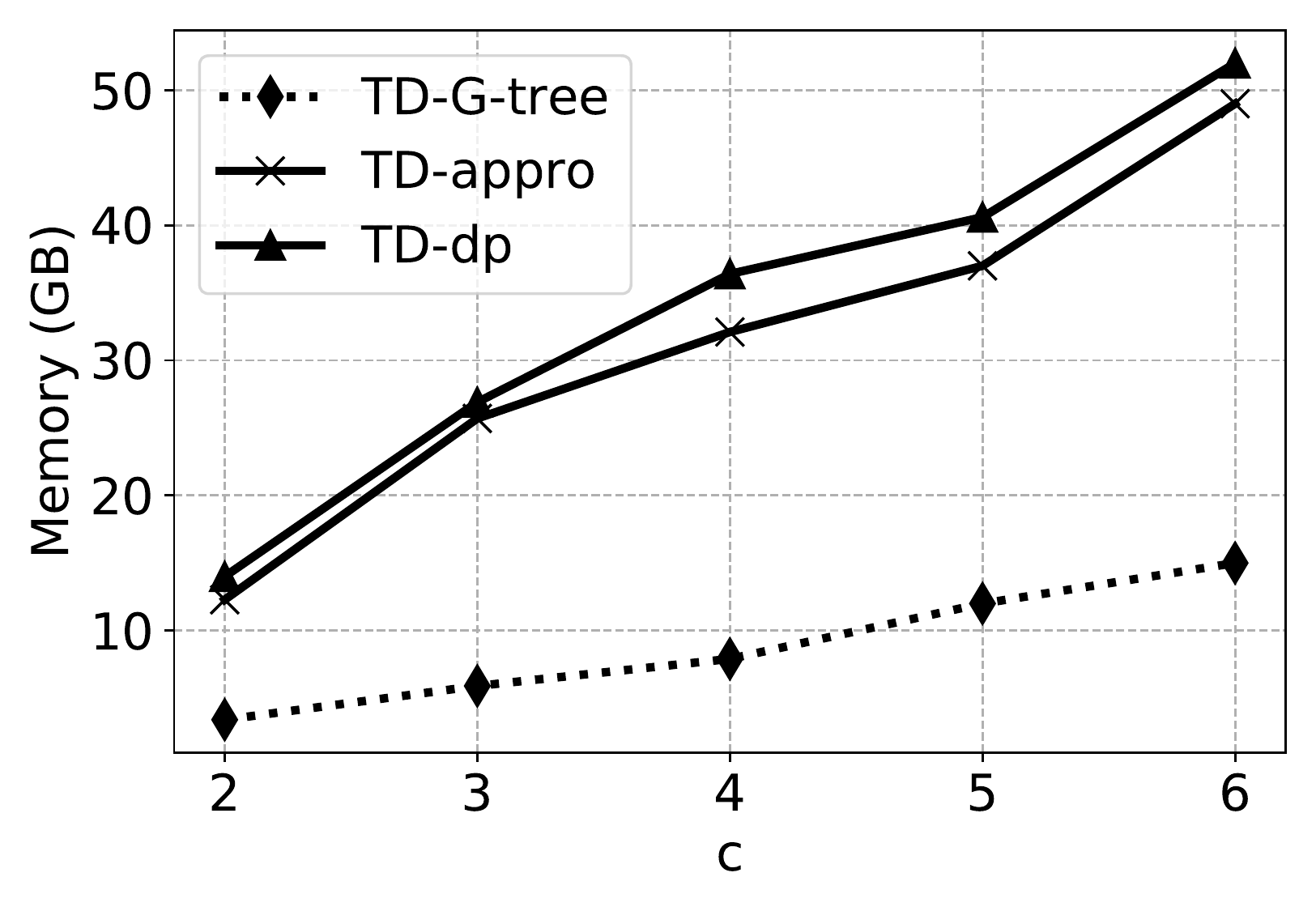}
		\vspace{-4ex}
		\caption{\footnotesize{Memory cost on $SF$}}
	\end{subfigure}
	
	\begin{subfigure}[b]{0.22\textwidth}
		\includegraphics[width=\textwidth]{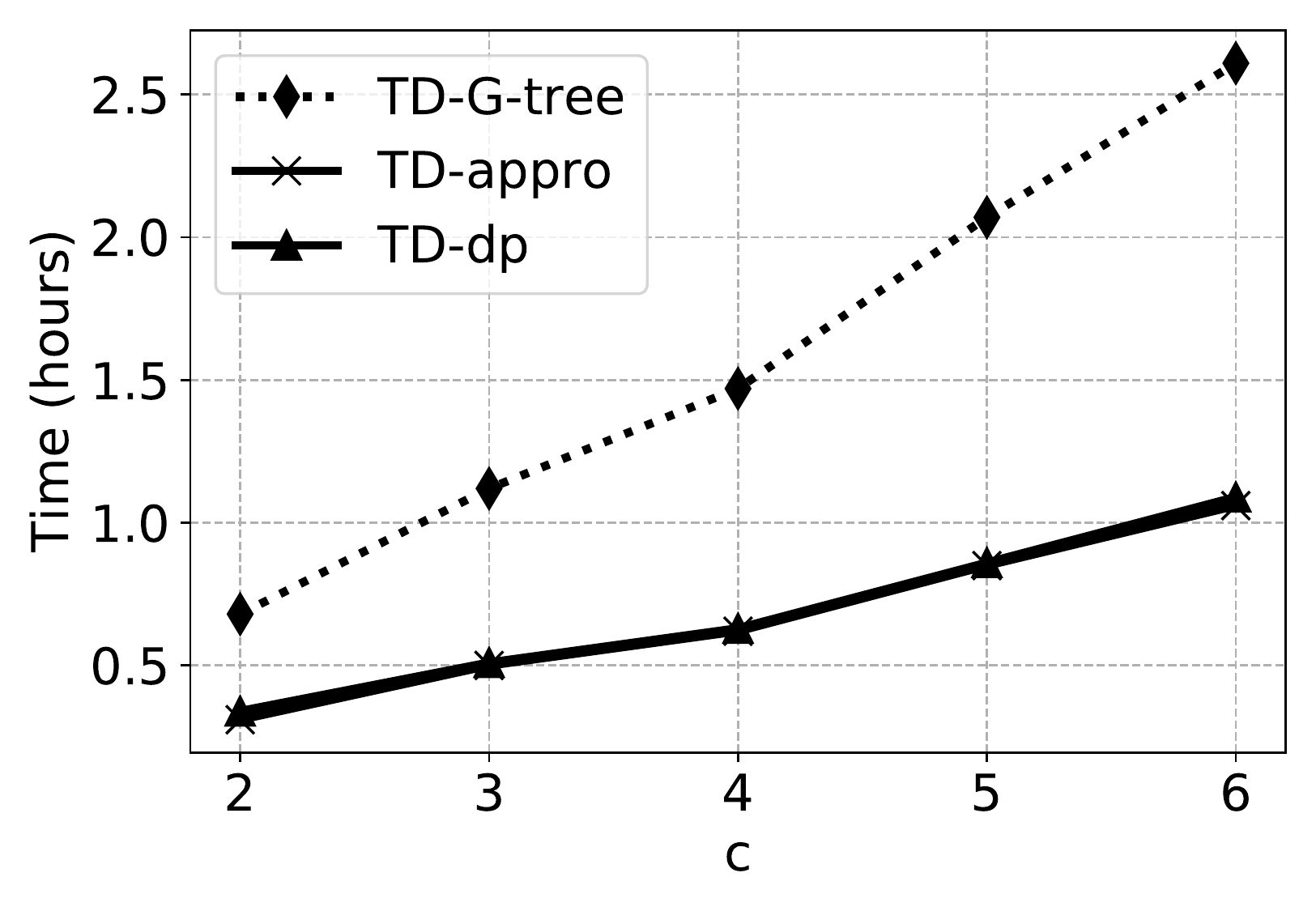}
		\vspace{-4ex}
		\caption{\footnotesize{Construction time on $COL$}}
	\end{subfigure}
	~~
    \begin{subfigure}[b]{0.22\textwidth}
		\includegraphics[width=\textwidth]{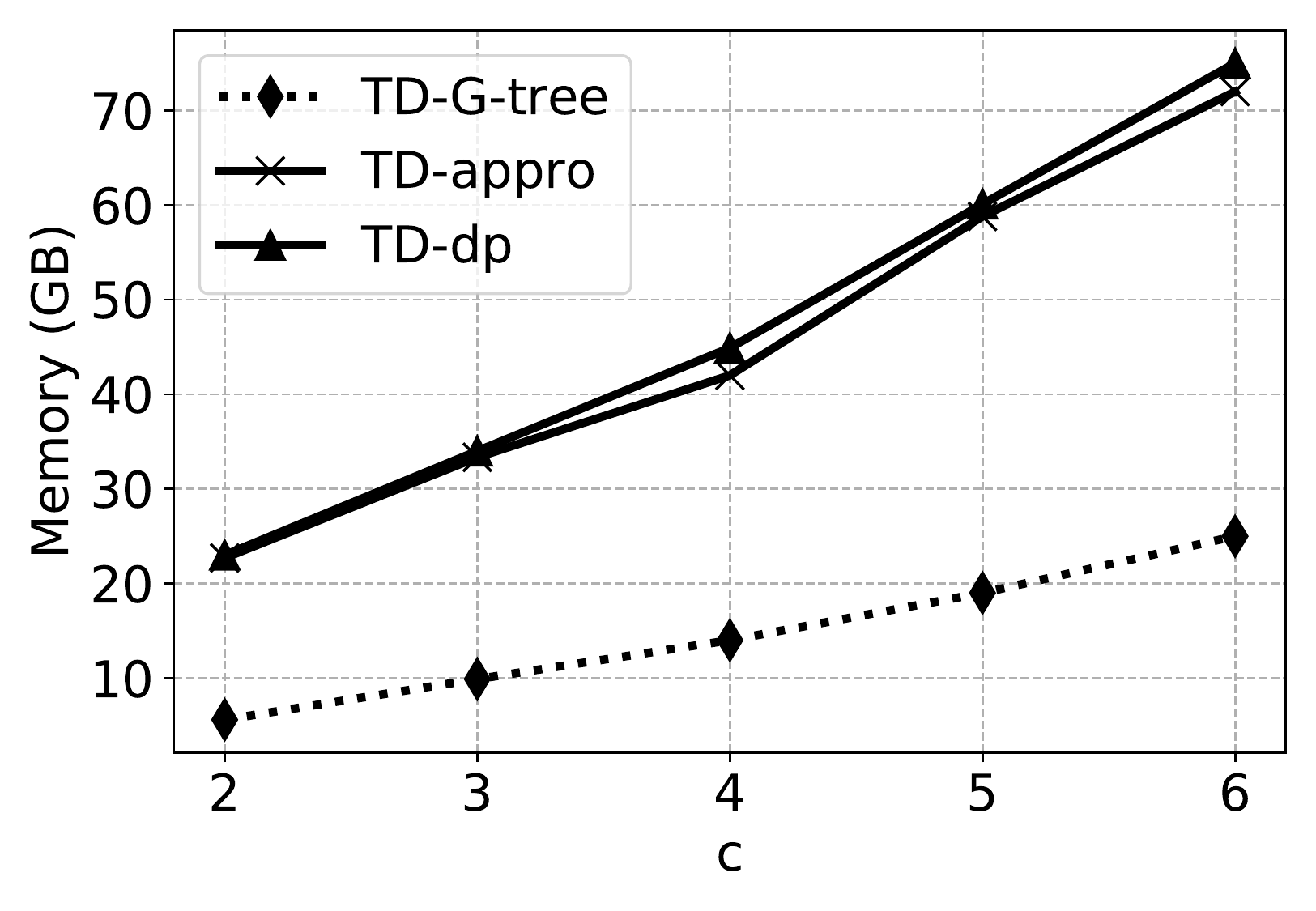}
		\vspace{-4ex}
		\caption{\footnotesize{Memory cost on $COL$}}
	\end{subfigure}	
	
	\begin{subfigure}[b]{0.22\textwidth}
		\includegraphics[width=\textwidth]{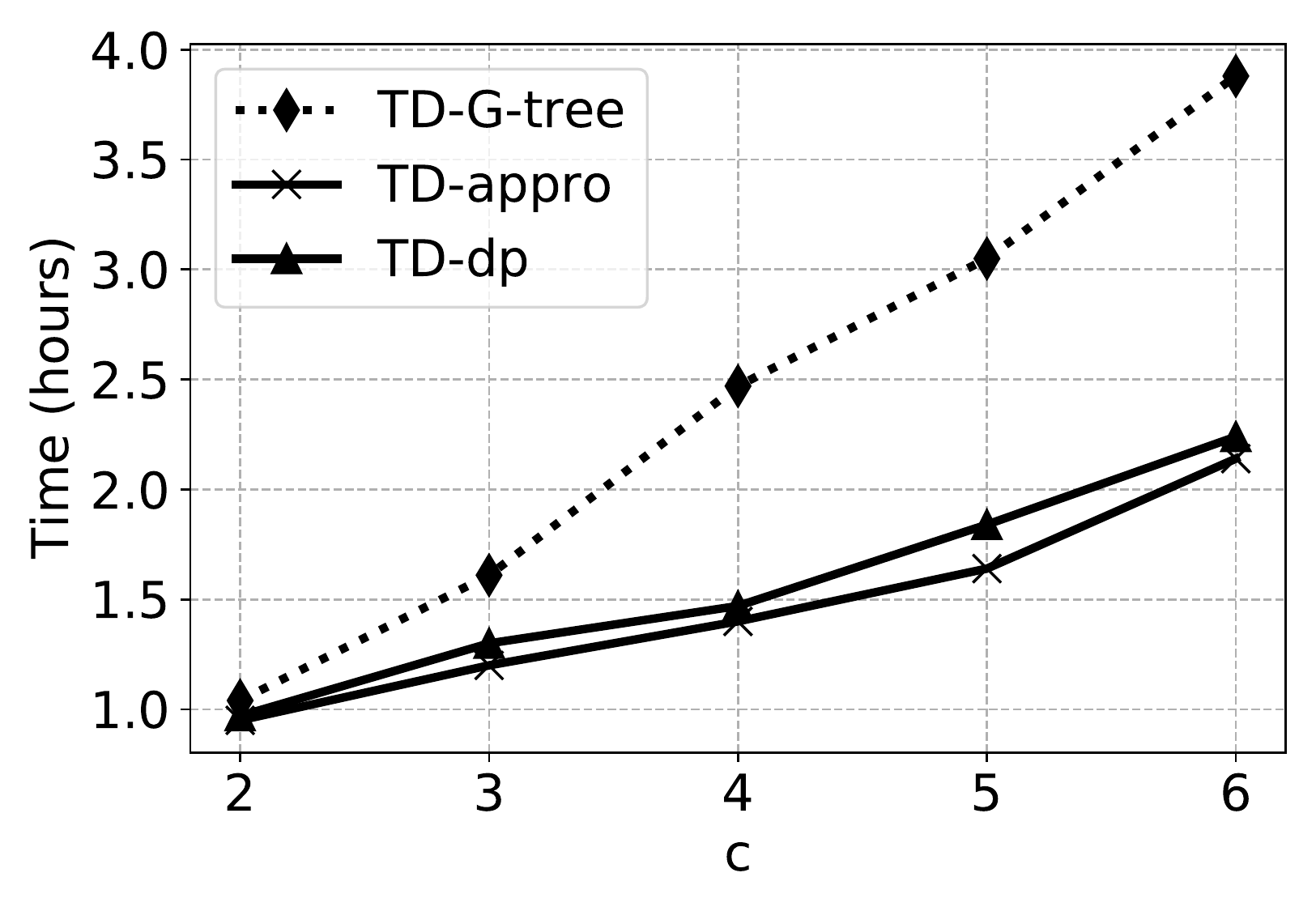}
		\vspace{-4ex}
		\caption{\footnotesize{Construction time on $FLA$}}
	\end{subfigure}
	~~
    \begin{subfigure}[b]{0.22\textwidth}
		\includegraphics[width=\textwidth]{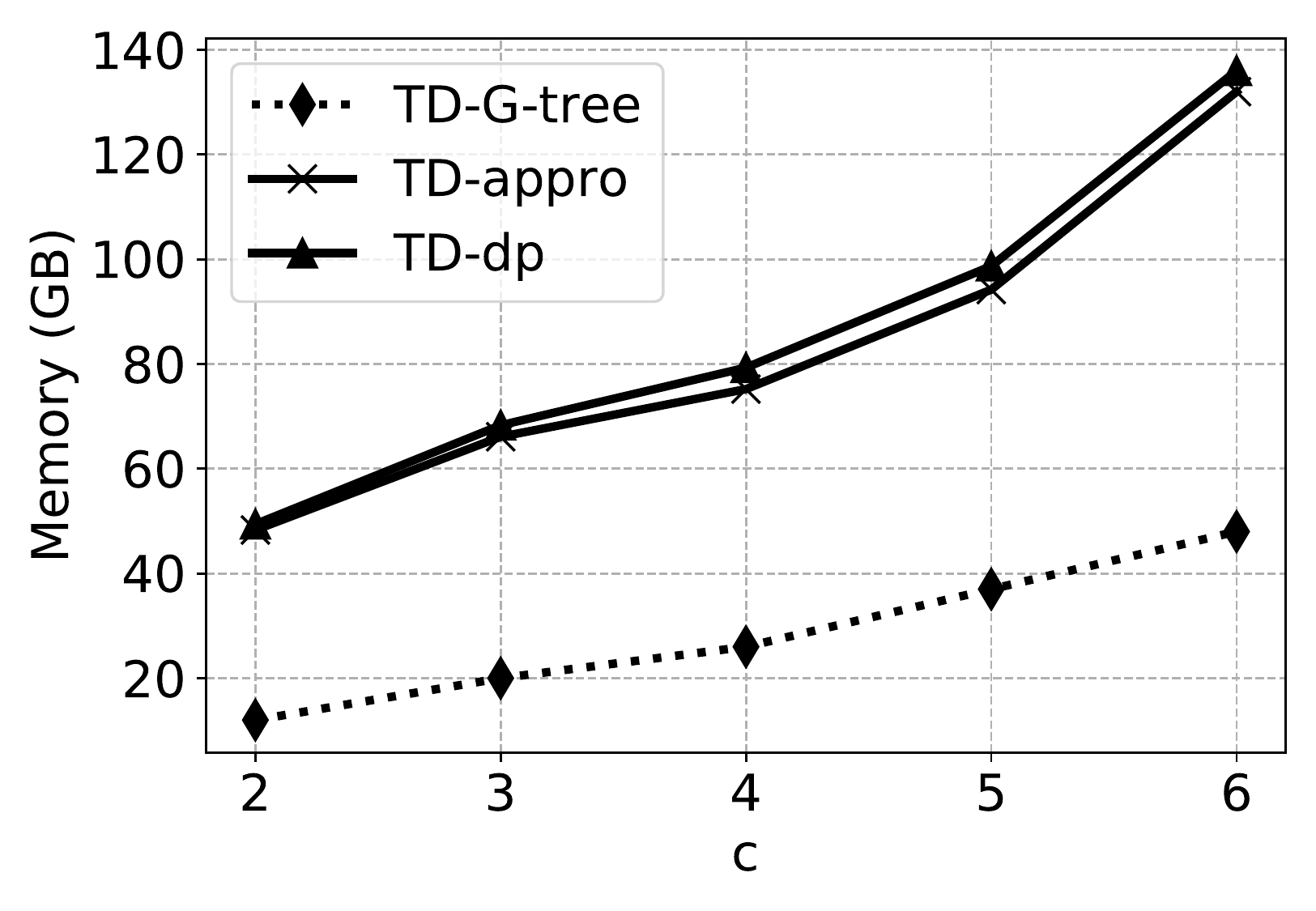}
		\vspace{-4ex}
		\caption{\footnotesize{Memory cost on $FLA$}}
	\end{subfigure}	
	
	\vspace{-2ex}
	\caption{Results of construction cost}
	\label{fig:exp2}
	\vspace{-3ex}
\end{figure}

\begin{figure}[ht]
    \centering
    \begin{minipage}[b]{0.45\linewidth}
    \includegraphics[width=\textwidth]{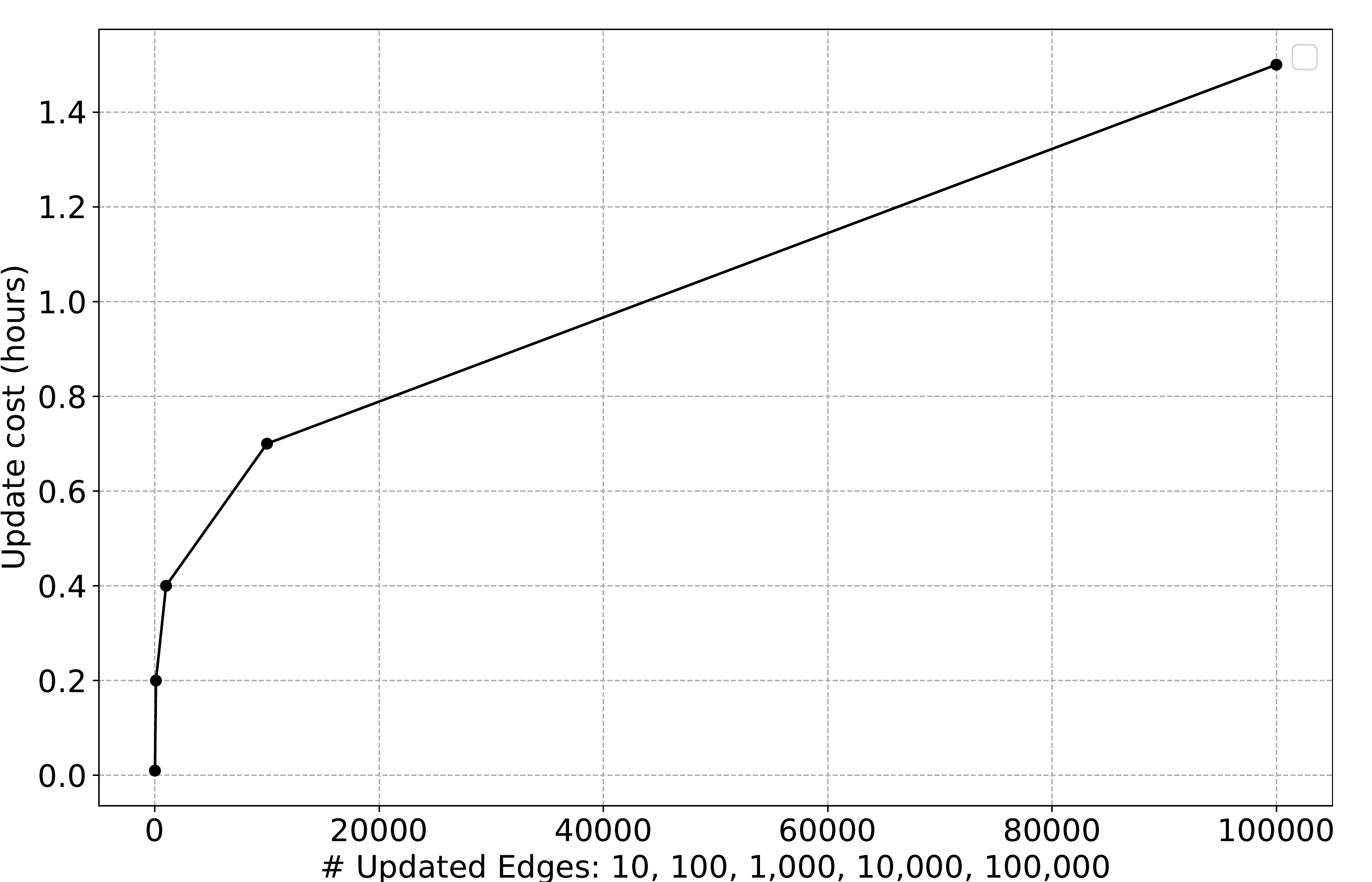}
    \caption{\footnotesize{Index update on $SF$}}
    \label{fig:exp3a}
    \end{minipage}
    \quad
    \begin{minipage}[b]{0.45\linewidth}
    \includegraphics[width=\textwidth]{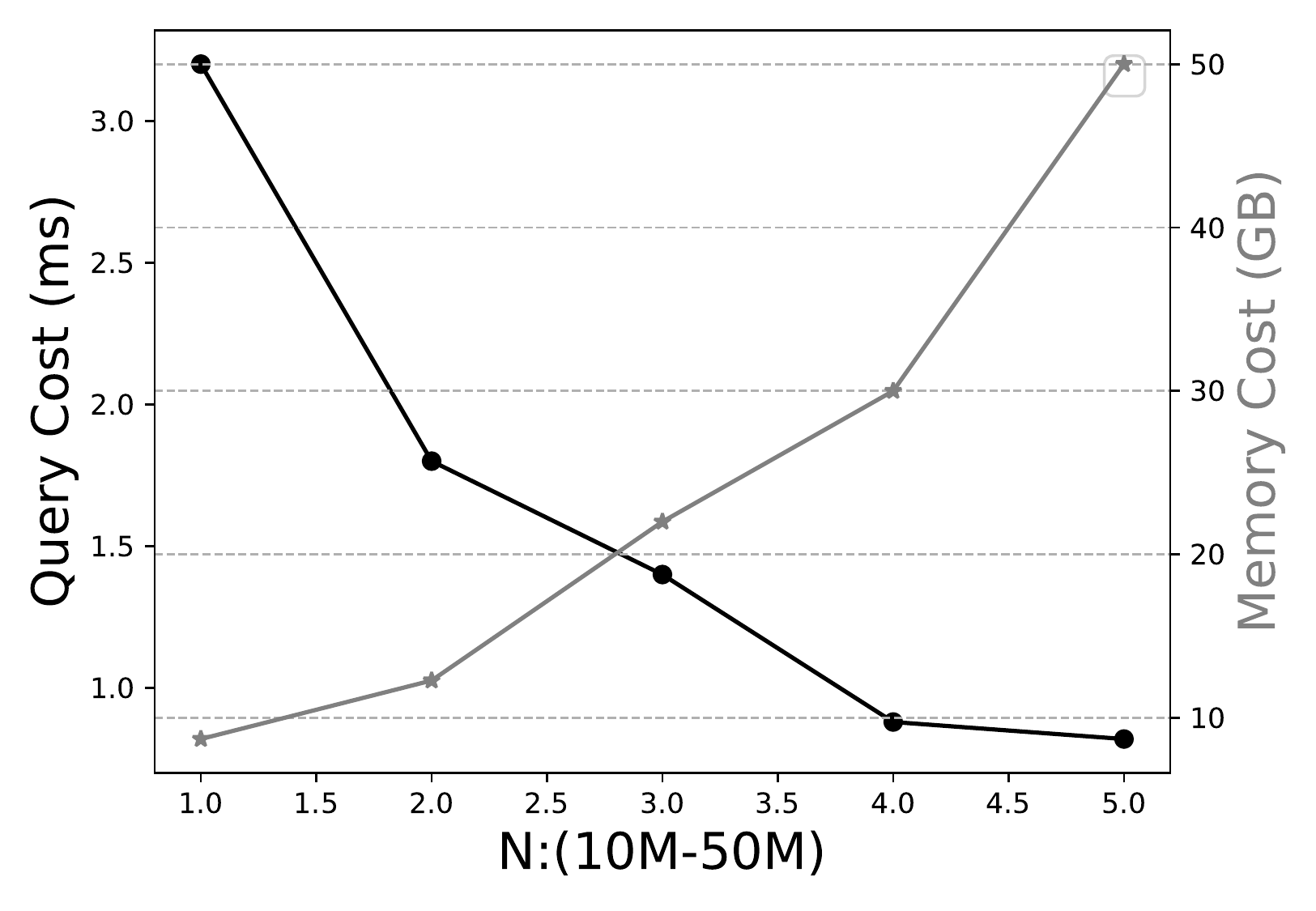}
    \caption{\footnotesize{Evaluation $N$ on $FLA$}}
    \label{fig:exp3b}
    \end{minipage}
    \label{fig:exp3}
    \vspace{-2ex}
\end{figure}

\subsection{Evaluation on Index Construction Cost}

\fakeparagraph{Evaluation on construction time} We evaluate the construction time of each index, and we compare our proposed methods with TD-G-tree and TD-H2H. \figref{fig:exp2} illustrates the index building time. We can note that the tree decomposition based methods can build the indexes much faster than TD-G-tree. The TD-basic takes no more than 1 second to build the index for dataset $CAL$. However, the index construction time grows dramatically as the parameter $c$ becomes larger, \eg{ the construction time of TD-G-tree increases from 3,761s to 13,972s when $c$ increases from 2 to 6}. Besides that, for the dataset $W-USA$ with more than 5,000,000 vertices, when $c$ is at least 3, the TD-G-tree takes more than 24 hours to build the index, therefore it is inefficient to adopt the TD-G-tree on the large-scale road network in the real world. For TD-dp and TD-appro, they have much more stable construction time under the different $c$ values compared with the baseline method. 

\fakeparagraph{Evaluation on memory cost}
The experimental
results for the memory cost are shown in \figref{fig:exp2}. We can observe that when the value of parameter $c$ increases, the index size of all approaches will increase. The index size of TD-H2H (5.7G) is at least 34 times larger than TD-G-tree (169MB) for dataset $CAL$. This is because, each node in TD-H2H, maintains not only auxiliary information related to its ancestors and also the shortest travel cost functions to all its ancestors. When the dataset becomes larger (\eg{ for $SF$ with more than 10,000 vertices}), TD-H2H occupies too much memory to support the query. However, the index sizes of TD-G-tree, TD-dp and TD-appro are comparable. This is because, we select the set of tree node pairs to build the shortcuts over the basic tree decomposition index TD-basic, the TD-basic has the smallest index size for all datasets (\eg{  TD-G-tree takes 5 times larger space than TD-basic}).

\fakeparagraph{Evaluation on Index Update} We vary the update ratio of the edges to evaluate the update efficiency of TD-appro, the results are shown in \figref{fig:exp3a}. For each dataset, we randomly choose 10, 100, 1,000, 10,000 and 100,000 of total number of edges, to update the weight functions. Then, we find the associated vertex $v$ for each chosen edge, and update the shortcuts in $G(v)$ based on the top-down manner in \textbf{Fact 1}. Finally, the total time cost is reported. We can see that the update time increased with the update ratio becomes larger, and the total update cost is acceptable.

\subsection{Evaluation on Constraint $N$} We vary the selection constraint $N$ to test the performance of TD-appro. The default value of $N$ is shown in \tabref{table:dataset}. The results are shown in \figref{fig:exp3b}, we can observe that the construction time and memory cost increase as the value $N$ increases. When $N$ becomes larger, the query processing time becomes shorter. This is because, if $N$ is larger, more shortcuts will be selected and built in the indexes TD-dp and TD-appro, which will take more time and space. However, more shortcuts could improve the efficiency of query processing.

\subsection{Summary of Experiments}
We summarize the results as follows:
\begin{itemize}
  \item
  Existing indexes are far from efficient to answer the shortest travel path query over the time-dependent road network. Especially,  when the road network is large-scale and the travel pattern changes frequently. For example, TD-G-tree takes more than 24 hours to build the index on $W-USA$. The TD-H2H occupies too much space to handle the road network with more than 300,000 vertexes (\eg{$SF$}). Therefore, it is impractical to extend the existing indexes to real-world applications.
  
  \item
   The proposed indexes TD-dp and TD-appro reduce 114 ms and 103 ms on $COL$, 170 ms and 95 ms on $FLA$ for the shortest travel cost function query respectively. As for the index construction, these two methods are nearly 2 times faster than the baseline, due to their construction based on the tree decomposition. 
  
  \item
    The TD-dp takes 0.01 to 0.2 hours more than TD-appro to build the index. The results prove the efficiency of the approximation shortcut selection algorithm. The query processing time of TD-dp is slightly smaller than TD-appro,  \eg{ no more than 30 ms}.
  
\end{itemize}

\section{Related Work}\label{sec:related}

\textbf{Time-dependent shortest path.} The time-dependent shortest path problem is first proposed in \cite{TDSP}, a recursion formula is utilized to model this problem. Dijkstra based algorithms are first designed in \cite{nonindex2} and \cite{DBLP:journals/algorithmica/DehneOS12}, the time complexity is $O((n \cdot logn+m)\cdot c)$. The algorithms can have bad performance in the large-scale time-dependent road networks.
In \cite{nonindex2}, the algorithm takes more than 10 seconds to answer the query over a graph with 10k vertices. $A^*$ algorithm is also extended to time-dependent road networks in \cite{DBLP:conf/icde/KanoulasDXZ06}. However in the worst case, the proposed method may take exponential running time to expand all possible shortest paths in the graph. Based on the $A^*$ algorithm, some improvement techniques like landmarks and bidirectional search scheme have also been implemented over the time-dependent scenario in  \cite{DBLP:conf/ssd/DemiryurekKSR11} and \cite{DBLP:journals/networks/NanniciniDSL12}. However these methods can not work well in the really large-scale road networks. 

There are also some hierarchy and index based studies to accelerate the shortest path query over time-dependent road networks. The algorithms proposed in \cite{DBLP:conf/esa/Delling08} and \cite{DBLP:conf/alenex/BatzDSV09} implement algorithms SHARC and CH to time-dependent road networks, which are two efficient methods in static road network. However, each vertex could take kilobyte memory \cite{DBLP:conf/alenex/BatzDSV09}, hence it is prohibitive to apply this method to the real world time-dependent road network with more than tens of millions vertices. For the index based method \cite{shortestquery}, it builds the tree based index by partitioning the road network in a hierarchical way, and the travel cost information between any pair of the border vertices in the partition is maintained. When there is a query, the index assembles the travel cost from bottom to up in the tree. The hierarchical index building scheme generates a lot of redundant travel cost information, which hinders the efficiency of implementing this method to the situation, where the road network is large-scale and the travel cost of road segments changes frequently.

\textbf{Tree decomposition.} Tree decomposition is an efficient technique which has been used in many applications to speed up certain computational problems over the graph structure \cite{graph}. 
Given a graph, it is NP-complete to determine if it has a treewidth which is at most a given value \cite{10.1137/0608024}. Same as the previous studies, we adopt a sub-optimal algorithm \cite{treedecomposition} in this paper. Although this algorithm cannot have a bounded approximation ratio to compute a minimum treewidth, it can achieve a reasonable constant treewidth even in the large real-world road networks. Consider the time complexity of our query algorithm is polynomial to the treewidth of the tree decomposition, this sub-optimal method is also practically applicable in the time-dependent road networks. 

Tree decomposition method has been widely used to improve the efficiency of shortest path computation over static road networks. In \cite{h2h}, the tree decomposition helps to overcome the shortcomings of both hierarchy-based solution and hop-based solution. For each tree node of the tree in \cite{h2h}, all ancestors have higher rank to this node, and this node maintains 1-hop shortest distances to all ancestors. Given a query source and destination, the algorithm achieves $O(w(T_G))$ time query by checking the 1-hop distances from source and destination to ancestors respectively. Motivated by \cite{h2h}, Ref. \cite{p2h} tries to improve the efficiency of querying shortest distance by reducing the label size of each tree node. Besides the shortest distance query, the tree decomposition method also supports other diverse queries over static road networks. In \cite{lsd}, the tree decomposition builds a new tree index which contains the label-constrained information of the edges. The tree decomposition based label-constrained shortest path query outperforms the baseline method significantly. For the shortest path counting problem, a novel index called TL-index is proposed in \cite{tlindex}. The TL-index is built based on the tree decomposition method, each tree node of the index preserves the shortest distance and the associated number of shortest path between it to its ancestors.

\section{conclusion}	\label{sec:conclusion}

In this paper, we study the shortest path query problem over large-scale time-dependent road networks. This operation has been extensively studied over static road networks. However, in many real-world applications, the travel cost of a path depends on the start time of the path. Therefore, in this paper, we first define the shortest path query problem over time-dependent road networks. Then we adopt the tree decomposition technique to build a tree over the graph to accelerate the shortest path processing. Further, based on the properties of the tree, we select and build some shortcuts to improve the efficiency of calculating the shortest path over the tree structure. Next, we model the shortcut selection problem over the tree and prove its NP-hardness. To utilize the limited main memory, we propose two selection algorithms to solve this problem. The first one is dynamic programming based selection algorithm, and the other is a greedy approximation based algorithm, and we prove that it has an approximation ratio of 0.5. Finally, we conduct extensive performance studies to demonstrate the effectiveness and efficiency of shortcut selection algorithms and query processing based on the selected shortcuts.


\bibliographystyle{ACM-Reference-Format}
\bibliography{sample}


\begin{thebibliography}{34}


\ifx \showCODEN    \undefined \def \showCODEN     #1{\unskip}     \fi
\ifx \showDOI      \undefined \def \showDOI       #1{#1}\fi
\ifx \showISBNx    \undefined \def \showISBNx     #1{\unskip}     \fi
\ifx \showISBNxiii \undefined \def \showISBNxiii  #1{\unskip}     \fi
\ifx \showISSN     \undefined \def \showISSN      #1{\unskip}     \fi
\ifx \showLCCN     \undefined \def \showLCCN      #1{\unskip}     \fi
\ifx \shownote     \undefined \def \shownote      #1{#1}          \fi
\ifx \showarticletitle \undefined \def \showarticletitle #1{#1}   \fi
\ifx \showURL      \undefined \def \showURL       {\relax}        \fi
\providecommand\bibfield[2]{#2}
\providecommand\bibinfo[2]{#2}
\providecommand\natexlab[1]{#1}
\providecommand\showeprint[2][]{arXiv:#2}

\bibitem[\protect\citeauthoryear{Arnborg, Corneil, and Proskurowski}{Arnborg
  et~al\mbox{.}}{1987}]%
        {10.1137/0608024}
\bibfield{author}{\bibinfo{person}{Stefan Arnborg}, \bibinfo{person}{Derek~G.
  Corneil}, {and} \bibinfo{person}{Andrzej Proskurowski}.}
  \bibinfo{year}{1987}\natexlab{}.
\newblock \showarticletitle{Complexity of Finding Embeddings in a K-Tree}.
\newblock \bibinfo{journal}{\emph{SIAM J. Algebraic Discrete Methods}}
  \bibinfo{volume}{8}, \bibinfo{number}{2} (\bibinfo{date}{apr}
  \bibinfo{year}{1987}), \bibinfo{pages}{277–284}.
\newblock
\showISSN{0196-5212}
\urldef\tempurl%
\url{https://doi.org/10.1137/0608024}
\showDOI{\tempurl}


\bibitem[\protect\citeauthoryear{Batz, Delling, Sanders, and Vetter}{Batz
  et~al\mbox{.}}{2009}]%
        {DBLP:conf/alenex/BatzDSV09}
\bibfield{author}{\bibinfo{person}{Gernot~Veit Batz}, \bibinfo{person}{Daniel
  Delling}, \bibinfo{person}{Peter Sanders}, {and} \bibinfo{person}{Christian
  Vetter}.} \bibinfo{year}{2009}\natexlab{}.
\newblock \showarticletitle{Time-Dependent Contraction Hierarchies}. In
  \bibinfo{booktitle}{\emph{Proceedings of the Eleventh Workshop on Algorithm
  Engineering and Experiments, {ALENEX} 2009, New York, New York, USA, January
  3, 2009}}, \bibfield{editor}{\bibinfo{person}{Irene Finocchi} {and}
  \bibinfo{person}{John Hershberger}} (Eds.). \bibinfo{publisher}{{SIAM}},
  \bibinfo{pages}{97--105}.
\newblock
\urldef\tempurl%
\url{https://doi.org/10.1137/1.9781611972894.10}
\showDOI{\tempurl}


\bibitem[\protect\citeauthoryear{Bondy and Murty}{Bondy and Murty}{2008}]%
        {graph}
\bibfield{author}{\bibinfo{person}{J.A. Bondy} {and} \bibinfo{person}{U.S.R
  Murty}.} \bibinfo{year}{2008}\natexlab{}.
\newblock \bibinfo{booktitle}{\emph{Graph Theory} (\bibinfo{edition}{1st}
  ed.)}.
\newblock \bibinfo{publisher}{Springer Publishing Company, Incorporated}.
\newblock
\showISBNx{1846289696}


\bibitem[\protect\citeauthoryear{Chen, Yuan, Du, Cheng, and Wang}{Chen
  et~al\mbox{.}}{2021b}]%
        {yuanyeicde2019}
\bibfield{author}{\bibinfo{person}{Di Chen}, \bibinfo{person}{Ye Yuan},
  \bibinfo{person}{Wenjin Du}, \bibinfo{person}{Yurong Cheng}, {and}
  \bibinfo{person}{Guoren Wang}.} \bibinfo{year}{2021}\natexlab{b}.
\newblock \showarticletitle{Online Route Planning over Time-Dependent Road
  Networks}. In \bibinfo{booktitle}{\emph{2021 IEEE 37th International
  Conference on Data Engineering (ICDE)}}. \bibinfo{pages}{325--335}.
\newblock
\urldef\tempurl%
\url{https://doi.org/10.1109/ICDE51399.2021.00035}
\showDOI{\tempurl}


\bibitem[\protect\citeauthoryear{Chen, Fu, Jiang, Lo, and Zhang}{Chen
  et~al\mbox{.}}{2021a}]%
        {p2h}
\bibfield{author}{\bibinfo{person}{Zitong Chen}, \bibinfo{person}{Ada Wai-Chee
  Fu}, \bibinfo{person}{Minhao Jiang}, \bibinfo{person}{Eric Lo}, {and}
  \bibinfo{person}{Pengfei Zhang}.} \bibinfo{year}{2021}\natexlab{a}.
\newblock \showarticletitle{P2H: Efficient Distance Querying on Road Networks
  by Projected Vertex Separators} \emph{(\bibinfo{series}{SIGMOD '21})}.
  \bibinfo{publisher}{Association for Computing Machinery},
  \bibinfo{address}{New York, NY, USA}, \bibinfo{pages}{313–325}.
\newblock
\showISBNx{9781450383431}
\urldef\tempurl%
\url{https://doi.org/10.1145/3448016.3459245}
\showDOI{\tempurl}


\bibitem[\protect\citeauthoryear{Cooke and Halsey}{Cooke and Halsey}{1966a}]%
        {nonindex1}
\bibfield{author}{\bibinfo{person}{Kenneth~L Cooke} {and} \bibinfo{person}{Eric
  Halsey}.} \bibinfo{year}{1966}\natexlab{a}.
\newblock \showarticletitle{The shortest route through a network with
  time-dependent internodal transit times}.
\newblock \bibinfo{journal}{\emph{Journal of mathematical analysis and
  applications}} \bibinfo{volume}{14}, \bibinfo{number}{3}
  (\bibinfo{year}{1966}), \bibinfo{pages}{493--498}.
\newblock


\bibitem[\protect\citeauthoryear{Cooke and Halsey}{Cooke and Halsey}{1966b}]%
        {TDSP}
\bibfield{author}{\bibinfo{person}{Kenneth~L Cooke} {and} \bibinfo{person}{Eric
  Halsey}.} \bibinfo{year}{1966}\natexlab{b}.
\newblock \showarticletitle{The shortest route through a network with
  time-dependent internodal transit times}.
\newblock \bibinfo{journal}{\emph{J. Math. Anal. Appl.}} \bibinfo{volume}{14},
  \bibinfo{number}{3} (\bibinfo{year}{1966}), \bibinfo{pages}{493--498}.
\newblock
\showISSN{0022-247X}
\urldef\tempurl%
\url{https://doi.org/10.1016/0022-247X(66)90009-6}
\showDOI{\tempurl}


\bibitem[\protect\citeauthoryear{Dehne, Omran, and Sack}{Dehne
  et~al\mbox{.}}{2012}]%
        {DBLP:journals/algorithmica/DehneOS12}
\bibfield{author}{\bibinfo{person}{Frank Dehne}, \bibinfo{person}{Masoud~T.
  Omran}, {and} \bibinfo{person}{J{\"{o}}rg{-}R{\"{u}}diger Sack}.}
  \bibinfo{year}{2012}\natexlab{}.
\newblock \showarticletitle{Shortest Paths in Time-Dependent {FIFO} Networks}.
\newblock \bibinfo{journal}{\emph{Algorithmica}} \bibinfo{volume}{62},
  \bibinfo{number}{1-2} (\bibinfo{year}{2012}), \bibinfo{pages}{416--435}.
\newblock
\urldef\tempurl%
\url{https://doi.org/10.1007/s00453-010-9461-6}
\showDOI{\tempurl}


\bibitem[\protect\citeauthoryear{Delling}{Delling}{2008}]%
        {DBLP:conf/esa/Delling08}
\bibfield{author}{\bibinfo{person}{Daniel Delling}.}
  \bibinfo{year}{2008}\natexlab{}.
\newblock \showarticletitle{Time-Dependent SHARC-Routing}. In
  \bibinfo{booktitle}{\emph{Algorithms - {ESA} 2008, 16th Annual European
  Symposium, Karlsruhe, Germany, September 15-17, 2008. Proceedings}}
  \emph{(\bibinfo{series}{Lecture Notes in Computer Science})},
  \bibfield{editor}{\bibinfo{person}{Dan Halperin} {and} \bibinfo{person}{Kurt
  Mehlhorn}} (Eds.), Vol.~\bibinfo{volume}{5193}.
  \bibinfo{publisher}{Springer}, \bibinfo{pages}{332--343}.
\newblock
\urldef\tempurl%
\url{https://doi.org/10.1007/978-3-540-87744-8\_28}
\showDOI{\tempurl}


\bibitem[\protect\citeauthoryear{Demiryurek, Kashani, Shahabi, and
  Ranganathan}{Demiryurek et~al\mbox{.}}{2011}]%
        {DBLP:conf/ssd/DemiryurekKSR11}
\bibfield{author}{\bibinfo{person}{Ugur Demiryurek},
  \bibinfo{person}{Farnoush~Banaei Kashani}, \bibinfo{person}{Cyrus Shahabi},
  {and} \bibinfo{person}{Anand Ranganathan}.} \bibinfo{year}{2011}\natexlab{}.
\newblock \showarticletitle{Online Computation of Fastest Path in
  Time-Dependent Spatial Networks}. In \bibinfo{booktitle}{\emph{Advances in
  Spatial and Temporal Databases - 12th International Symposium, {SSTD} 2011,
  Minneapolis, MN, USA, August 24-26, 2011, Proceedings}}
  \emph{(\bibinfo{series}{Lecture Notes in Computer Science})},
  \bibfield{editor}{\bibinfo{person}{Dieter Pfoser}, \bibinfo{person}{Yufei
  Tao}, \bibinfo{person}{Kyriakos Mouratidis}, \bibinfo{person}{Mario~A.
  Nascimento}, \bibinfo{person}{Mohamed~F. Mokbel}, \bibinfo{person}{Shashi
  Shekhar}, {and} \bibinfo{person}{Yan Huang}} (Eds.),
  Vol.~\bibinfo{volume}{6849}. \bibinfo{publisher}{Springer},
  \bibinfo{pages}{92--111}.
\newblock
\urldef\tempurl%
\url{https://doi.org/10.1007/978-3-642-22922-0\_7}
\showDOI{\tempurl}


\bibitem[\protect\citeauthoryear{Ding, Yu, and Qin}{Ding et~al\mbox{.}}{2008}]%
        {nonindex2}
\bibfield{author}{\bibinfo{person}{Bolin Ding}, \bibinfo{person}{Jeffrey~Xu
  Yu}, {and} \bibinfo{person}{Lu Qin}.} \bibinfo{year}{2008}\natexlab{}.
\newblock \showarticletitle{Finding Time-Dependent Shortest Paths over Large
  Graphs}. In \bibinfo{booktitle}{\emph{Proceedings of the 11th International
  Conference on Extending Database Technology: Advances in Database
  Technology}} (Nantes, France) \emph{(\bibinfo{series}{EDBT '08})}.
  \bibinfo{publisher}{Association for Computing Machinery},
  \bibinfo{address}{New York, NY, USA}, \bibinfo{pages}{205–216}.
\newblock
\showISBNx{9781595939265}
\urldef\tempurl%
\url{https://doi.org/10.1145/1353343.1353371}
\showDOI{\tempurl}


\bibitem[\protect\citeauthoryear{Du, Tong, Zhou, Tao, and Zhou}{Du
  et~al\mbox{.}}{2018}]%
        {DBLP:conf/kdd/DuTZTZ18}
\bibfield{author}{\bibinfo{person}{Bowen Du}, \bibinfo{person}{Yongxin Tong},
  \bibinfo{person}{Zimu Zhou}, \bibinfo{person}{Qian Tao}, {and}
  \bibinfo{person}{Wenjun Zhou}.} \bibinfo{year}{2018}\natexlab{}.
\newblock \showarticletitle{Demand-Aware Charger Planning for Electric Vehicle
  Sharing}. In \bibinfo{booktitle}{\emph{{KDD}}}. \bibinfo{pages}{1330--1338}.
\newblock


\bibitem[\protect\citeauthoryear{Gao, Tong, She, Song, Chen, and Xu}{Gao
  et~al\mbox{.}}{2016}]%
        {DBLP:conf/waim/GaoTSSCX16}
\bibfield{author}{\bibinfo{person}{Dawei Gao}, \bibinfo{person}{Yongxin Tong},
  \bibinfo{person}{Jieying She}, \bibinfo{person}{Tianshu Song},
  \bibinfo{person}{Lei Chen}, {and} \bibinfo{person}{Ke Xu}.}
  \bibinfo{year}{2016}\natexlab{}.
\newblock \showarticletitle{Top-k Team Recommendation in Spatial
  Crowdsourcing}. In \bibinfo{booktitle}{\emph{{WAIM}}}.
  \bibinfo{pages}{191--204}.
\newblock


\bibitem[\protect\citeauthoryear{Huang, Wang, Zhao, and Li}{Huang
  et~al\mbox{.}}{2021}]%
        {ICDE2021}
\bibfield{author}{\bibinfo{person}{Shuai Huang}, \bibinfo{person}{Yong Wang},
  \bibinfo{person}{Tianyu Zhao}, {and} \bibinfo{person}{Guoliang Li}.}
  \bibinfo{year}{2021}\natexlab{}.
\newblock \showarticletitle{A Learning-based Method for Computing Shortest Path
  Distances on Road Networks}. In \bibinfo{booktitle}{\emph{2021 IEEE 37th
  International Conference on Data Engineering (ICDE)}}.
  \bibinfo{pages}{360--371}.
\newblock
\urldef\tempurl%
\url{https://doi.org/10.1109/ICDE51399.2021.00038}
\showDOI{\tempurl}


\bibitem[\protect\citeauthoryear{Kanoulas, Du, Xia, and Zhang}{Kanoulas
  et~al\mbox{.}}{2006}]%
        {DBLP:conf/icde/KanoulasDXZ06}
\bibfield{author}{\bibinfo{person}{Evangelos Kanoulas}, \bibinfo{person}{Yang
  Du}, \bibinfo{person}{Tian Xia}, {and} \bibinfo{person}{Donghui Zhang}.}
  \bibinfo{year}{2006}\natexlab{}.
\newblock \showarticletitle{Finding Fastest Paths on {A} Road Network with
  Speed Patterns}. In \bibinfo{booktitle}{\emph{Proceedings of the 22nd
  International Conference on Data Engineering, {ICDE} 2006, 3-8 April 2006,
  Atlanta, GA, {USA}}}, \bibfield{editor}{\bibinfo{person}{Ling Liu},
  \bibinfo{person}{Andreas Reuter}, \bibinfo{person}{Kyu{-}Young Whang}, {and}
  \bibinfo{person}{Jianjun Zhang}} (Eds.). \bibinfo{publisher}{{IEEE} Computer
  Society}, \bibinfo{pages}{10}.
\newblock
\urldef\tempurl%
\url{https://doi.org/10.1109/ICDE.2006.71}
\showDOI{\tempurl}


\bibitem[\protect\citeauthoryear{Kaufman and Smith}{Kaufman and Smith}{1993}]%
        {nonindex4}
\bibfield{author}{\bibinfo{person}{David~E Kaufman} {and}
  \bibinfo{person}{Robert~L Smith}.} \bibinfo{year}{1993}\natexlab{}.
\newblock \showarticletitle{Fastest paths in time-dependent networks for
  intelligent vehicle-highway systems application}.
\newblock \bibinfo{journal}{\emph{Journal of Intelligent Transportation
  Systems}} \bibinfo{volume}{1}, \bibinfo{number}{1} (\bibinfo{year}{1993}),
  \bibinfo{pages}{1--11}.
\newblock


\bibitem[\protect\citeauthoryear{Li, Ni, He, Li, Xia, and Zhou}{Li
  et~al\mbox{.}}{2022}]%
        {TD-KNN}
\bibfield{author}{\bibinfo{person}{Jiajia Li}, \bibinfo{person}{Cancan Ni},
  \bibinfo{person}{Dan He}, \bibinfo{person}{Lei Li}, \bibinfo{person}{Xiufeng
  Xia}, {and} \bibinfo{person}{Xiaofang Zhou}.}
  \bibinfo{year}{2022}\natexlab{}.
\newblock \showarticletitle{Efficient kNN query for moving objects on
  time-dependent road networks}.
\newblock \bibinfo{journal}{\emph{The VLDB Journal}} (\bibinfo{year}{2022}).
\newblock
\showISSN{0196-5212}
\urldef\tempurl%
\url{https://doi.org/10.1007/s00778-022-00758-w}
\showDOI{\tempurl}


\bibitem[\protect\citeauthoryear{Li, Zhang, Hua, and Zhou}{Li
  et~al\mbox{.}}{2020}]%
        {ICDE2020}
\bibfield{author}{\bibinfo{person}{Lei Li}, \bibinfo{person}{Mengxuan Zhang},
  \bibinfo{person}{Wen Hua}, {and} \bibinfo{person}{Xiaofang Zhou}.}
  \bibinfo{year}{2020}\natexlab{}.
\newblock \showarticletitle{Fast Query Decomposition for Batch Shortest Path
  Processing in Road Networks}. In \bibinfo{booktitle}{\emph{2020 IEEE 36th
  International Conference on Data Engineering (ICDE)}}.
  \bibinfo{pages}{1189--1200}.
\newblock
\urldef\tempurl%
\url{https://doi.org/10.1109/ICDE48307.2020.00107}
\showDOI{\tempurl}


\bibitem[\protect\citeauthoryear{Li, Chen, and Wang}{Li et~al\mbox{.}}{2019}]%
        {GstarTree}
\bibfield{author}{\bibinfo{person}{Zijian Li}, \bibinfo{person}{Lei Chen},
  {and} \bibinfo{person}{Yue Wang}.} \bibinfo{year}{2019}\natexlab{}.
\newblock \showarticletitle{G*-Tree: An Efficient Spatial Index on Road
  Networks}. In \bibinfo{booktitle}{\emph{35th {IEEE} International Conference
  on Data Engineering, {ICDE} 2019, Macao, China, April 8-11, 2019}}.
  \bibinfo{publisher}{{IEEE}}, \bibinfo{pages}{268--279}.
\newblock
\urldef\tempurl%
\url{https://doi.org/10.1109/ICDE.2019.00032}
\showDOI{\tempurl}


\bibitem[\protect\citeauthoryear{Nannicini, Delling, Schultes, and
  Liberti}{Nannicini et~al\mbox{.}}{2012}]%
        {DBLP:journals/networks/NanniciniDSL12}
\bibfield{author}{\bibinfo{person}{Giacomo Nannicini}, \bibinfo{person}{Daniel
  Delling}, \bibinfo{person}{Dominik Schultes}, {and} \bibinfo{person}{Leo
  Liberti}.} \bibinfo{year}{2012}\natexlab{}.
\newblock \showarticletitle{Bidirectional \emph{A}* search on time-dependent
  road networks}.
\newblock \bibinfo{journal}{\emph{Networks}} \bibinfo{volume}{59},
  \bibinfo{number}{2} (\bibinfo{year}{2012}), \bibinfo{pages}{240--251}.
\newblock
\urldef\tempurl%
\url{https://doi.org/10.1002/net.20438}
\showDOI{\tempurl}


\bibitem[\protect\citeauthoryear{Ouyang, Qin, Chang, Lin, Zhang, and
  Zhu}{Ouyang et~al\mbox{.}}{2018}]%
        {h2h}
\bibfield{author}{\bibinfo{person}{Dian Ouyang}, \bibinfo{person}{Lu Qin},
  \bibinfo{person}{Lijun Chang}, \bibinfo{person}{Xuemin Lin},
  \bibinfo{person}{Ying Zhang}, {and} \bibinfo{person}{Qing Zhu}.}
  \bibinfo{year}{2018}\natexlab{}.
\newblock \showarticletitle{When Hierarchy Meets 2-Hop-Labeling: Efficient
  Shortest Distance Queries on Road Networks}. In
  \bibinfo{booktitle}{\emph{Proceedings of the 2018 International Conference on
  Management of Data}} (Houston, TX, USA) \emph{(\bibinfo{series}{SIGMOD
  '18})}. \bibinfo{publisher}{Association for Computing Machinery},
  \bibinfo{address}{New York, NY, USA}, \bibinfo{pages}{709–724}.
\newblock
\showISBNx{9781450347037}
\urldef\tempurl%
\url{https://doi.org/10.1145/3183713.3196913}
\showDOI{\tempurl}


\bibitem[\protect\citeauthoryear{Qiao, Cheng, Chang, and Yu}{Qiao
  et~al\mbox{.}}{2012}]%
        {ICDE2012}
\bibfield{author}{\bibinfo{person}{Miao Qiao}, \bibinfo{person}{Hong Cheng},
  \bibinfo{person}{Lijun Chang}, {and} \bibinfo{person}{Jeffrey~Xu Yu}.}
  \bibinfo{year}{2012}\natexlab{}.
\newblock \showarticletitle{Approximate Shortest Distance Computing: A
  Query-Dependent Local Landmark Scheme}. In \bibinfo{booktitle}{\emph{2012
  IEEE 28th International Conference on Data Engineering (ICDE)}}.
  \bibinfo{pages}{462--473}.
\newblock
\urldef\tempurl%
\url{https://doi.org/10.1109/ICDE.2012.53}
\showDOI{\tempurl}


\bibitem[\protect\citeauthoryear{Qiu, Wen, Qin, Li, Li, and Zhang}{Qiu
  et~al\mbox{.}}{2022}]%
        {tlindex}
\bibfield{author}{\bibinfo{person}{Yu{-}Xuan Qiu}, \bibinfo{person}{Dong Wen},
  \bibinfo{person}{Lu Qin}, \bibinfo{person}{Wentao Li},
  \bibinfo{person}{Ronghua Li}, {and} \bibinfo{person}{Ying Zhang}.}
  \bibinfo{year}{2022}\natexlab{}.
\newblock \showarticletitle{Efficient Shortest Path Counting on Large Road
  Networks}.
\newblock \bibinfo{journal}{\emph{Proc. {VLDB} Endow.}} \bibinfo{volume}{15},
  \bibinfo{number}{10} (\bibinfo{year}{2022}), \bibinfo{pages}{2098--2110}.
\newblock
\urldef\tempurl%
\url{https://www.vldb.org/pvldb/vol15/p2098-qiu.pdf}
\showURL{%
\tempurl}


\bibitem[\protect\citeauthoryear{Robertson and Seymour}{Robertson and
  Seymour}{1984}]%
        {treedecomposition}
\bibfield{author}{\bibinfo{person}{Neil Robertson} {and}
  \bibinfo{person}{Paul~D. Seymour}.} \bibinfo{year}{1984}\natexlab{}.
\newblock \showarticletitle{Graph minors. {III.} Planar tree-width}.
\newblock \bibinfo{journal}{\emph{J. Comb. Theory, Ser. {B}}}
  \bibinfo{volume}{36}, \bibinfo{number}{1} (\bibinfo{year}{1984}),
  \bibinfo{pages}{49--64}.
\newblock
\urldef\tempurl%
\url{https://doi.org/10.1016/0095-8956(84)90013-3}
\showDOI{\tempurl}


\bibitem[\protect\citeauthoryear{She, Tong, Chen, and Song}{She
  et~al\mbox{.}}{2017}]%
        {DBLP:conf/sigmod/SheT0S17}
\bibfield{author}{\bibinfo{person}{Jieying She}, \bibinfo{person}{Yongxin
  Tong}, \bibinfo{person}{Lei Chen}, {and} \bibinfo{person}{Tianshu Song}.}
  \bibinfo{year}{2017}\natexlab{}.
\newblock \showarticletitle{Feedback-Aware Social Event-Participant
  Arrangement}. In \bibinfo{booktitle}{\emph{{SIGMOD}}}.
  \bibinfo{pages}{851--865}.
\newblock


\bibitem[\protect\citeauthoryear{Tong, Yuan, Cheng, Chen, and Wang}{Tong
  et~al\mbox{.}}{2017}]%
        {TongJoS17}
\bibfield{author}{\bibinfo{person}{Yongxin Tong}, \bibinfo{person}{Ye Yuan},
  \bibinfo{person}{Yurong Cheng}, \bibinfo{person}{Lei Chen}, {and}
  \bibinfo{person}{Guoren Wang}.} \bibinfo{year}{2017}\natexlab{}.
\newblock \showarticletitle{Survey on spatiotemporal crowdsourced data
  management techniques}.
\newblock \bibinfo{journal}{\emph{{J. Softw.}}} \bibinfo{volume}{28},
  \bibinfo{number}{1} (\bibinfo{year}{2017}), \bibinfo{pages}{35--58}.
\newblock


\bibitem[\protect\citeauthoryear{Tong, Zeng, Zhou, Chen, and Xu}{Tong
  et~al\mbox{.}}{2022}]%
        {DBLP:journals/tods/TongZZCX22}
\bibfield{author}{\bibinfo{person}{Yongxin Tong}, \bibinfo{person}{Yuxiang
  Zeng}, \bibinfo{person}{Zimu Zhou}, \bibinfo{person}{Lei Chen}, {and}
  \bibinfo{person}{Ke Xu}.} \bibinfo{year}{2022}\natexlab{}.
\newblock \showarticletitle{Unified Route Planning for Shared Mobility: An
  Insertion-based Framework}.
\newblock \bibinfo{journal}{\emph{{ACM} Trans. Database Syst.}}
  \bibinfo{volume}{47}, \bibinfo{number}{1} (\bibinfo{year}{2022}),
  \bibinfo{pages}{2:1--2:48}.
\newblock


\bibitem[\protect\citeauthoryear{Tong, Zeng, Zhou, Chen, Ye, and Xu}{Tong
  et~al\mbox{.}}{2018}]%
        {DBLP:journals/pvldb/TongZZCYX18}
\bibfield{author}{\bibinfo{person}{Yongxin Tong}, \bibinfo{person}{Yuxiang
  Zeng}, \bibinfo{person}{Zimu Zhou}, \bibinfo{person}{Lei Chen},
  \bibinfo{person}{Jieping Ye}, {and} \bibinfo{person}{Ke Xu}.}
  \bibinfo{year}{2018}\natexlab{}.
\newblock \showarticletitle{A Unified Approach to Route Planning for Shared
  Mobility}.
\newblock \bibinfo{journal}{\emph{Proc. {VLDB} Endow.}} \bibinfo{volume}{11},
  \bibinfo{number}{11} (\bibinfo{year}{2018}), \bibinfo{pages}{1633--1646}.
\newblock


\bibitem[\protect\citeauthoryear{Wang, Li, and Tang}{Wang
  et~al\mbox{.}}{2019}]%
        {shortestquery}
\bibfield{author}{\bibinfo{person}{Yong Wang}, \bibinfo{person}{Guoliang Li},
  {and} \bibinfo{person}{Nan Tang}.} \bibinfo{year}{2019}\natexlab{}.
\newblock \showarticletitle{Querying Shortest Paths on Time Dependent Road
  Networks}.
\newblock \bibinfo{journal}{\emph{Proc. VLDB Endow.}} \bibinfo{volume}{12},
  \bibinfo{number}{11} (\bibinfo{date}{jul} \bibinfo{year}{2019}),
  \bibinfo{pages}{1249–1261}.
\newblock
\showISSN{2150-8097}
\urldef\tempurl%
\url{https://doi.org/10.14778/3342263.3342265}
\showDOI{\tempurl}


\bibitem[\protect\citeauthoryear{Wang, Yuan, Wang, Zhou, Mu, and Wang}{Wang
  et~al\mbox{.}}{2021}]%
        {yuanyeicde2021}
\bibfield{author}{\bibinfo{person}{Yishu Wang}, \bibinfo{person}{Ye Yuan},
  \bibinfo{person}{Hao Wang}, \bibinfo{person}{Xiangmin Zhou},
  \bibinfo{person}{Congcong Mu}, {and} \bibinfo{person}{Guoren Wang}.}
  \bibinfo{year}{2021}\natexlab{}.
\newblock \showarticletitle{Constrained Route Planning over Large Multi-Modal
  Time-Dependent Networks}. In \bibinfo{booktitle}{\emph{2021 IEEE 37th
  International Conference on Data Engineering (ICDE)}}.
  \bibinfo{publisher}{IEEE Computer Society}, \bibinfo{address}{Los Alamitos,
  CA, USA}, \bibinfo{pages}{313--324}.
\newblock
\urldef\tempurl%
\url{https://doi.org/10.1109/ICDE51399.2021.00034}
\showDOI{\tempurl}


\bibitem[\protect\citeauthoryear{Zeng, Tong, Song, and Chen}{Zeng
  et~al\mbox{.}}{2020}]%
        {DBLP:journals/pvldb/ZengTSC20}
\bibfield{author}{\bibinfo{person}{Yuxiang Zeng}, \bibinfo{person}{Yongxin
  Tong}, \bibinfo{person}{Yuguang Song}, {and} \bibinfo{person}{Lei Chen}.}
  \bibinfo{year}{2020}\natexlab{}.
\newblock \showarticletitle{The Simpler The Better: An Indexing Approach for
  Shared-Route Planning Queries}.
\newblock \bibinfo{journal}{\emph{Proc. {VLDB} Endow.}} \bibinfo{volume}{13},
  \bibinfo{number}{13} (\bibinfo{year}{2020}), \bibinfo{pages}{3517--3530}.
\newblock


\bibitem[\protect\citeauthoryear{Zhang, Yuan, Li, Qin, and Zhang}{Zhang
  et~al\mbox{.}}{2021}]%
        {lsd}
\bibfield{author}{\bibinfo{person}{Junhua Zhang}, \bibinfo{person}{Long Yuan},
  \bibinfo{person}{Wentao Li}, \bibinfo{person}{Lu Qin}, {and}
  \bibinfo{person}{Ying Zhang}.} \bibinfo{year}{2021}\natexlab{}.
\newblock \showarticletitle{Efficient Label-Constrained Shortest Path Queries
  on Road Networks: A Tree Decomposition Approach}.
\newblock \bibinfo{journal}{\emph{Proc. VLDB Endow.}} \bibinfo{volume}{15},
  \bibinfo{number}{3} (\bibinfo{date}{nov} \bibinfo{year}{2021}),
  \bibinfo{pages}{686–698}.
\newblock
\showISSN{2150-8097}
\urldef\tempurl%
\url{https://doi.org/10.14778/3494124.3494148}
\showDOI{\tempurl}


\bibitem[\protect\citeauthoryear{Zhong, Li, Tan, Zhou, and Gong}{Zhong
  et~al\mbox{.}}{2015}]%
        {GTree}
\bibfield{author}{\bibinfo{person}{Ruicheng Zhong}, \bibinfo{person}{Guoliang
  Li}, \bibinfo{person}{Kian-Lee Tan}, \bibinfo{person}{Lizhu Zhou}, {and}
  \bibinfo{person}{Zhiguo Gong}.} \bibinfo{year}{2015}\natexlab{}.
\newblock \showarticletitle{G-Tree: An Efficient and Scalable Index for Spatial
  Search on Road Networks}.
\newblock \bibinfo{journal}{\emph{IEEE Transactions on Knowledge and Data
  Engineering}} \bibinfo{volume}{27}, \bibinfo{number}{8}
  (\bibinfo{year}{2015}), \bibinfo{pages}{2175--2189}.
\newblock
\urldef\tempurl%
\url{https://doi.org/10.1109/TKDE.2015.2399306}
\showDOI{\tempurl}


\bibitem[\protect\citeauthoryear{Zhu, Ma, Xiao, Luo, Tang, and Zhou}{Zhu
  et~al\mbox{.}}{2013}]%
        {SIGMOD2013}
\bibfield{author}{\bibinfo{person}{Andy~Diwen Zhu}, \bibinfo{person}{Hui Ma},
  \bibinfo{person}{Xiaokui Xiao}, \bibinfo{person}{Siqiang Luo},
  \bibinfo{person}{Youze Tang}, {and} \bibinfo{person}{Shuigeng Zhou}.}
  \bibinfo{year}{2013}\natexlab{}.
\newblock \showarticletitle{Shortest Path and Distance Queries on Road
  Networks: Towards Bridging Theory and Practice}. In
  \bibinfo{booktitle}{\emph{Proceedings of the 2013 ACM SIGMOD International
  Conference on Management of Data}} (New York, New York, USA)
  \emph{(\bibinfo{series}{SIGMOD '13})}. \bibinfo{publisher}{Association for
  Computing Machinery}, \bibinfo{address}{New York, NY, USA},
  \bibinfo{pages}{857–868}.
\newblock
\showISBNx{9781450320375}
\urldef\tempurl%
\url{https://doi.org/10.1145/2463676.2465277}
\showDOI{\tempurl}


\end{thebibliography}

\end{document}